\documentclass[letterpaper,twocolumn,10pt]{article}
\usepackage{zhanggroup}

\usepackage{hyperref}
\usepackage{tikz}
\usepackage{amsfonts}
\usepackage{xspace} 
\usepackage{mathrsfs}
\usepackage{amsmath}
\usepackage{amsthm}
\usepackage{subcaption}
\usepackage{booktabs}
\usepackage{multirow}
\usepackage{graphicx} 
\usepackage{caption,subcaption}
\usepackage{makecell}
\usepackage{bbm}
\usepackage{float}
\hypersetup{
  colorlinks,
  linkcolor={blue!70!green},
  citecolor={green!70!blue},
  urlcolor={orange!70!red}
}

\newcommand{\mypara}[1]{\noindent{\bf {#1}.}\xspace}
\newcommand{\mpre}{$\mathcal{M}$\xspace}
\newcommand{\samplef}{$s_{\mathcal{P}}(\mathcal{D}, \Phi, N)$\xspace}
\newcommand{\dt}{$\mathcal{D}_{\textit{target}}$\xspace}

\newcommand{\ds}{$\mathcal{D}_{\textit{shadow}}$\xspace}
\newcommand{\dv}{$\mathcal{D}_{\textit{validation}}$\xspace}
\newcommand{\dstr}{$\mathcal{D}_{\textit{shadow}}^{\textit{train}}$\xspace}
\newcommand{\dste}{$\mathcal{D}_{\textit{shadow}}^{\textit{test}}$\xspace}
\newcommand{\dttr}{$\mathcal{D}_{\textit{target}}^{\textit{train}}$\xspace}
\newcommand{\dtte}{$\mathcal{D}_{\textit{target}}^{\textit{test}}$\xspace}
\newcommand{\Deltat}{$\Delta_{\textit{target}}$\xspace}
\newcommand{\Deltas}{$\Delta_{\textit{shadow}}$\xspace}
\newcommand{\deltat}{$\delta_{\textit{target}}$\xspace}
\newcommand{\deltas}{$\delta_{\textit{shadow}}$\xspace}
\newcommand{\trad}{NN-based\xspace}
\newcommand{\metric}{metric-based\xspace}
\newcommand{\grad}{gradient-based\xspace}
\newcommand{\tones}{$T_1^{\textit{size}}$\xspace}
\newcommand{\ttwos}{$T_2^{\textit{size}}$\xspace}
\newcommand{\ttwog}{$T_2^{\textit{male}}$\xspace}
\newcommand{\ttwoa}{$T_2^{\textit{youth}}$\xspace}
\newcommand{\tthrees}{$T_3^{\textit{size}}$\xspace}
\newcommand{\tthreeg}{$T_3^{\textit{male}}$\xspace}
\newcommand{\tthreea}{$T_3^{\textit{youth}}$\xspace}
\newcommand{\tfours}{$T_4^{\textit{size}}$\xspace}
\newcommand{\tfourg}{$T_4^{\textit{male}}$\xspace}

\begin{document}

\title{Quantifying Privacy Risks of Prompts in Visual Prompt Learning}

\author{
\rm Yixin Wu\textsuperscript{1}\ \
Rui Wen\textsuperscript{1}\ \
Michael Backes\textsuperscript{1}\ \
Pascal Berrang\textsuperscript{2}\ \
Mathias Humbert\textsuperscript{3}\ \
Yun Shen\textsuperscript{4}\ \
Yang Zhang\textsuperscript{1}
\\
\\
\textsuperscript{1}\textit{CISPA Helmholtz Center for Information Security}\ \ \ 
\\
\textsuperscript{2}\textit{University of Birmingham}\ \ \ 
\textsuperscript{3}\textit{University of Lausanne}\ \ \ 
\textsuperscript{4}\textit{Netapp}\ \ \
}

\date{}

\maketitle

\begin{abstract}

Large-scale pre-trained models are increasingly adapted to downstream tasks through a new paradigm called prompt learning.
In contrast to fine-tuning, prompt learning does not update the pre-trained model's parameters.
Instead, it only learns an input perturbation, namely prompt, to be added to the downstream task data for predictions.
Given the fast development of prompt learning, a well-generalized prompt inevitably becomes a valuable asset as significant effort and proprietary data are used to create it.
This naturally raises the question of whether a prompt may leak the proprietary information of its training data.
In this paper, we perform the first comprehensive privacy assessment of prompts learned by visual prompt learning through the lens of property inference and membership inference attacks.
Our empirical evaluation shows that the prompts are vulnerable to both attacks.
We also demonstrate that the adversary can mount a successful property inference attack with limited cost.
Moreover, we show that membership inference attacks against prompts can be successful with relaxed adversarial assumptions.
We further make some initial investigations on the defenses and observe that our method can mitigate the membership inference attacks with a decent utility-defense trade-off but fails to defend against property inference attacks.
We hope our results can shed light on the privacy risks of the popular prompt learning paradigm.
To facilitate the research in this direction, we will share our code and models with the community.\footnote{\url{https://github.com/yxoh/prompt_leak_usenix2024/}.}

\end{abstract}

\section{Introduction}
\label{section:introduction}

Recent research has provided ample evidence that increasing the size of machine learning (ML) models, i.e., the number of parameters, is a pivotal factor in enhancing their overall performance~\cite{DCLT19,BMRSKDNSSAAHKHCRZWWHCSLGCCBMRSA20,RSRLNMZLL20}.
One of the commonly employed strategies for adapting such large-scale pre-trained ML models to downstream tasks is fine-tuning~\cite{ZQDXZZXH19}, which updates model parameters for specific downstream tasks via back-propagation.
Fine-tuning, however, suffers from two main drawbacks.
First, it leads to high computational costs because all model parameters need to be updated.
In addition, it is storage inefficient since a separate copy of the fine-tuned model needs to be stored for each downstream task.

\begin{figure}[!t]
\centering
\includegraphics[width=\columnwidth]{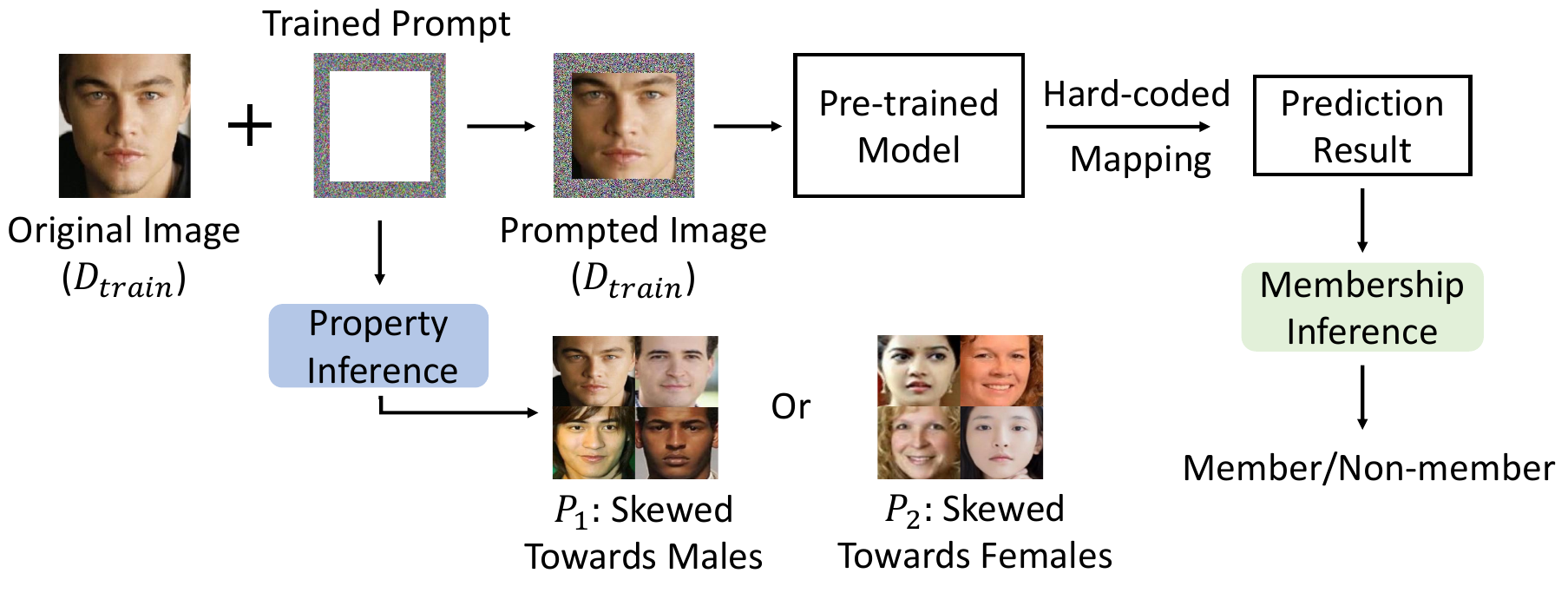}
\caption{Overview of prompt usage and inference attacks.
The prompt is a pixel patch.
The prompted image is an original image with an added prompt.
Property inference infers sensitive properties of the target prompt's training dataset that the PaaS provider does not intend to disclose.
Membership inference infers whether a given sample was in the target prompt's training dataset.}
\label{figure:overview}
\end{figure}

In order to address these limitations, researchers have proposed prompt learning as an alternative to fine-tuning~\cite{LAC21,BJSI22,LSPC23,ZLCWL23,HDCZWHY23,BGDGE22,LL21,RM21,LYFJHN23}.
Prompt learning involves learning an input perturbation, referred to as a \emph{prompt}, that enables shifting downstream task data to the original data distribution.
The pre-trained model generates a task-specific output based on this prompt.
It is important to note that, during prompt learning, the pre-trained model remains frozen, leading to a significant decrease in the number of learned parameters compared to fine-tuning (see~\autoref{section:preliminaries}).
In recent years, prompt learning has been extensively validated and shown to be effective in the domains of computer vision (CV)~\cite{LSPC23, ZLCWL23, HDCZWHY23, BGDGE22, BJSI22} and natural language processing (NLP)~\cite{ LAC21, LL21, RM21, LYFJHN23}.
It is expected that prompt as a service (PaaS) will gain popularity.\footnote{\url{https://twitter.com/AndrewYNg/status/1650938079027548160}.} 
In this scenario, a user can request a prompt for a downstream task from the PaaS provider without the need for arduous fine-tuning.
The user then combines their data with the prompt and inputs them into the pre-trained model to obtain the predictions, as depicted in~\autoref{figure:overview}.
In this way, the user can run the pre-trained model and keep their data on-premise, while the PaaS provider can reuse a single pre-trained model to support multiple downstream tasks.
These benefits differentiate PaaS from machine learning as a service (MLaaS)~\cite{RGC15}.
As a result, a well-generalized prompt becomes a valuable asset for PaaS providers, as they invest significant efforts and use proprietary data to develop it.

Previous research has demonstrated that ML models are vulnerable to various privacy attacks, such as property inference attacks~\cite{GWYGB18,ZCSZ22} and membership inference attacks~\cite{SSSS17,SZHBFB19,LZ21}, which can disclose sensitive information about the training data used to create the models.
Such data leakage can severely damage the provider's privacy as well as intellectual property.
However, to the best of our knowledge, previous research about such privacy risks has focused on ML models at the model level and has not yet been explored on prompts at the input level.
As the number of learned parameters is significantly reduced in prompt learning, it is natural to assume that this paradigm would compress the proprietary information of its training data, leading to less effective privacy attacks (see~\autoref{section:diff}).
This motivates us to investigate whether a prompt also leaks the proprietary information of its training data that the PaaS provider does not intend to disclose, especially when such prompts are generated from images containing sensitive private information.

\mypara{Contributions}
In this paper, we conduct the first privacy risk assessment of prompts learned by prompt learning.
We focus on prompt learning for image classification tasks~\cite{BJSI22}, which represents one of the most promising directions in computer vision research~\cite{LSPC23, ZLCWL23, HDCZWHY23, BGDGE22, BJSI22, KKR23, XWCZLWZ23, ZLZHL22, CLYCL22}.
Our primary objective is to determine \emph{to what extent a visual prompt possesses the potential to disclose confidential information.} 
Specifically, we perform property inference and membership inference, two dominant privacy attacks against ML models~\cite{CAOJTU23,GWYGB18,MGC22,SSSS17}, where the former aims to deduce sensitive properties of the dataset used to train the target prompt, and the latter determines if a given data sample is part of the target prompt's training dataset.
We adopt the existing attack methodologies for property inference~\cite{GWYGB18, AMSVVF15} and membership inference (neural network-based attacks~\cite{SSSS17, SZHBFB19}, \metric attacks~\cite{SM21}, and \grad attacks~\cite{LF20, NSH19}).
Note that our goal is not to develop new property inference attacks or membership inference attacks against prompts.
Instead, we aim to use existing methods with well-established threat models to systematically assess the privacy risks of prompt learning.
The overview of our study is depicted in~\autoref{figure:overview}.

The empirical evaluation shows prompts are susceptible to property inference attacks across multiple datasets and pre-trained models.
For example, we can achieve at least 81\% accuracy in inferring different target properties from prompts learned for CelebA~\cite{LLWT15}.
Moreover, when inferring the training dataset size of the prompt, we can achieve 100\% test accuracy in all cases.
We also conduct a cost analysis to show that the adversary can either train the shadow prompts for fewer epochs or use fewer shadow prompts to minimize their cost while maintaining decent attack performance.

Our study also provides empirical evidence that membership inference poses a practical threat to prompts.
The experimental results demonstrate that existing attack methodologies are effective across a range of datasets and pre-trained models.
In particular, the \metric attack with modified prediction entropy is the most effective one, e.g., achieving 93\% membership inference accuracy on the AFAD dataset~\cite{NZWGH16}.
The \grad attacks follow closely behind and outperform the neural network-based (\trad) attacks.
We further investigate factors that may affect membership inference from both the victim's and the adversary's perspectives.
Specifically, from the victim's side, we conduct a detailed analysis of the relationship between the overfitting levels of prompts and attack success~\cite{YGFJ18}.
The results indicate that the attack success is positively correlated with the overfitting level.
Moreover, excessive training epochs and inadequate training data increase overfitting levels, exacerbating the privacy threat posed by membership inference attacks.
From an adversarial perspective, we demonstrate that the adversary can relax the assumption that the shadow dataset has the same distribution as the target prompt's training dataset.
This finding further exemplifies the membership privacy risks of prompts learned by prompt learning.

We also conduct preliminary investigations into mitigating privacy risks associated with prompt learning.
In particular, we explore the effectiveness of adding Gaussian noise to prompts, as proposed in prior research~\cite{ZLH19,ZLDLZY23,HRF19}.
Our experiments demonstrate that there exists a decent utility-defense trade-off when mitigating both naive and adaptive membership inference attacks.
However, when defending against property inference attacks, we need higher Gaussian noise to reduce the attack performance, leading to unacceptable utility deterioration.
Our findings indicate that the statistical information of the training dataset in the target prompts is harder to hide than individual information, i.e., membership.
Our results highlight the need for further research into more effective defense mechanisms for mitigating property inference attacks in prompt learning.

\mypara{Impact}
This study presents an exploration of the privacy risks associated with prompt learning, an emerging machine-learning paradigm.
Our investigation represents the first of its kind in this area.
Our findings indicate that prompts learned through prompt learning are susceptible to privacy breaches.
We hope our study will increase the awareness of the stakeholders when deploying prompt learning in real-world applications.
Moreover, to facilitate research in the field, we will share our code and models.

\section{Preliminaries}
\label{section:preliminaries}

\subsection{Prompt Learning}
\label{section:prompt_learning}

\mypara{Overview}
Prompt learning is a new machine-learning paradigm introduced to address the limitations of fine-tuning~\cite{LAC21,BJSI22,LSPC23,ZLCWL23,HDCZWHY23,BGDGE22,LL21,RM21,LYFJHN23}.
It aims at learning a task-specific prompt that can be added to the input data while keeping the pre-trained model's parameters frozen.
With this new paradigm, the service provider can share the same pre-trained model across various downstream tasks with different prompts in a space- and computation-efficient manner.
In this paper, we focus on prompt learning in computer vision, i.e., \emph{visual prompt learning (VPL)}~\cite{BJSI22}.
It is generally composed of two stages: input transformation and output transformation.

\begin{figure}[!t]
\centering
\includegraphics[width=\columnwidth]{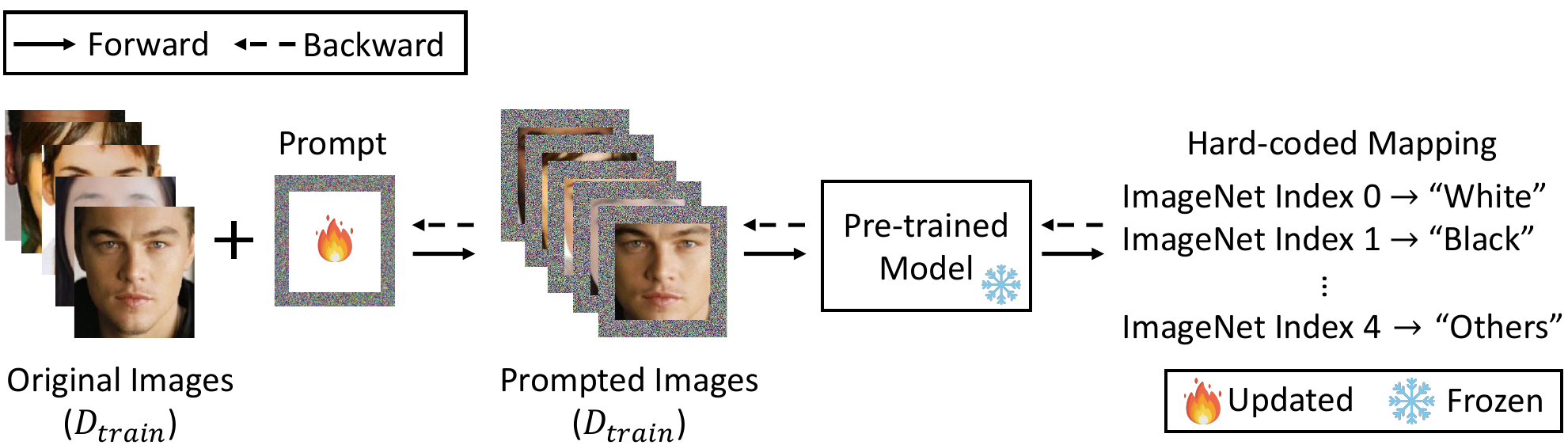}
\caption{Overview of visual prompt learning (VPL).
We learn an input prompt via back-propagation~\cite{BJSI22} at the input transformation stage.
We apply hard-coded mapping~\cite{EGS19} to map the pre-trained model's outputs into the target labels at the output transformation stage.}
\label{figure:train_overview}
\end{figure}

\mypara{Input Transformation}
As shown in~\autoref{figure:train_overview}, the goal of input transformation is to learn an input prompt $\delta$ in the pixel space, \textit{i.e., in the form of a single image}, via back-propagation.
Given a dataset $\mathcal{D} = (\mathcal{X}, \mathcal{Y})$, a pre-trained model \mpre parameterized by $\omega$, and a prompt $\delta$ parameterized by $\theta$, the prompt generation process $q(\mathcal{D},\mathcal{M})$ uses~\autoref{equation:prompt_tuning} to maximize the likelihood of $\mathcal{Y}$:
\begin{equation}
\label{equation:prompt_tuning}
\max\limits_{\theta} P_{\theta;\omega} (\mathcal{Y}|\mathcal{X} + \delta_{\theta}),
\end{equation}
where the prompt parameters $\theta$ are learned via back-propagation and the model parameters $\omega$ are frozen.
Note that the prompt can be any visual template chosen by the users, e.g., padding~\cite{BJSI22}.
At inference time, the learned prompt $\delta$ is added to each test image $x$ to specify the task.

\mypara{Output Transformation}
Usually, the pre-trained model has a different number of classes from the downstream tasks.
To accomplish the downstream task, the prompt owner supplies a label mapping scheme $\tau$ to map the model's outputs into the target labels.
As shown in~\autoref{figure:train_overview}, a commonly used scheme is hard-coded mapping~\cite{EGS19}.
It consists of mapping the first $n$ pre-trained model class indices to the downstream class indices, where $n$ is the number of classes in the downstream task.
The unassigned pre-trained classes are left out for the loss computation.
We rely on hard-coded mapping due to its simplicity and proven effectiveness~\cite{BJSI22}.

\subsection{Prompt Learning vs.\ Fine-Tuning}
\label{section:diff}

\mypara{Training Time} 
The fine-tuning paradigm updates all parameters of the pre-trained model via back-propagation.
However, as shown in~\autoref{figure:train_overview},
VPL learns a visual prompt, \textit{i.e., in the form of a single image}, on the training dataset $\mathcal{D_{\textit{train}}} = (\mathcal{X}, \mathcal{Y})$.
During the back-propagation, the pre-trained model is frozen, and only the parameters of the visual prompt are updated.
In this way, prompt learning dramatically lowers the bar for users adapting large-scale vision models for real-world applications.
Prompt learning saves significant training resources and storage space, especially when a pre-trained model serves multiple downstream tasks.
For example, the Vision Transformer (ViT-B)~\cite{KBZPYGH20} we use in later experiments has 86,567,656 parameters, and the visual prompt, i.e., a padding template with a prompt size of 30, has 69,840 parameters.
For each downstream task, the fine-tuning paradigm updates the entire model (86,567,656 parameters), whereas, in prompt learning, a single prompt, i.e., a single image with only 69,840 parameters, is updated.
The number of parameters updated by prompt learning is only 0.08\% of those of fine-tuning, so it is natural to assume that prompt learning would heavily compress the training dataset information, leading to less effective privacy attacks.
However, we show that the prompts are still susceptible to two privacy attacks in later experiments.

\mypara{Inference Time}
As shown in~\autoref{figure:overview}, both the trained prompt and pre-trained models are involved in the inference process.
Given a test image $x$, the user gets the prompted image, i.e., adding the trained prompt $\delta$ to $x$, and then feeds the prompted image into the pre-trained model \mpre to get the prediction result.
In the fine-tuning approach, the user directly feeds the given test image $x$ into the fine-tuned model to get the prediction result.

\subsection{Application Scenario}

Taking a medical researcher as an example, they aim to classify CT images for COVID-19 diagnosis.
Instead of hiring a computer vision expert to fine-tune a model, the researcher can request a prompt from a PaaS provider.
They can either opt for a publicly available pre-trained model or allow the PaaS provider to select a suitable one for them.
The provider uses their proprietary data, e.g., CT images with explicit consent, to learn a customized prompt and return it to them.
At inference time, the researcher simply combines their testing data with the prompt and feeds them to the pre-trained model to get the predictions.
In this way, users minimize their effort in developing a well-generalized prompt and keep their data on-premise, while the PaaS provider can reuse a single pre-trained model to support multiple downstream tasks.
Meanwhile, the user can adapt to different tasks, e.g., clinical decision support, by trivially switching to different prompts.
These benefits differentiate PaaS from machine learning as a service (MLaaS)~\cite{RGC15}.

\section{Property Inference Attacks}
\label{section:pia}

We first measure the privacy risks of prompt learning through the lens of property inference attacks.
Our objective here is not to devise novel attacks for prompts but rather to leverage well-established threat models and existing techniques to gauge the privacy implications of prompts.

\subsection{Threat Model}
\label{section:pia_threat_model}

\mypara{Attack Scenario}
The PaaS provider is a resourceful entity that uses a pre-trained model \mpre and their private dataset \dt to create well-generalized prompts $\Delta$ for downstream tasks.
The adversary can be any legitimate user of this PaaS provider and can obtain a prompt $\delta$ for a target downstream task together with the white-box access to \mpre.
Note that a pre-defined label mapping $\tau$ is also provided by the PaaS provider (see~\autoref{section:prompt_learning}).
The adversary runs the target downstream task locally and does not interact with the PaaS provider.

\mypara{Adversary's Goal}
Given a target prompt \deltat, the goal of the adversary is to infer confidential macro-level properties of the training dataset \dt, which the PaaS provider does not intend to share.
Taking a prompt \deltat for facial recognition as an example, the adversary may intend to infer the confidential properties of the private dataset \dt, such as the proportion of males and the proportion of youth.
The adversary considers such confidential properties as targets, causing real-world harm to the PasS provider, e.g., reputation damages, if the adversary can infer that certain classes of people, such as minorities, are underrepresented in the training data~\cite{GWYGB18}.
For simplicity, we focus on binary properties, such as if the proportion of males in the training dataset is 30\% or 70\%, in most of our experiments, following previous work~\cite{GWYGB18}.
We later show that our attack can be generalized to properties with multiple choices (see~\autoref{section:pia_results}).

\mypara{Adversary's Knowledge and Capability}
We assume that the adversary has white-box access to the pre-trained model \mpre and the label mapping $\tau$.
Note that the white-box access in the paper is more restricted than conventional white-box access, as the latter can retrieve all information about the model, such as model parameters and intermediate outputs.
In this paper, the adversary only needs to know the architecture and version of the pre-trained model from the PaaS provider, and such knowledge is often disclosed by the PaaS provider for marketing purposes.
We also assume that the adversary has a shadow dataset \ds of similar distribution as \dt.
For instance, in our evaluation (see~\autoref{section:pia_setting}), we select both \ds and \dt from the same dataset CelebA~\cite{LLWT15}.
These two subsets are disjoint and may have different statistical properties, such as gender/race/age ratios.
We emphasize that previous property inference attacks also make the same assumption~\cite{GWYGB18, ZCSZ22}.

\subsection{Measurement Methodology}
\label{section:pia_method}

\mypara{Shadow Prompt Generation}
Given a shadow dataset \ds and associated data properties $\mathcal{P}=\{p^1, ..., p^k\}$, the adversary uses~\autoref{equation:shadow_prompt_generation} to generate the shadow prompts:
\begin{equation}
\label{equation:shadow_prompt_generation}
\Delta_{\textit{shadow}} = \{ q(s_{\mathcal{P}}(\mathcal{D}_{\textit{shadow}}, \Phi_i, N_i), \mathcal{M})  \}_{i=1}^m,   
\end{equation}
where $s_{\mathcal{P}}$ denotes a sampling function that samples $N_i$ data points from $\mathcal{D}_{\textit{shadow}}$ without replacement  and the distribution of sampled data with properties $\mathcal{P}$ satisfying the conditions $\Phi_i$.
Note that, $\Phi_i= \{\phi^1_i, \phi^2_i, ..., \phi^k_i\}$, $\phi^k_i$ is the actual value of $p^k$ in round $i$, $m$ denotes the number of shadow prompts, and $N_i$ denotes the size of the sampled dataset from \ds in round $i$.
In previous work~\cite{GWYGB18,ZCSZ22,MGC22}, apart from the targeted property, say $p^1$ (and associated $\phi^1$), they tend to use a fixed set of other conditions, i.e., $\{\phi^2, ..., \phi^k\}$, and $N$.
For example, the target property is the proportion of males.
They tend to keep the training data size the same when training all target prompts and shadow prompts in their evaluation.
However, the training data size of target prompts and shadow prompts is likely to be different in a realistic scenario.
If the target prompt is trained on 500 samples with 70\% males, but shadow prompts are trained on 2000 samples with 70\% males.
Such discrepancies in training dataset sizes, e.g., 500 and 2000, may influence the attack performance.
In contrast to those approaches, we consider a \emph{mixed setting} by design.
As we can see in~\autoref{equation:shadow_prompt_generation}, in every round $i$, we generate a prompt $\delta_i$ from a subset sampled by $s_{\mathcal{P}}(\mathcal{D}_{\textit{shadow}}, \Phi_i, N_i)$ with properties $\mathcal{P}$ satisfying different $\Phi_i$.
For instance, given $\mathcal{P} = \{\textit{youth}, \textit{male}\}$, $\Phi=\{70\%, 70\%\}$, and $N=2000$, \samplef samples 2000 data points from $\mathcal{D}$ to train the prompt.
Among them, 980 data points are \textit{young males}, 420 data points are \textit{young females}, 420 data points are \textit{old males}, and 180 data points are \textit{old females}.
As such, our approach can guarantee a more realistic shadow prompt generation and fairer evaluation.

\mypara{Attack Model Training}
After obtaining the shadow prompts \Deltas, we can build the attack model for each property $p^k$:
\begin{equation}
\mathcal{A}: \Delta_{\textit{shadow}} \rightarrow y^k.
\end{equation}
We train the attack model $\mathcal{A}$ by optimizing the following loss function:
\begin{equation}
\mathcal{L} [\mathcal{A} (\Delta_{\textit{shadow}}), y^k ],
\end{equation}
where $\mathcal{L}$ is a cross-entropy loss function in this paper.
Concretely, the attack model takes $\delta_i \in \Delta_{\textit{shadow}}$ as input.
To incorporate the input size of the attack model, we use zero value to pad it to an image of size 224 $\times$ 224, with RGB channels.
This means the attack model is an image classifier.
The adversary then treats the corresponding condition value $\phi^k_i$ of the target property $p^k$ as the class labels $y^k_i$.
To infer the target property $p^k$ of \dt, the adversary queries the attack model $\mathcal{A}$ with \deltat and obtains the corresponding prediction result, i.e., the exact condition value of $p^k$.

\begin{table*}[!t]
\centering
\caption{Experimental settings of the property inference attacks with the corresponding attack performance.}
\label{table:pia_setting_results}
\renewcommand{\arraystretch}{1.1}
\scalebox{0.6}{
\begin{tabular}{c|c|c|c|c|ccc}
\toprule
Inference & \multirow{2}{*}{Dataset} & Downstream & Target & Inference & \multicolumn{3}{c}{Test Accuracy} \\
Task & & Task & Property & Labels & RN18 & BiT-M & ViT-B \\
\midrule
$T_1$ & CIFAR10 & Image Classification & Size (\tones) & \{500, 2000\} & 100.00 & 100.00 & 100.00 \\
\midrule
$T_2$ & CelebA & Multi-Atrribute Classification & \begin{tabular}[c]{@{}c@{}}Size (\ttwos)\\ Proportion of Males (\ttwog)\\ Proportion of Youth (\ttwoa)\end{tabular} & \begin{tabular}[c]{@{}c@{}}\{500, 2000\}\\ \{30\%, 70\%\}\\ \{30\%, 70\%\}\end{tabular} & \begin{tabular}[c]{@{}c@{}} 100.00 \\ 99.75 \\93.00 \\ \end{tabular} & \begin{tabular}[c]{@{}c@{}} 100.00 \\ 99.25 \\90.75 \\ \end{tabular}  &
\begin{tabular}[c]{@{}c@{}} 100.00 \\ 93.00 \\81.00 \\ \end{tabular}\\
\midrule 
$T_3$ & UTKFace & Race Classification & \begin{tabular}[c]{@{}c@{}}Size (\tthrees)\\ Proportion of Males (\tthreeg)\\ Proportion of Youth (\tthreea)\end{tabular} & \begin{tabular}[c]{@{}c@{}}\{500, 2000\}\\ \{30\%, 70\%\}\\ \{30\%, 70\%\}\end{tabular} & \begin{tabular}[c]{@{}c@{}} 100.00 \\ 80.50 \\81.75 \\ \end{tabular} & \begin{tabular}[c]{@{}c@{}} 100.00 \\ 80.50 \\87.50 \\ \end{tabular} & \begin{tabular}[c]{@{}c@{}} 100.00 \\ 82.00 \\84.00 \\ \end{tabular} \\
\midrule
$T_4$ & AFAD & Age Classification & \begin{tabular}[c]{@{}c@{}}Size (\tfours)\\ Proportion of Males (\tfourg)\end{tabular} & \begin{tabular}[c]{@{}c@{}}\{500, 2000\}\\ \{30\%, 70\%\}\end{tabular} & \begin{tabular}[c]{@{}c@{}} 100.00 \\ 80.75 \\ \end{tabular} & \begin{tabular}[c]{@{}c@{}} 100.00 \\ 78.00 \\ \end{tabular} & \begin{tabular}[c]{@{}c@{}} 100.00 \\ 72.25 \\ \end{tabular} \\
\bottomrule
\end{tabular}}
\end{table*}

\subsection{Measurement Settings}
\label{section:pia_setting}

\mypara{Datasets and Downstream Tasks}
We use four datasets in our study, including CIFAR10~\cite{CIFAR}, CelebA~\cite{LLWT15}, UTKFace~\cite{ZSQ17}, and AFAD~\cite{NZWGH16}.
These datasets contain sensitive properties (the proportion of males, the proportion of youth, etc.) and are widely used to evaluate the performance of property inference attacks~\cite{GWYGB18, ZCSZ22}.
The introduction of these datasets and corresponding downstream tasks are as follows.

\begin{itemize}
\item \textbf{CIFAR10} is a benchmark dataset for image classification that contains 60K images in 10 classes.
In this paper, the downstream task is a 10-class image classification.
\item \textbf{CelebA} is a large-scale facial attribute dataset containing more than 200K facial images with 40 binary attributes.
We pick three attributes, including \textit{MouthSlightlyOpen}, \textit{Attractive}, and \textit{WearingLipstick}, and use their combinations to create an 8-class attribute classification as the downstream task.
\item \textbf{UTKFace} has about 23K facial images.
Each image has three attributes: \textit{gender}, \textit{race}, and \textit{age}.
We consider race classification, i.e., White, Black, Asian, Indian, and Others, as the downstream task.
\item \textbf{AFAD} is short for Asian Face Age Dataset.
It contains more than 160K facial images, each with \textit{age} and \textit{gender} attributes.
In this paper, we consider age classification as the downstream task.
Specifically, we divide the values of \textit{age} attribute into five bins: $15 \leq \textit{age} < 20$, $20 \leq \textit{age} < 25$, $25 \leq \textit{age} < 30$, $30 \leq \textit{age} < 35$, and $35 \leq \textit{age} < 40$, leading to a 5-class image classification.
\end{itemize}

\mypara{Property Inference Task Configurations}
For each task, we split the dataset into three disjoint subsets \dt, \ds, and \dv in the ratio of $0.475: 0.475: 0.05$.
\dt and \ds are used to develop the target prompt set \Deltat and shadow prompt set \Deltas, respectively.
We evaluate the utility of all prompts on \dv.
We train 2000 shadow prompts to construct the attack training dataset and 400 target prompts to build the attack testing dataset in our experiments.
Our property inference targets include training dataset size, proportion of males, and proportion of youth.
Note that recent research demonstrates that the size of the training dataset significantly affects the performance of the model, necessitating substantial efforts to identify the optimal values~\cite{ZVFR16,LHL07}.
Consequently, we also view the training dataset size as confidential information and as one of our inference objectives.
We outline the details of all inference tasks below and summarize them in~\autoref{table:pia_setting_results}.
\begin{itemize}
\item \textbf{Inference Task on CIFAR10 ($T_1$).}
For CIFAR10, we only consider the size of the prompt training dataset $N$ as the property inference target (\tones).
We focus on two training data sizes, i.e., $y^1 \in \{500, 2000\}$, and run the sampling function (see~\autoref{equation:shadow_prompt_generation}) 1000 times on \ds to generate 1000 shadow prompts for each training data size.
Meanwhile, we generate 200 target prompts in the same manner for each training data size.
\item \textbf{Inference Task on CelebA ($T_2$).}
For CelebA, we consider the size of the prompt training dataset (\ttwos), the proportion of males (\ttwog), and the proportion of youth (\ttwoa) of the data samples used to train the target prompts as the property inference targets.
\ttwog is based on the \textit{male} attribute, and \ttwoa is based on the \textit{young} attribute.
Both attributes are binary.
The inference labels of each property are: $y^1 \in \{500, 2000\}$, $y^2 \in \{30\%, 70\%\}$, and $y^3 \in \{30\%, 70\%\}$.
Recall that we consider a mixed data sample strategy.
Given these three properties, we end up with eight sampling functions in total.
We run each sampling function 250 times on \ds and 50 times on \dt to generate the shadow prompt set \Deltas and target prompt set \Deltat, respectively.
\item \textbf{Inference Task on UTKFace ($T_3$).}
For UTKFace, we also consider the size of the prompt training dataset (\tthrees), the proportion of males (\tthreeg), and the proportion of youth (\tthreea) as the property inference targets.
Note that \tthreeg is based on the \textit{gender} attribute, and \tthreea is based on the \textit{age} attribute.
Specifically, we use the median of \textit{age} values from all images, i.e., 30, as the threshold.
We then label samples with $0 \leq \textit{age} \leq 30$ as Young and $30 \leq \textit{age} \leq 116$ as Old.
The inference labels of each property are the same as those of CelebA.
Thus, we follow the same sampling settings as those of $T_2$ to generate the shadow and target prompts.
\item \textbf{Inference Task on AFAD ($T_4$).}
For AFAD, we consider the size of the prompt training dataset (\tfours) and the proportion of males (\tfourg) as the property inference targets.
\tfourg is based on the \textit{gender} attribute.
The inference labels of each property are: $y^1 \in \{500, 2000\}$ and $y^2 \in \{30\%, 70\%\}$.
We use four sampling functions to generate the shadow and target prompts.
We run each sampling function 500 times on \ds and 100 times on \dt to generate the shadow prompt set \Deltas and target prompt set \Deltat, respectively.
\end{itemize}

\mypara{Metric}
As the attack training/testing dataset is balanced in terms of class distribution, we use test accuracy as the main metric to evaluate the prompt utility and the property inference attacks.

\mypara{Pre-trained Models}
We select three representative vision models in our experiments, including ResNet-18 (RN18)~\cite{HZRS16}, Big Transfer (BiT-M)~\cite{KBZPYGH20}, and Vision Transformer (ViT-B)~\cite{DBKWZUDMHGUH21}.
More details can be found in~\autoref{appendix:detail_models}.

\mypara{Prompts}
We follow the default training settings~\cite{BJSI22} to train prompts on the above vision models.
Specifically, we choose the padding template with a prompt size of 30.
The number of parameters for each prompt is calculated as $2 \times C \times p \times (H+W-2p)$, where $p$, $C$, $H$, and $W$ are prompt size, image channels, height, and width, respectively.
All images are resized to 224 $\times$ 224 to match the input of the pre-trained models.
The number of parameters for each prompt is 69,840.
We leverage the same hard-coded mapping method~\cite{BJSI22} to map the first $n$ indices of the pre-trained model's outputs to the target labels, where $n$ is the number of target classes.
We adopt cross-entropy as the loss function and SGD as the optimizer with a learning rate of 40 and the cosine scheduler.
In our property inference attacks, we train all prompts for 50 epochs for efficiency.

\mypara{Attack Models}
We leverage the pre-trained RN18~\cite{HZRS16} as the backbone of the attack model $\mathcal{A}$.
We fit a linear classifier on top of the pre-trained RN18 to infer the property labels.
We employ cross-entropy as the loss function and Adam as the optimizer with a learning rate of 1e-5.
The attack model is trained on the shadow prompt set \Deltas for 100 epochs.

\subsection{Measurement Results}
\label{section:pia_results}

\mypara{Property Inference Privacy Risks}
We report the main results on four datasets and three pre-trained models in~\autoref{table:pia_setting_results}.
In general, we observe that proposed attacks achieve good performance across different pre-trained models and datasets.
For example, on CelebA and RN18, we achieve at least 93.00\% accuracy in inferring target properties.
Furthermore, we achieve maximum performance (100.00\%) on all datasets, considering the size of the prompt training dataset as the target property.
Additionally, we observe that the pre-trained model has a moderate influence on the attack performance.
Specifically, when inferring the proportion of youth on UTKFace (\tthreea in \autoref{table:pia_setting_results}), the test accuracy is 81.75\% on RN18, 87.50\% on BiT-M, and 84.00\% on ViT-B.
Although the test accuracy varies across pre-trained models and datasets, the proposed attacks are generally effective, indicating prompts are indeed vulnerable to property inference attacks.

\mypara{Extension to Multi-Class Property Inference}
In the above experiments, we treat property inference as a binary classification task.
Here, we extend it to multi-class classification and explore if the adversary can infer finer granularity information from the prompts.
To this end, we adjust the condition values of the proportion of males in CelebA to \{10\%, 30\%, 50\%, 70\%, 90\%\} and use RN18 as the pre-trained model.
Note that the condition values of the training dataset size and the proportion of youth remain the same.
In turn, we have 20 sampling functions in total.
We keep the sizes of \Deltas and \Deltat unchanged and run each sampling function 100 times on \ds and 20 times on \dt to generate the shadow prompt set \Deltas and target prompt set \Deltat, respectively.
We further adjust the condition values of the training dataset size in CIFAR10 to \{500, 1000, 1500, 1750, 1800, 2000\} and use RN18 as the pre-trained model to explore the performance of the property inference attack when the range of options for the dataset size is closer together.
To this end, we have 6 sampling functions in total and run each sampling function 400 times on \ds and 80 times on \dt to generate the shadow prompt set \Deltas and target prompt set \Deltat, respectively.
The test accuracy for the proportion of males is 90.25\%, while for training dataset size is 95.40\%, demonstrating that property inference attacks can successfully infer finer granularity information from prompts.

\mypara{Takeaways}
We show that the property inference attacks achieve remarkable performance on diverse datasets and pre-trained models.
Moreover, the proposed attacks can be extended to multi-class classification, providing evidence that the adversary can infer fine granularity property information from prompts.

\begin{figure}[!t]
\centering
\begin{subfigure}{0.4\columnwidth}
\includegraphics[width=\columnwidth]{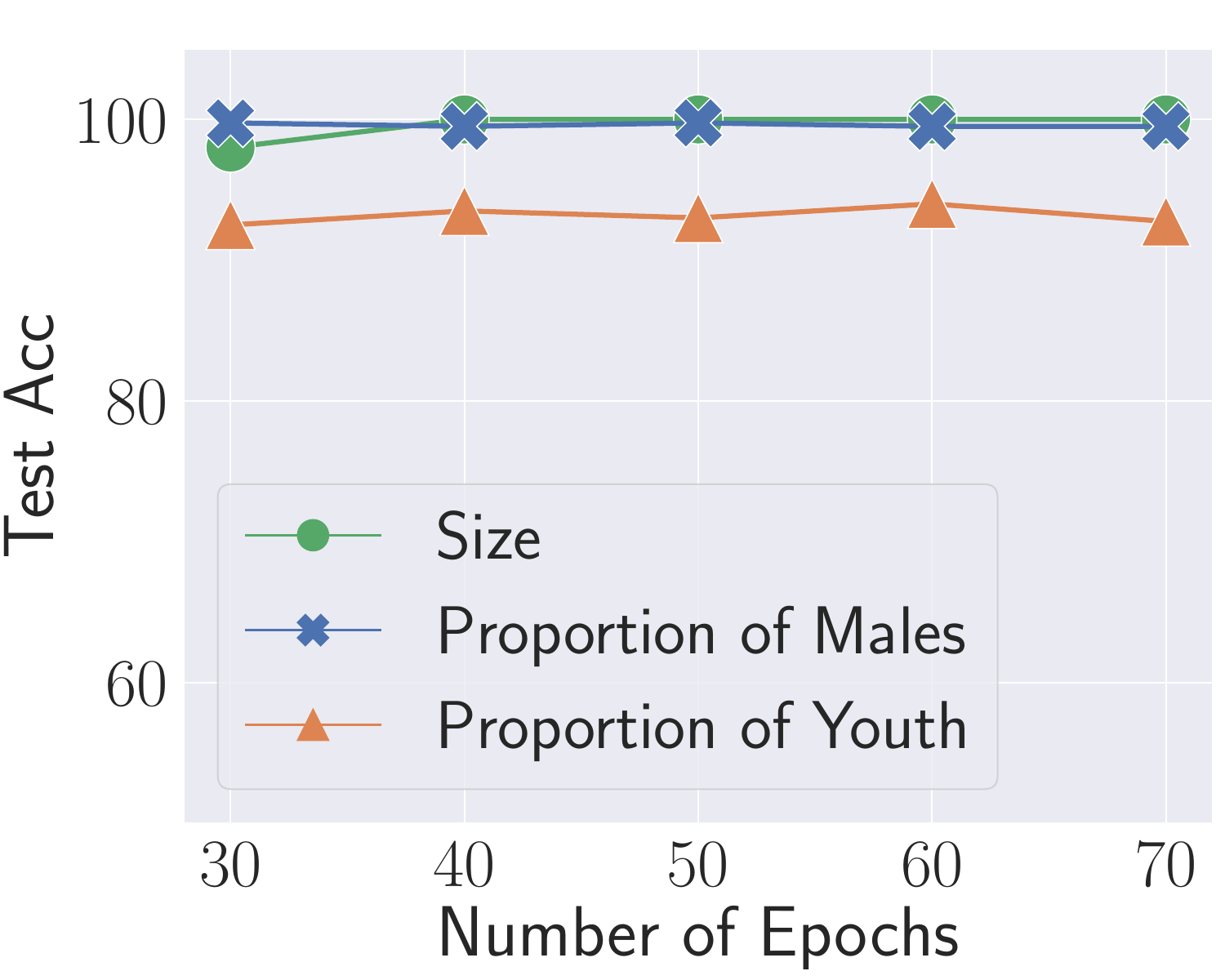}
\caption{Epoch Effect}
\label{figure:ablation_epoch_effect}
\end{subfigure}
\begin{subfigure}{0.4\columnwidth}
\includegraphics[width=\columnwidth]{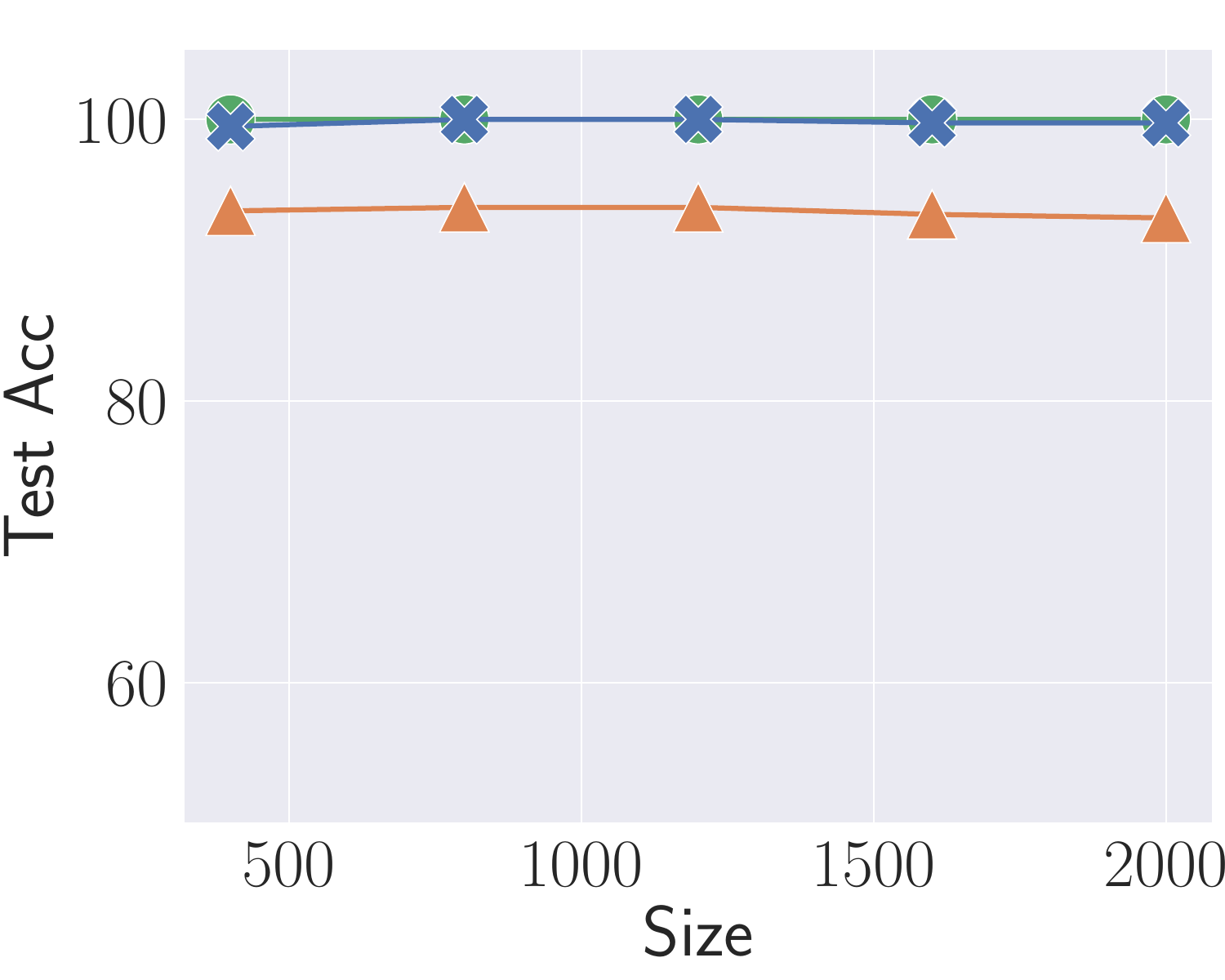}
\caption{Size Effect}
\label{figure:ablation_size_effect}
\end{subfigure}
\caption{Attack performance of the proposed property inference attacks on CelebA with (a) different numbers of epochs for training shadow prompts and (b) different sizes of the attack training dataset, using RN18 as the pre-trained model.}
\label{figure:ablation}
\end{figure}

\begin{figure*}[!t]
\centering
\begin{subfigure}{0.4\columnwidth}
\includegraphics[width=\columnwidth]{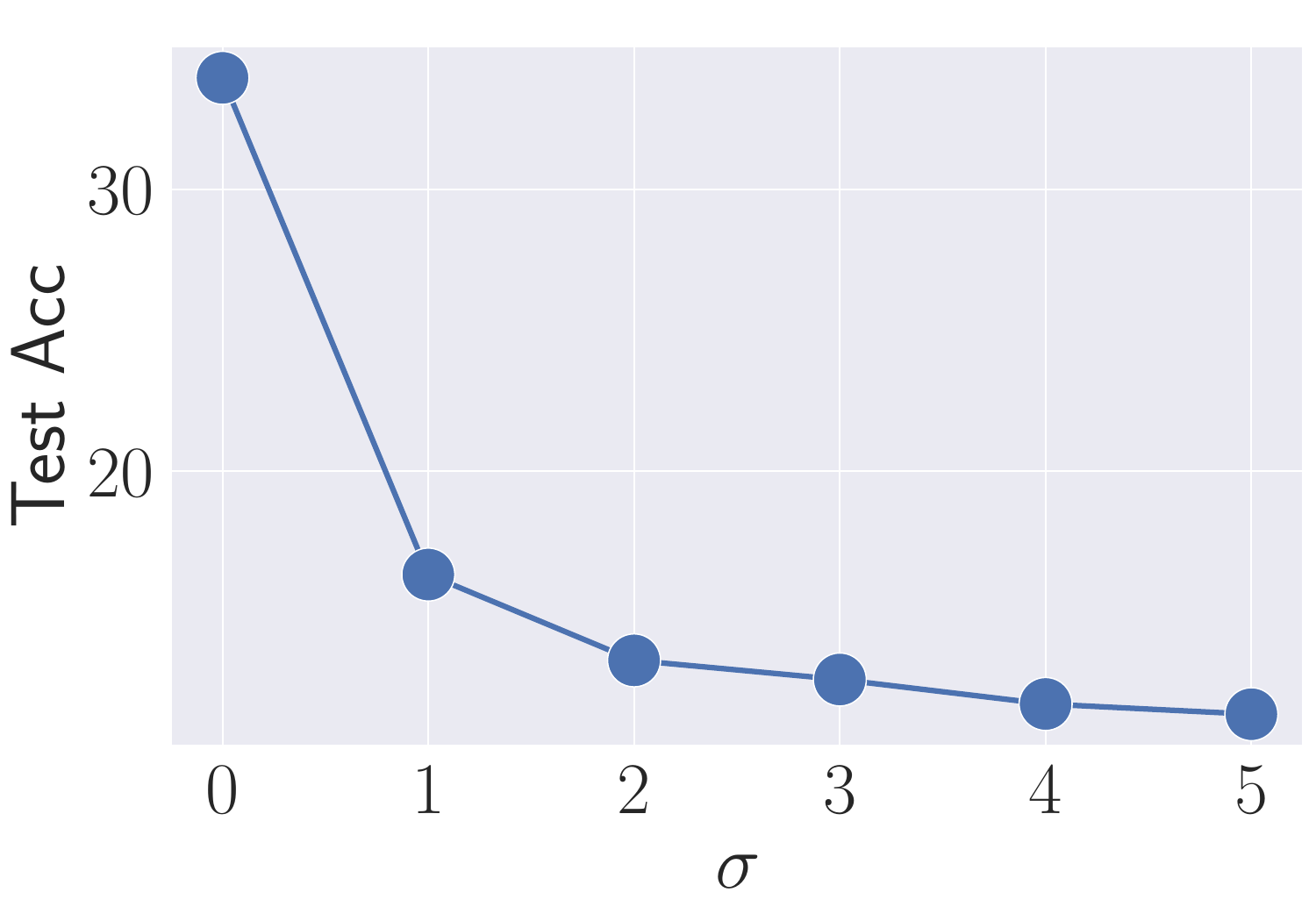}
\caption{CIFAR10}
\label{figure:utility_cifar10}
\end{subfigure}
\begin{subfigure}{0.4\columnwidth}
\includegraphics[width=\columnwidth]{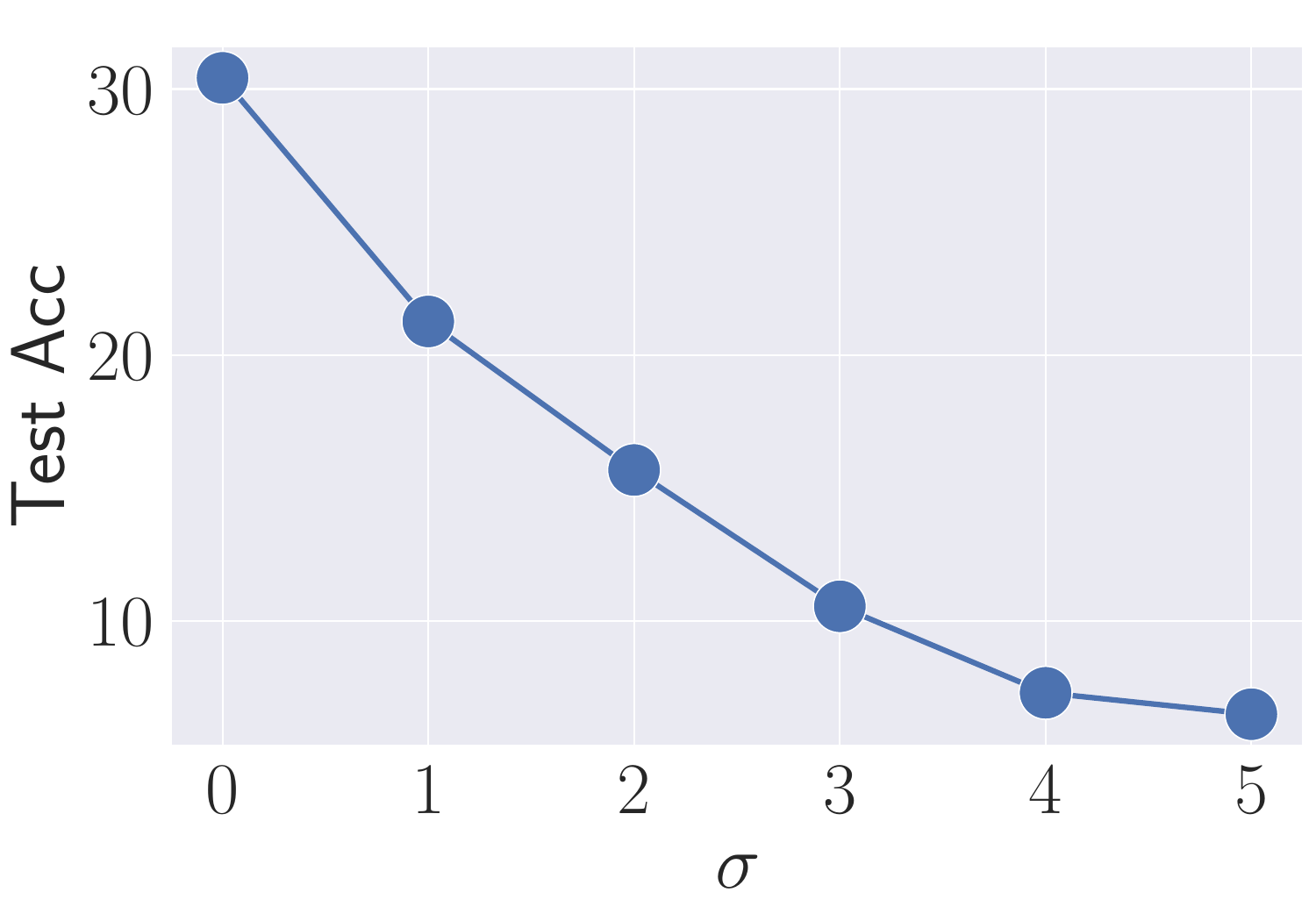}
\caption{CelebA}
\label{figure:utility_celeba}
\end{subfigure}
\begin{subfigure}{0.4\columnwidth}
\includegraphics[width=\columnwidth]{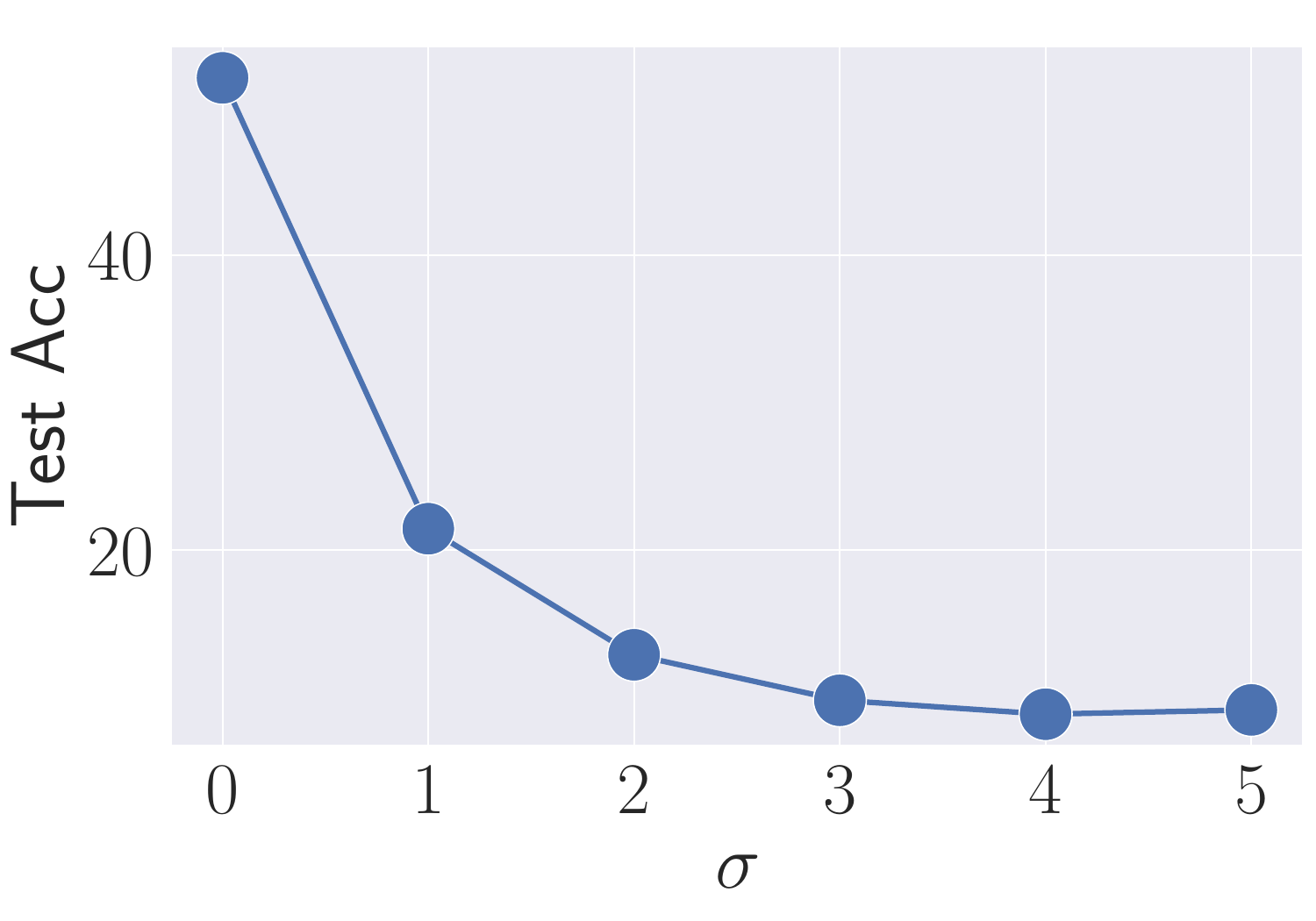}
\caption{UTKFace}
\label{figure:utility_utkface}
\end{subfigure}
\begin{subfigure}{0.4\columnwidth}
\includegraphics[width=\columnwidth]{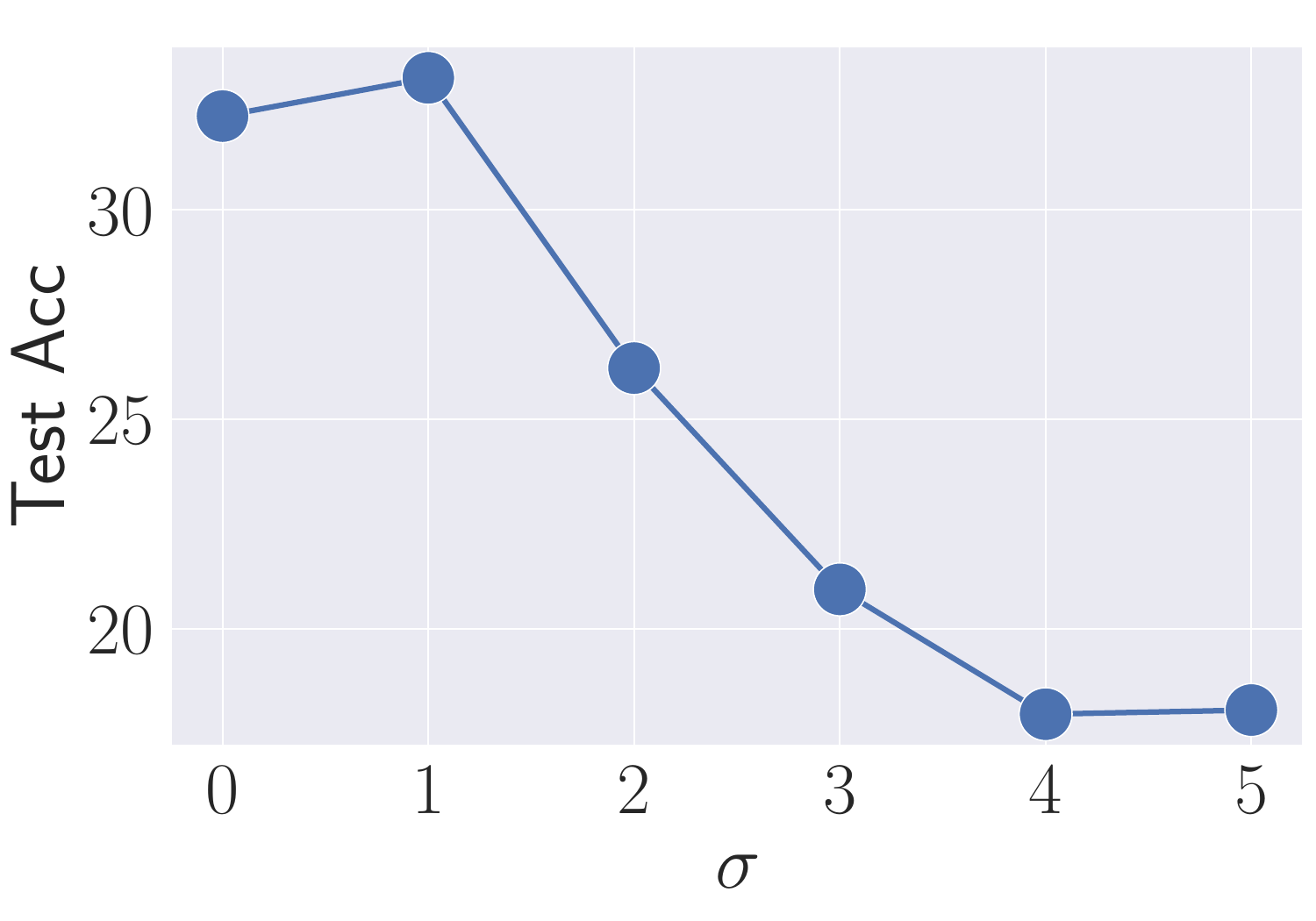}
\caption{AFAD}
\label{figure:utility_afad}
\end{subfigure}
\caption{Target performance on four datasets.
The x-axis denotes the magnitude of the Gaussian noise, from 0 to 5, where 0 means the proposed defense mechanism is not implemented.
The y-axis represents the target performance on the downstream tasks with respect to the average test accuracy of all target prompts in the attack testing dataset.}
\label{figure:pia_utility}
\end{figure*}

\begin{figure*}[!t]
\centering
\begin{subfigure}{0.8\columnwidth}
\includegraphics[width=\columnwidth]{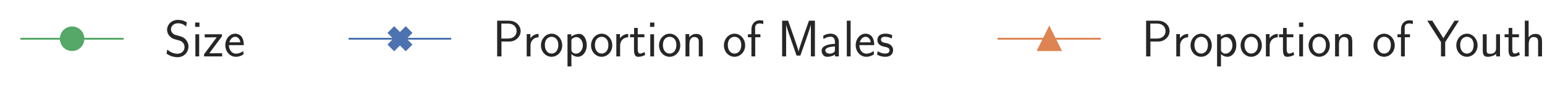}
\end{subfigure}
\begin{subfigure}{1.60\columnwidth}
\includegraphics[width=\columnwidth]{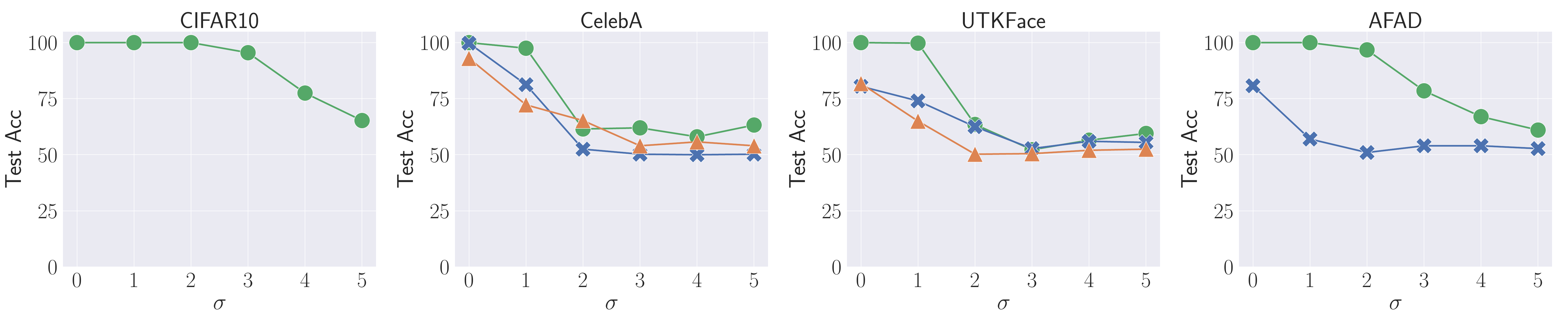}
\caption{Naive Attacks}
\label{figure:defense}
\end{subfigure}
\begin{subfigure}{1.60\columnwidth}
\includegraphics[width=\columnwidth]{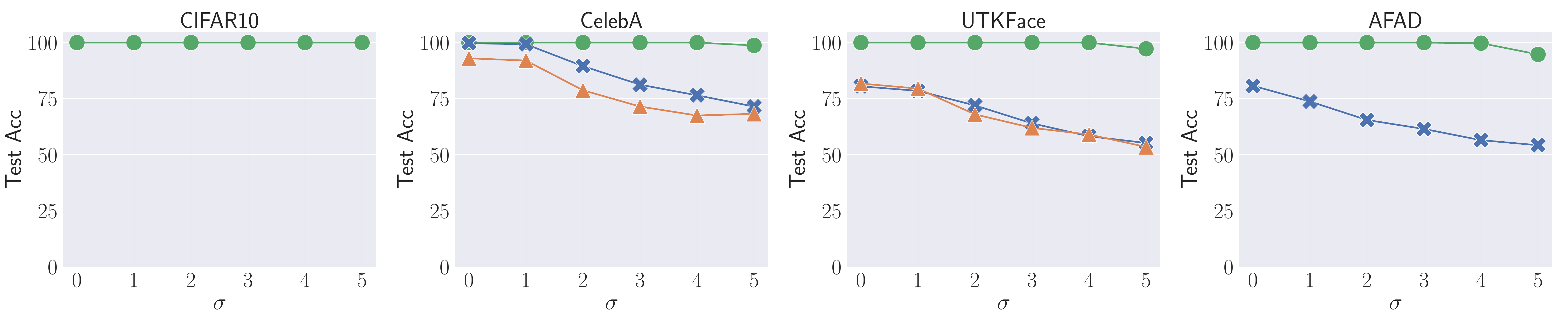}
\caption{Adaptive Attacks}
\label{figure:ad_defense}
\end{subfigure}
\caption{Attack performance of (a) naive attacks where the adversary is not aware of the proposed defense and (b) adaptive attacks on four datasets.
The x-axis denotes the magnitude of Gaussian noise, from 0 to 5, where 0 means the proposed defense mechanism is not implemented.
The y-axis represents the attack performance with respect to the test accuracy.}
\label{figure:pia_defense}
\end{figure*}

\subsection{Factors Affecting Property Inference}

We conduct an empirical analysis to investigate the factors that may influence the performance and cost of property inference attacks on prompts.

\mypara{Number of Epochs}
Previously, we train both shadow prompts and target prompts for 50 epochs.
However, it is likely that the adversary has no knowledge about the number of epochs for target prompts.
Next, we investigate whether the number of epochs in the training process of shadow prompts must match that of the target prompts in order to maintain a strong attack performance.
Concretely, we vary the number of epochs for shadow prompts from 30 to 70 while fixing the number of epochs for target prompts to 50.
The minimum number of epochs is 30 because the prompt starts to outperform the pre-trained model solely on the downstream task at this point.
We show the attack performance on CelebA in~\autoref{figure:ablation_epoch_effect}.
In general, the proposed attacks work similarly well when the number of epochs for target and shadow prompts do not match.
The results also show that the proposed attacks can achieve comparable performance even with fewer epochs, e.g., 30 epochs, to train shadow prompts.
In addition, increasing the number of epochs for shadow prompts does not improve the attack performance.
For example, the test accuracy for inferring the proportion of youth is between 92.50\% and 94.00\% depending on the number of epochs.
This implies that the proposed attacks are robust to variations in the number of epochs for training shadow prompts, making them more practical and efficient.

\mypara{Attack Training Dataset Size}
So far, we have assumed the adversary can rely on an attack training dataset containing 2000 shadow prompts.
However, creating such a dataset costs considerable computational resources.
Hence, we investigate the influence of the attack training dataset size on the attack performance.
Specifically, we randomly sample balanced subsets from the original attack training dataset on CelebA with different sizes \{400, 800, 1200, 1600, 2000\}.
The size of the attack testing dataset remains the same as for the previous experiments, i.e., 400 target prompts.
As shown in~\autoref{figure:ablation_size_effect}, the size of the training dataset only has negligible influence on the attack performance, indicating that a relatively small number of training samples, e.g., 400 shadow prompts, are sufficient to launch the property inference attacks against prompts.
This finding implies that the cost of the proposed attack can be further reduced.

\mypara{Takeaways}
We demonstrate that the proposed attacks can be performed with cost-efficiency by training the shadow prompts with fewer epochs or a smaller number of shadow prompts.
We further show that to achieve a good attack performance, the adversary must have a shadow dataset of similar distribution as the target dataset and must have access to the same pre-trained model.
The results are displayed in~\autoref{appendix:pia_append}.

\subsection{Defense}
\label{section:pia_defense}

\mypara{Gaussian Noise as Defense}
We propose a defense mechanism~\cite{ZLH19,ZLDLZY23, HRF19}.
Specifically, the PaaS provider adds Gaussian noise $\mathcal{N}(0, \sigma^2I)$ to the released prompts, resulting in noised target prompts $\Delta_{\textit{target}}^{\prime} = \{\delta_i + \epsilon_i \mid \forall \delta_i \in \Delta_{\textit{target}}, \epsilon_i \sim \mathcal{N}(0, \sigma^2I) \}$.
The magnitude of the noise is controlled by the value of $\sigma$, with larger values corresponding to higher noise.
We examine the effectiveness of the proposed defense with $\sigma \in \{1, 2, 3, 4, 5\}$.
We first report the target performance, i.e., prompt utility, on all datasets in~\autoref{figure:pia_utility}.
The evaluation metric is the average test accuracy of all target prompts in the attack testing dataset on specific downstream tasks.
In general, we observe that the target performance decreases on all datasets by a large margin with the increase of $\sigma$.
For example, the prompt utility decreases from 33.95\% to 10.55\%, which is even lower than random guess (12.50\%), meaning the prompt is no longer usable.
We present the attack performance in~\autoref{figure:defense}.
We can observe that the effectiveness of the proposed attack significantly declines with the increase of $\sigma$.
The test accuracy on CelebA and UTKFace drops to almost random guess when $\sigma \ge 2$.
The attacks on CIFAR10 are more robust to the defense, but the performance still starts decreasing when $\sigma=3$.

\mypara{Adaptive Attacks}
We further consider an adaptive adversary~\cite{JSBZG19} who is aware of the defense mechanism, i.e., that Gaussian noise has been added to the target prompts.
They can construct their attack training dataset with noised shadow prompts.
Similarly, we set $\sigma \in \{1,2,3,4,5\}$ for both shadow and target prompts.
We report the performance of adaptive attacks on all datasets in~\autoref{figure:ad_defense}.
The results show that the attack performance declines less and more slowly.
For example, the attack performance barely decreases when $\sigma=1$ on all datasets.
In addition, when considering the size of the prompt training dataset as the target property, the attack performance only has negligible degradation with the growth of Gaussian noise.
For example, the attack performance has almost no drop even with $\sigma=5$ on all datasets.

\mypara{Takeaways}
These findings indicate that adding Gaussian noise as a defense mechanism can ostensibly decrease the attack performance.
But the defender suffers from unacceptable prompt utility degradation.
Moreover, this defense can be bypassed by the adaptive attack.
We leave it as future work to investigate more effective defenses.
We later show that the proposed defense can achieve a decent utility-defense trade-off by using a smaller $\sigma$, e.g., $\sigma=0.6$, indicating that the statistical information of the training dataset in the target prompts is harder to hide than individual information, i.e., membership (see~\autoref{section:mia_defense}).

\section{Membership Inference Attacks}
\label{section:mia}

In this section, we leverage the membership inference attacks to quantify the privacy risks of prompts.

\subsection{Threat Model}
\label{section:mia_threat_model}

\mypara{Adversary's Goal}
In membership inference, the goal of the adversary is to infer whether a given data sample $x$ is in the training dataset of the target prompt \deltat.

\mypara{Adversary's Knowledge and Capability}
Similar to the property inference attack, the adversary can query the PaaS service to get \deltat and has white-box access to the pre-trained model \mpre.
The adversary has a shadow dataset \ds that is from the same distribution as \dt to train the shadow prompt \deltas.
We later demonstrate that the adversary can operate in a data-free manner, i.e., leveraging \ds that comes from a different distribution than \dt.

\subsection{Measurement Methodology}
\label{section:mia_method}

\begin{figure*}[!t]
\centering
\begin{subfigure}{0.4\columnwidth}
\includegraphics[width=\columnwidth]{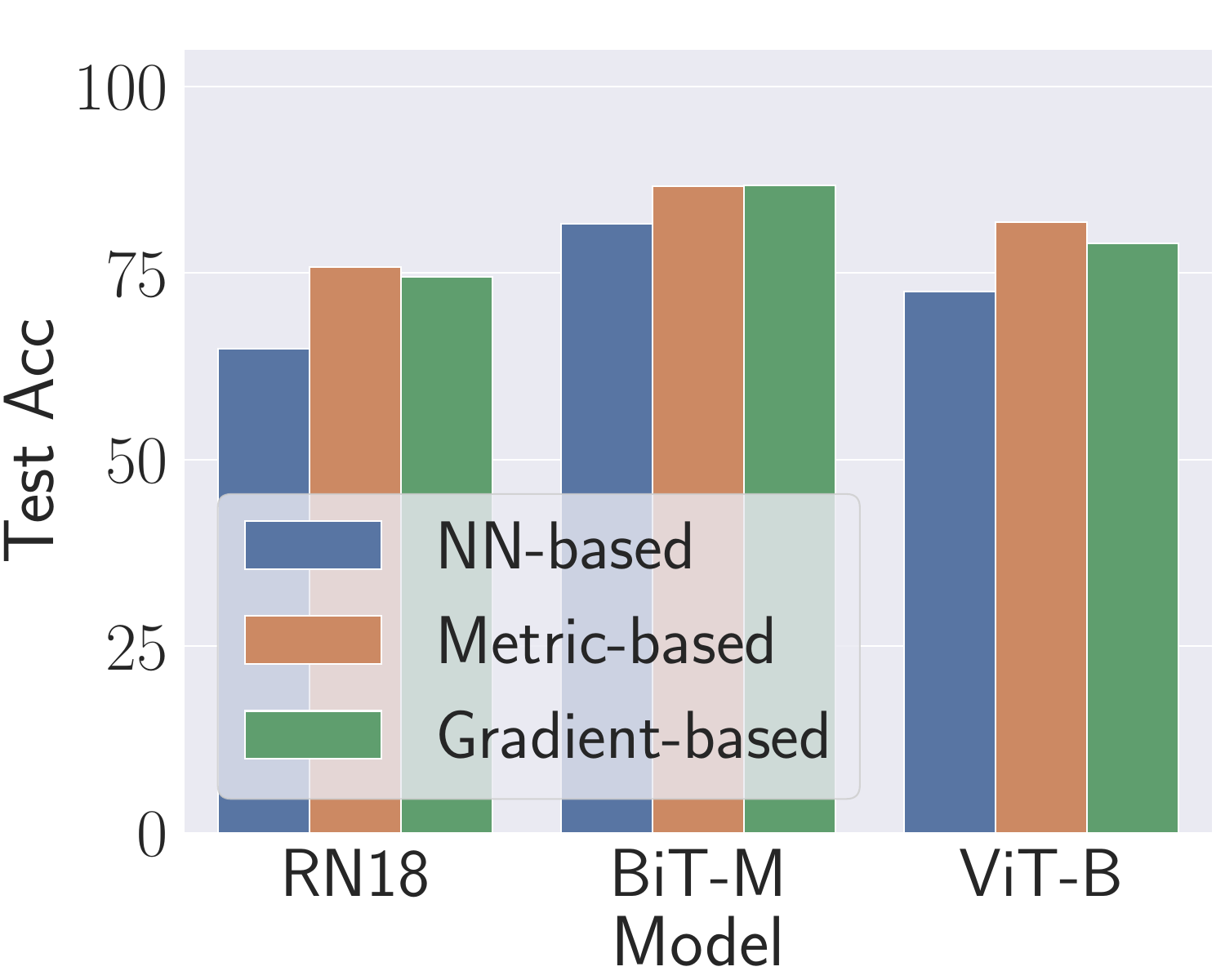}
\caption{CIFAR10}
\label{figure:main_result_cifar10}
\end{subfigure}
\begin{subfigure}{0.4\columnwidth}
\includegraphics[width=\columnwidth]{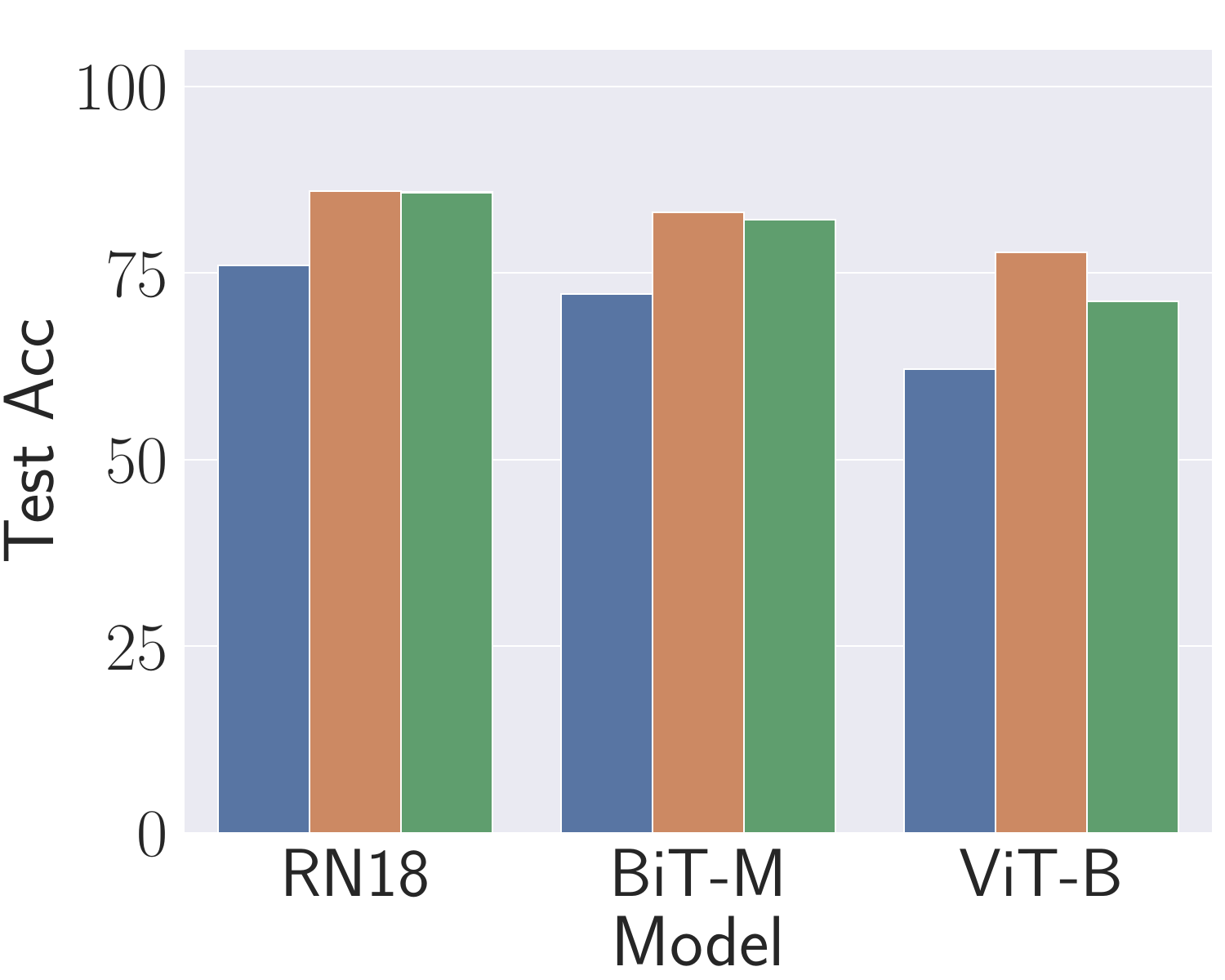}
\caption{CelebA}
\label{figure:main_result_celeba}
\end{subfigure}
\begin{subfigure}{0.4\columnwidth}
\includegraphics[width=\columnwidth]{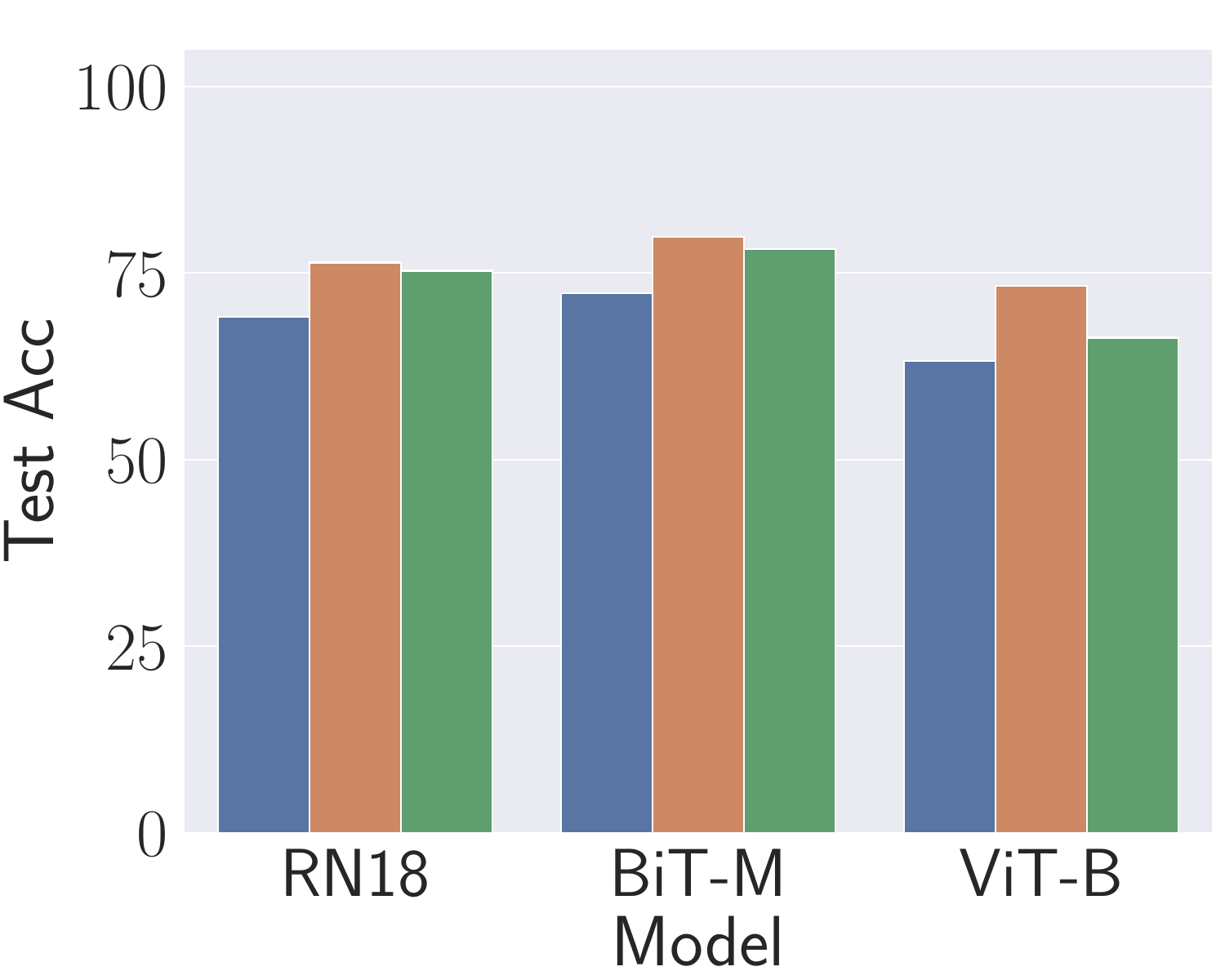}
\caption{UTKFace}
\label{figure:main_result_utkface}
\end{subfigure}
\begin{subfigure}{0.4\columnwidth}
\includegraphics[width=\columnwidth]{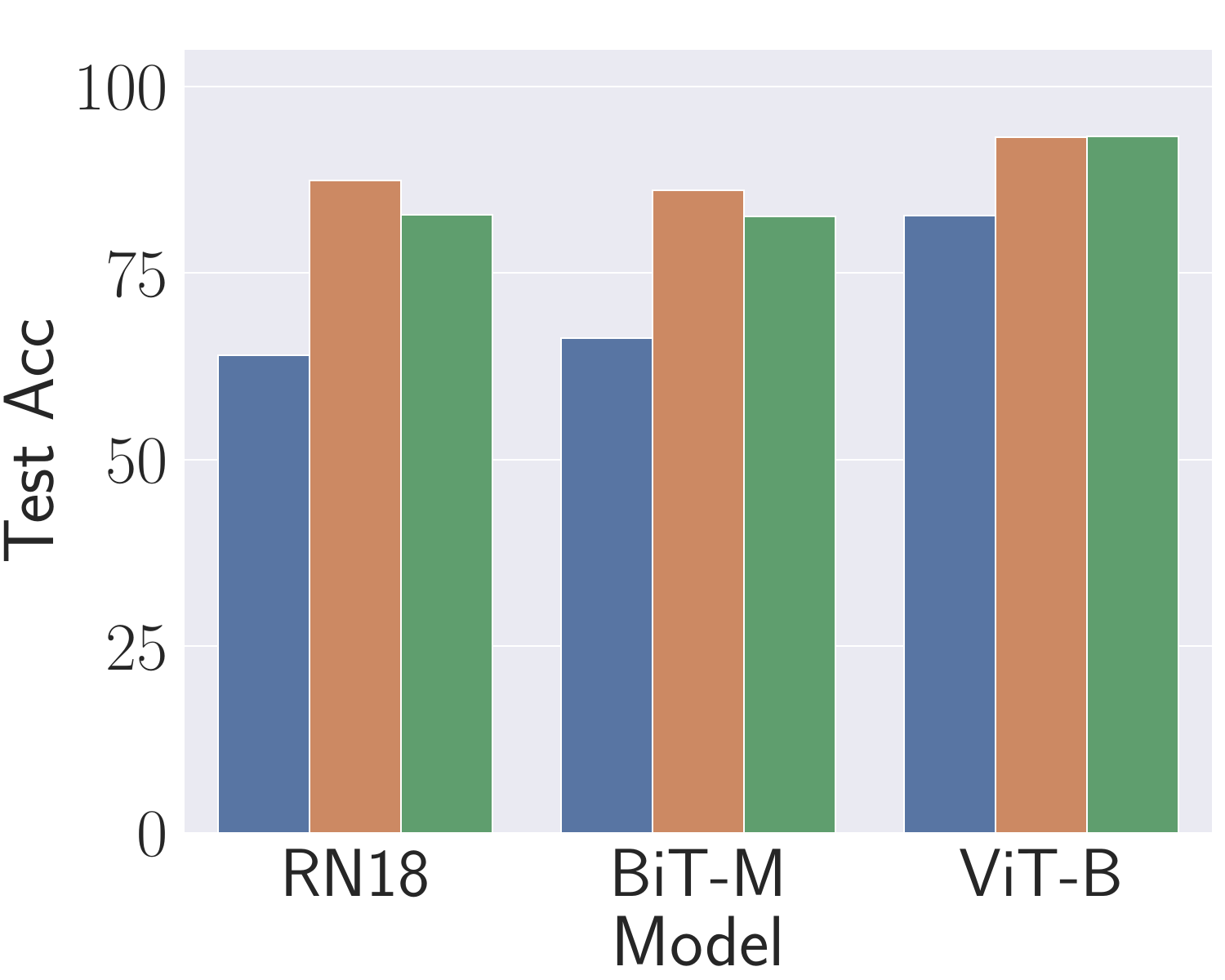}
\caption{AFAD}
\label{figure:main_result_afad}
\end{subfigure}
\caption{Attack performance of three membership inference attacks on four datasets.}
\label{figure:mia_main_result}
\end{figure*}

\mypara{Attack Setup}
The adversary first divides the shadow dataset into two disjoint subsets: \dstr, referred to as the member split, and \dste, referred to as the non-member split.
The member split is then utilized for training the shadow prompt \deltas, which mimics the behavior of \deltat.

\mypara{Attack Descriptions}
We adopt three types of membership inference attacks, i.e., neural network-based (\trad) attacks~\cite{SZHBFB19}, \metric attacks~\cite{SM21}, and \grad attacks~\cite{LF20, NSH19}.
We outline their technical details below.

\noindent \emph{NN-based Attacks~\cite{SZHBFB19}.}
The adversary constructs the attack training dataset on \ds.
Specifically, they combine each sample in \ds with the shadow prompt trained on \dstr and query the corresponding pre-trained model to get the top-5 posteriors as attack input features.
Then, for each sample in the member split, the adversary labels the corresponding top-5 posteriors as ``member.'' 
For samples that belong to the non-member split, their top-5 posteriors are labeled as ``non-member.''
At inference time, the adversary queries the pre-trained model with the given data sample $x$ and \deltat to obtain the top-5 posteriors and feeds them to the attack model to obtain its membership prediction.

\noindent \emph{Metric-based Attacks~\cite{SM21}.}
Song and Mittal propose metric-based attacks using four metrics, i.e., prediction correctness (metric-corr), prediction confidence (metric-conf), prediction entropy (metric-ent), and modified prediction entropy (metric-ment).
Unlike \trad attacks where a neural network is trained to make membership predictions, metric-based attacks first calculate class-wise thresholds over \deltas.
Then, at inference time, the adversary calculates the metric values and compares them with the pre-calculated thresholds to determine the membership status for given data samples.
It is worth noting that in scenarios where the adversary possesses data from a different distribution than the target dataset, we calculate an overall threshold for all classes.
This is because certain classes present in the target dataset may not be represented in the shadow dataset, so class-specific thresholds would not be applicable.

\noindent \emph{Gradient-based Attacks~\cite{NSH19}.}
Nasr et al.\ propose \grad attacks on the basis of the \trad attacks by incorporating augmented input information.
Specifically, the adversary has white-box access to the pre-trained model and target prompt with its intermediate computations, e.g., gradients.
They combine each sample $x$ with the prompt and input resulting data into the pre-trained model to obtain top-5 posteriors, the loss incurred during the forward pass, the gradient of the prompt during the backward pass, and an indicator that denotes the correctness of the prediction.
These obtained data are treated as the attack input for the attack model.

\subsection{Measurement Settings}
\label{section:mia_setting}

\mypara{Datasets and Downstream Tasks}
We reuse CIFAR10, CelebA, UTKFace, and AFAD to evaluate membership inference attacks.
The downstream tasks for all datasets are the same as those for property inference attacks (see~\autoref{section:pia_setting}).
We randomly sample 8000 data samples for each dataset in the main experiments and then evenly split each dataset into four disjoint sets, i.e., \dttr, \dtte, \dstr, and \dste.
\dttr is used to develop the target prompt \deltat, and \dtte is the evaluation set.
\dstr and \dste are used to build the attack model as discussed in \autoref{section:mia_method}.

\mypara{Attack Configurations}
All experimental settings of the pre-trained models and target prompts are the same as those for property inference attacks except for the number of epochs.
We follow the default setting to train all prompts for 1000 epochs.
For attacks that leverage neural networks as the attack model, we employ the cross-entropy loss function and optimize it using Adam optimizer.
We conduct a grid search on \{1e-2, 1e-3, 1e-4, 1e-5\} to determine the optimal learning rate for each attack, and all attack models are trained for 100 epochs.
For the \trad attacks, we use a 2-layer MLP as the attack model and set the size of the hidden layer to 32.
For the \grad attacks, we utilize an attack model composed of four sub-networks, each corresponding to one attack information (gradient, top-5 posteriors, loss, and indicator), and the outputs of these sub-networks are concatenated to form the final input of a 2-layer MLP.

\mypara{Metric}
Following the convention~\cite{SSSS17, HZ21}, we use test accuracy as the main metric to evaluate the attack performance.

\begin{figure}[!t]
\centering
\begin{subfigure}{0.4\columnwidth}
\includegraphics[width=\columnwidth]{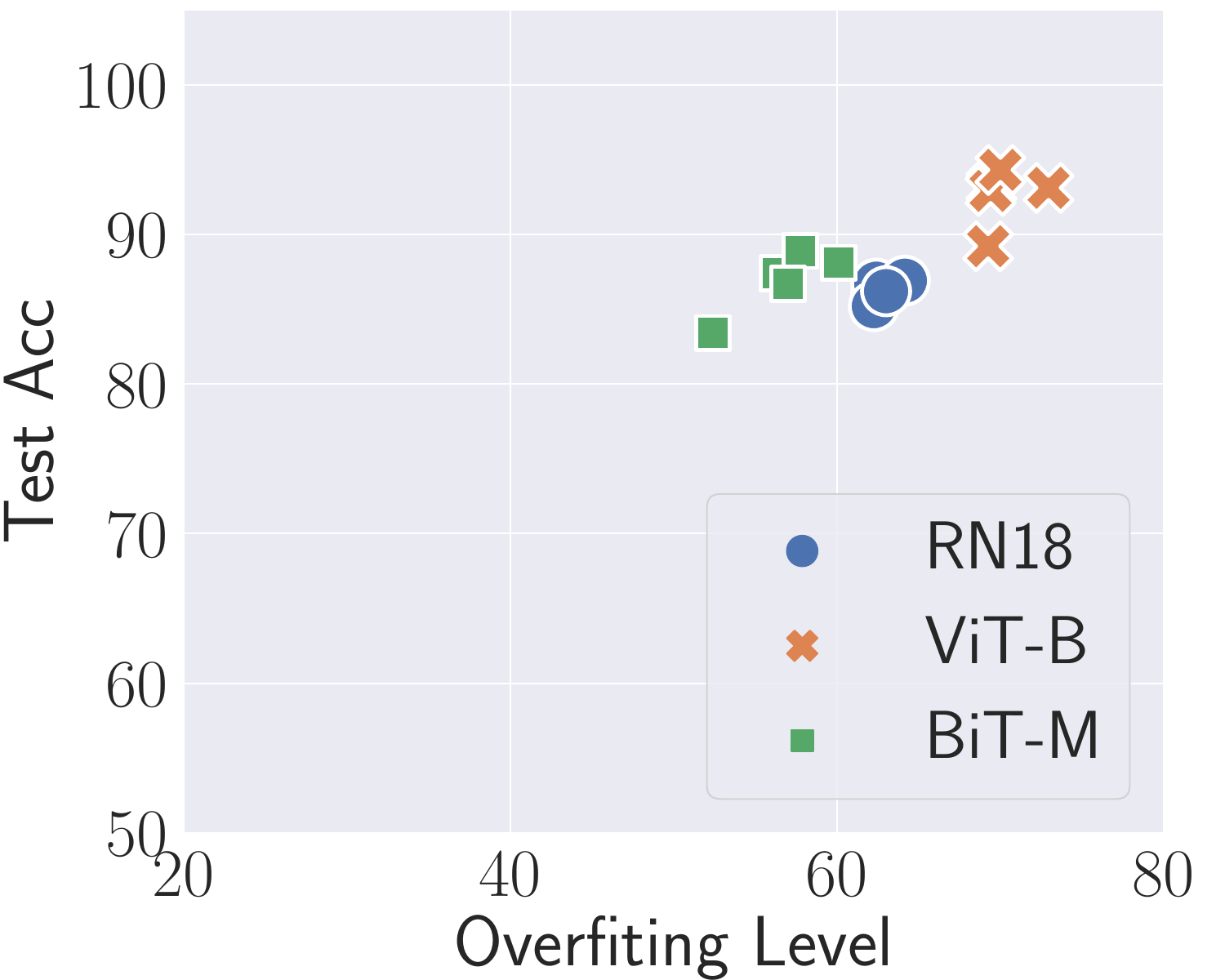}
\caption{Pre-trained Models}
\label{figure:overfit_model}
\end{subfigure}
\begin{subfigure}{0.4\columnwidth}
\includegraphics[width=\columnwidth]{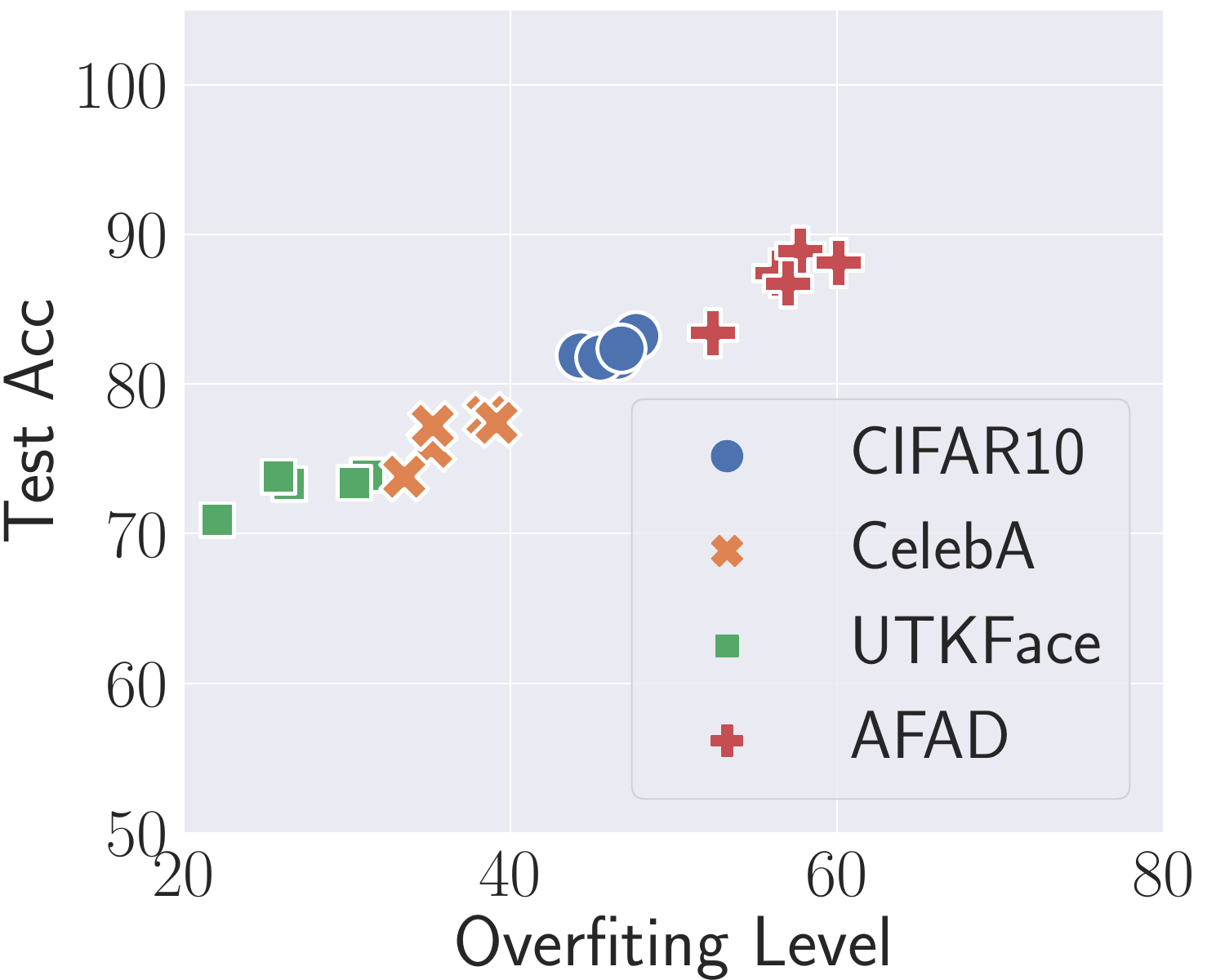}
\caption{Datasets}
\label{figure:overfit_dataset}
\end{subfigure}
\caption{Overfitting levels of target prompts across (a) different pre-trained models on AFAD and (b) different datasets using BiT-M as the pre-trained model.
Different points with the same marker denote different runs of the same pre-trained model/dataset using different random seeds.}
\label{figure:overfit}
\end{figure}

\begin{figure}[!t]
\centering
\begin{subfigure}{0.4\columnwidth}
\includegraphics[width=\columnwidth]{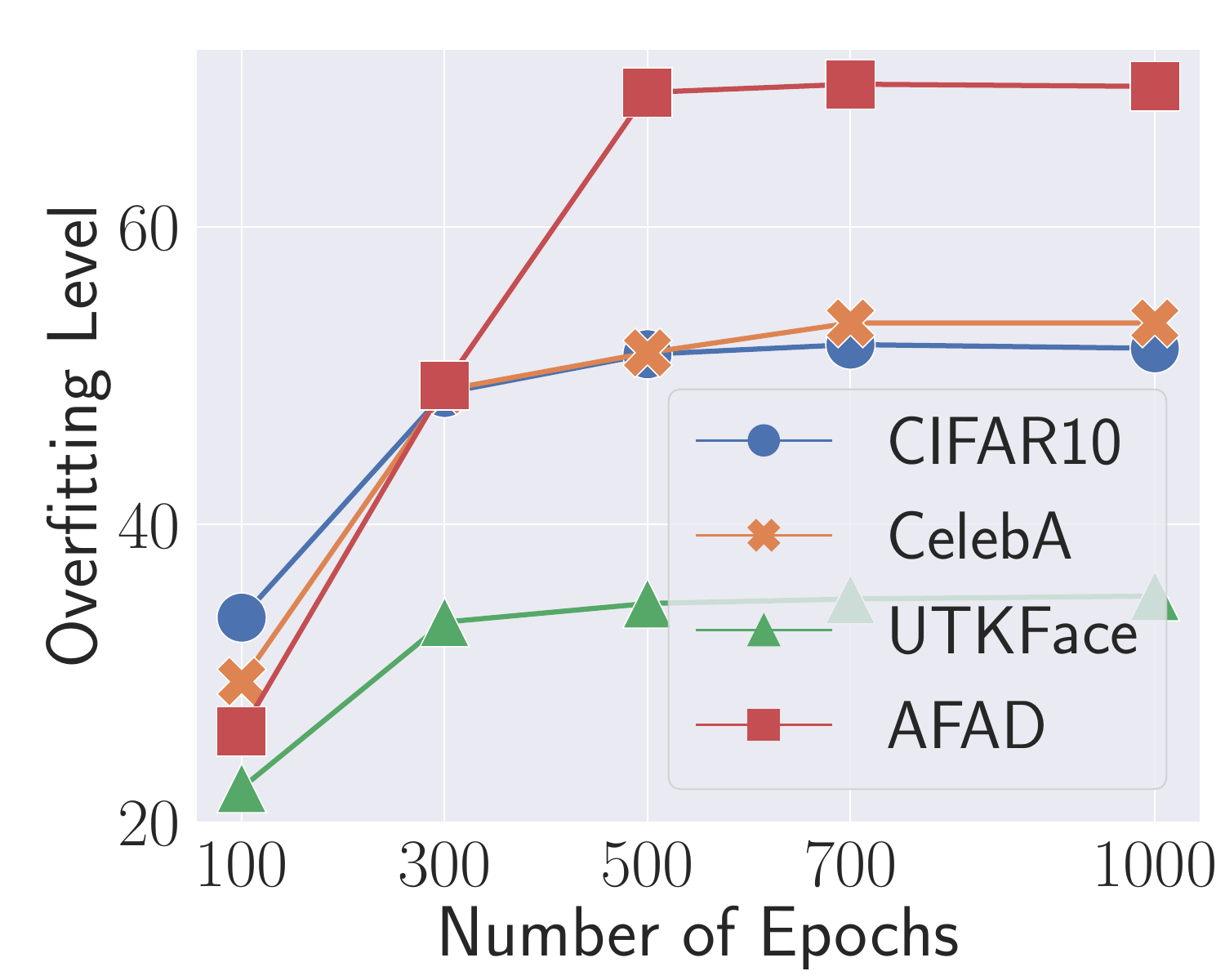}
\caption{Epoch Effect}
\label{figure:overfit_epoch_effect}
\end{subfigure}
\begin{subfigure}{0.4\columnwidth}
\includegraphics[width=\columnwidth]{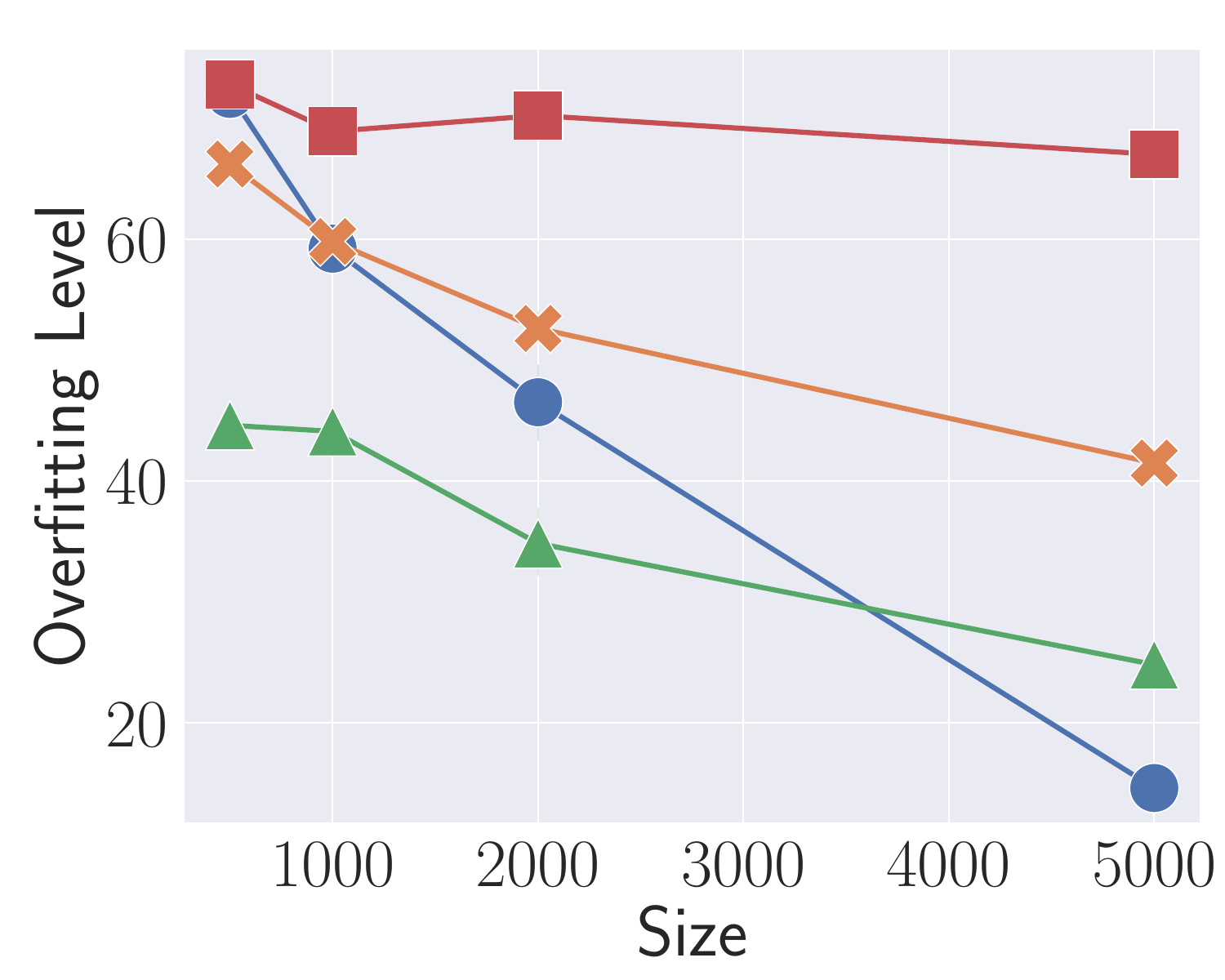}
\caption{Size Effect}
\label{figure:overfit_size_effect}
\end{subfigure}
\caption{Overfitting levels of target prompts with (a) different numbers of epochs and (b) different sizes of the training dataset, using ViT-B as the pre-trained model.}
\label{figure:overfit_effect}
\end{figure}

\begin{figure*}[!t]
\centering
\begin{subfigure}{0.4\columnwidth}
\includegraphics[width=\columnwidth]{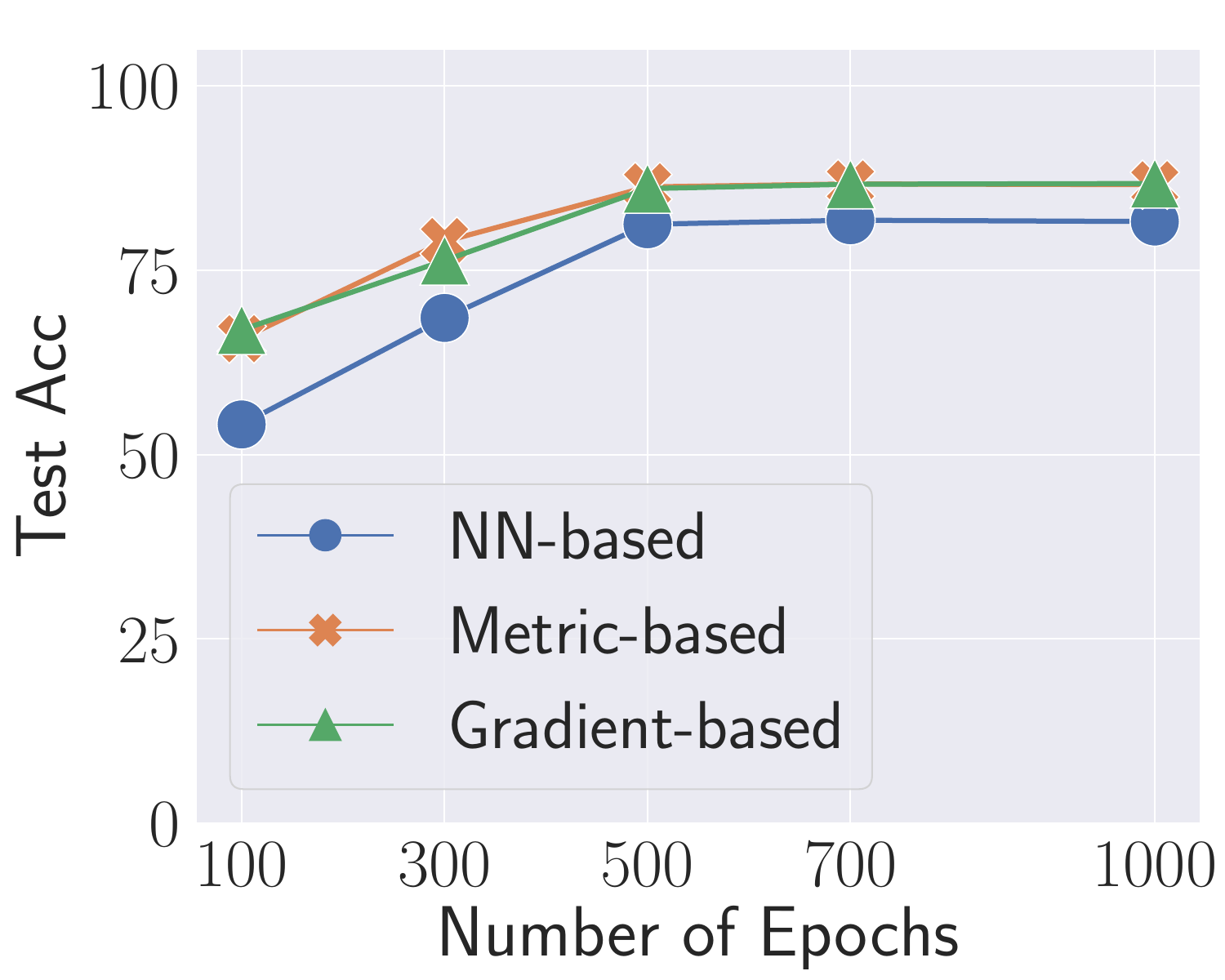}
\caption{CIFAR10}
\label{figure:epoch_effect_cifar10}
\end{subfigure}
\begin{subfigure}{0.4\columnwidth}
\includegraphics[width=\columnwidth]{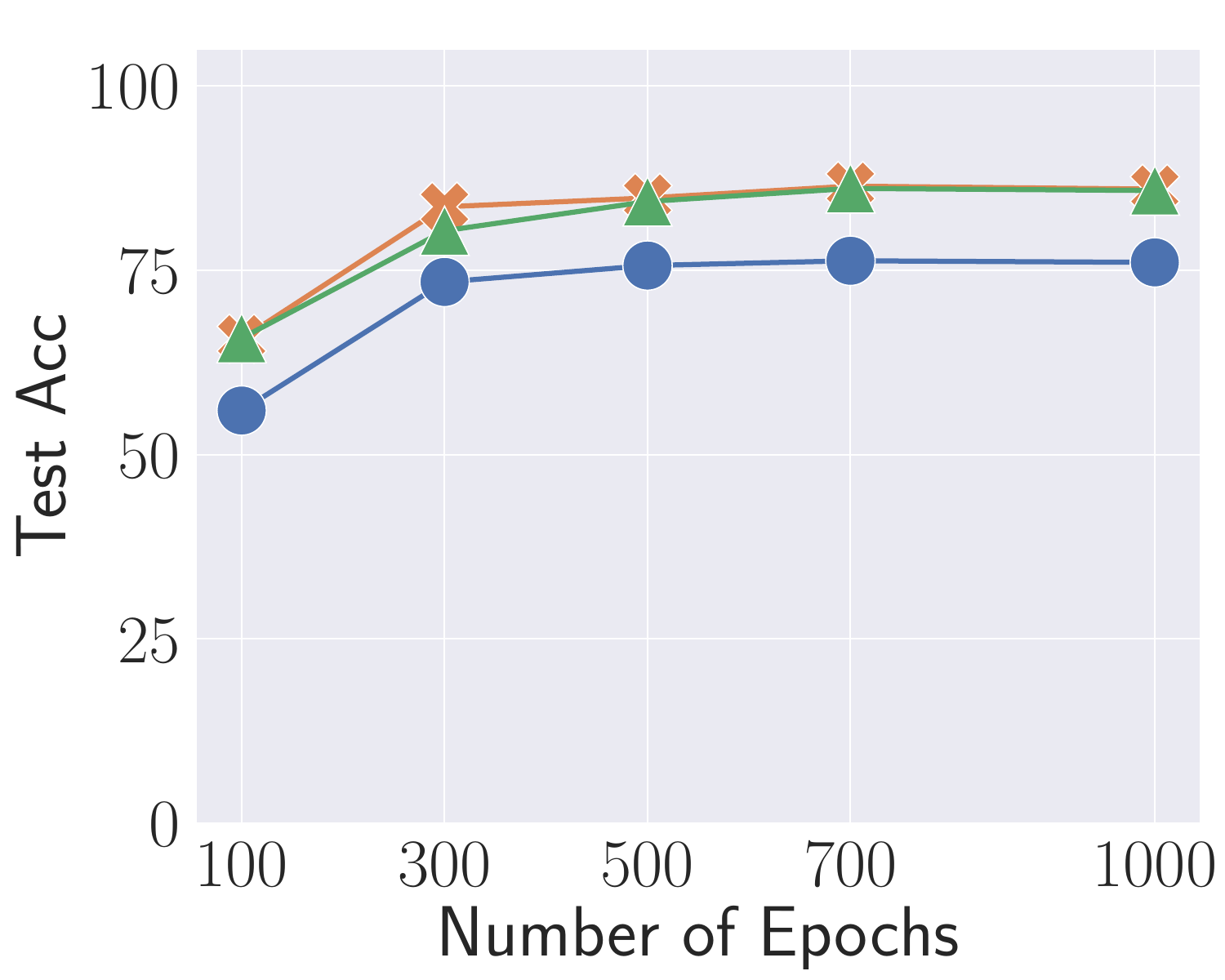}
\caption{CelebA}
\label{figure:epoch_effect_celeba}
\end{subfigure}
\begin{subfigure}{0.4\columnwidth}
\includegraphics[width=\columnwidth]{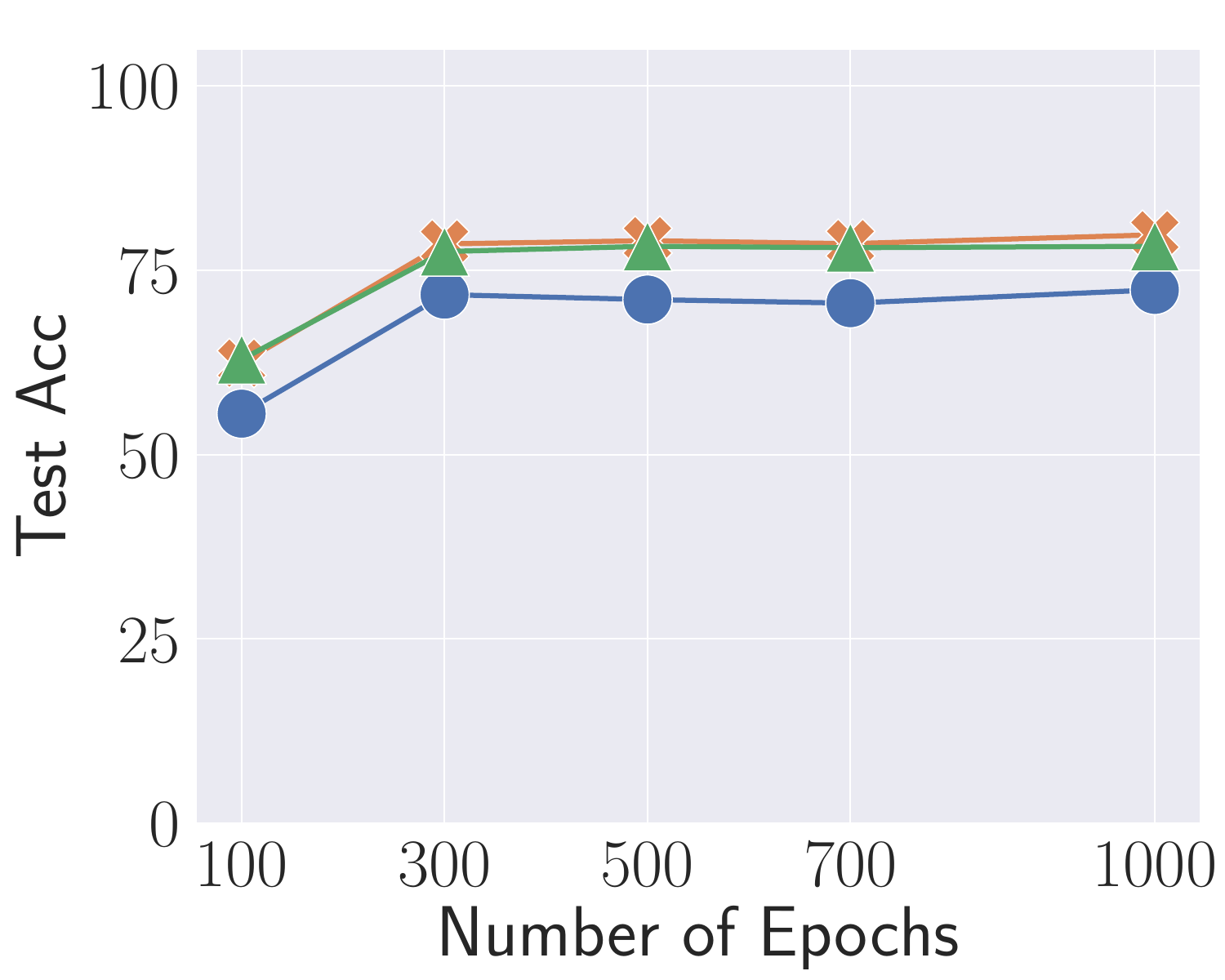}
\caption{UTKFace}
\label{figure:epoch_effect_utkface}
\end{subfigure}
\begin{subfigure}{0.4\columnwidth}
\includegraphics[width=\columnwidth]{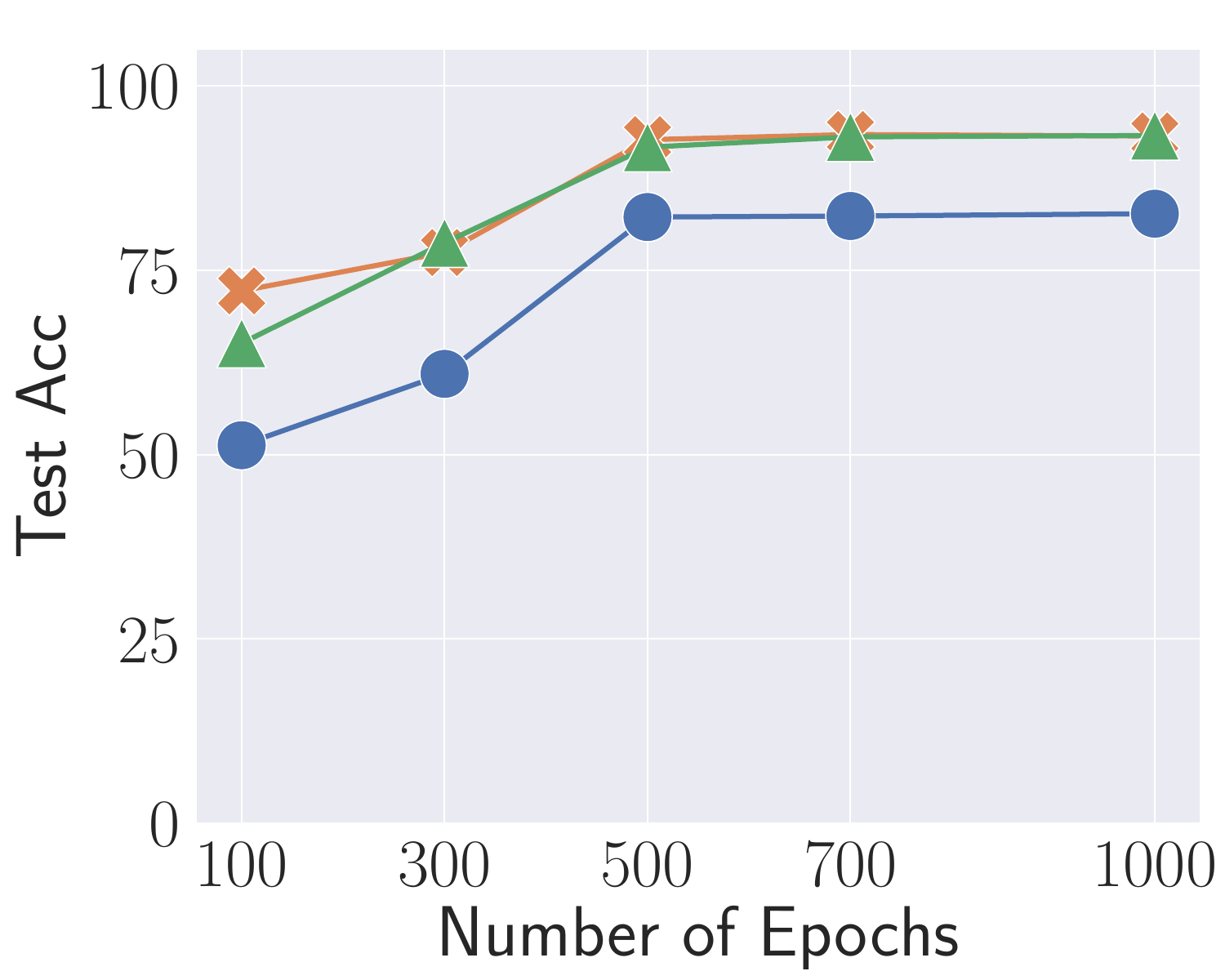}
\caption{AFAD}
\label{figure:epoch_effect_afad}
\end{subfigure}
\caption{Attack performance of three membership attacks with varying number of epochs, using ViT-B as the pre-trained model.}
\label{figure:mia_epoch}
\end{figure*}

\begin{figure*}[!t]
\centering
\begin{subfigure}{0.4\columnwidth}
\includegraphics[width=\columnwidth]{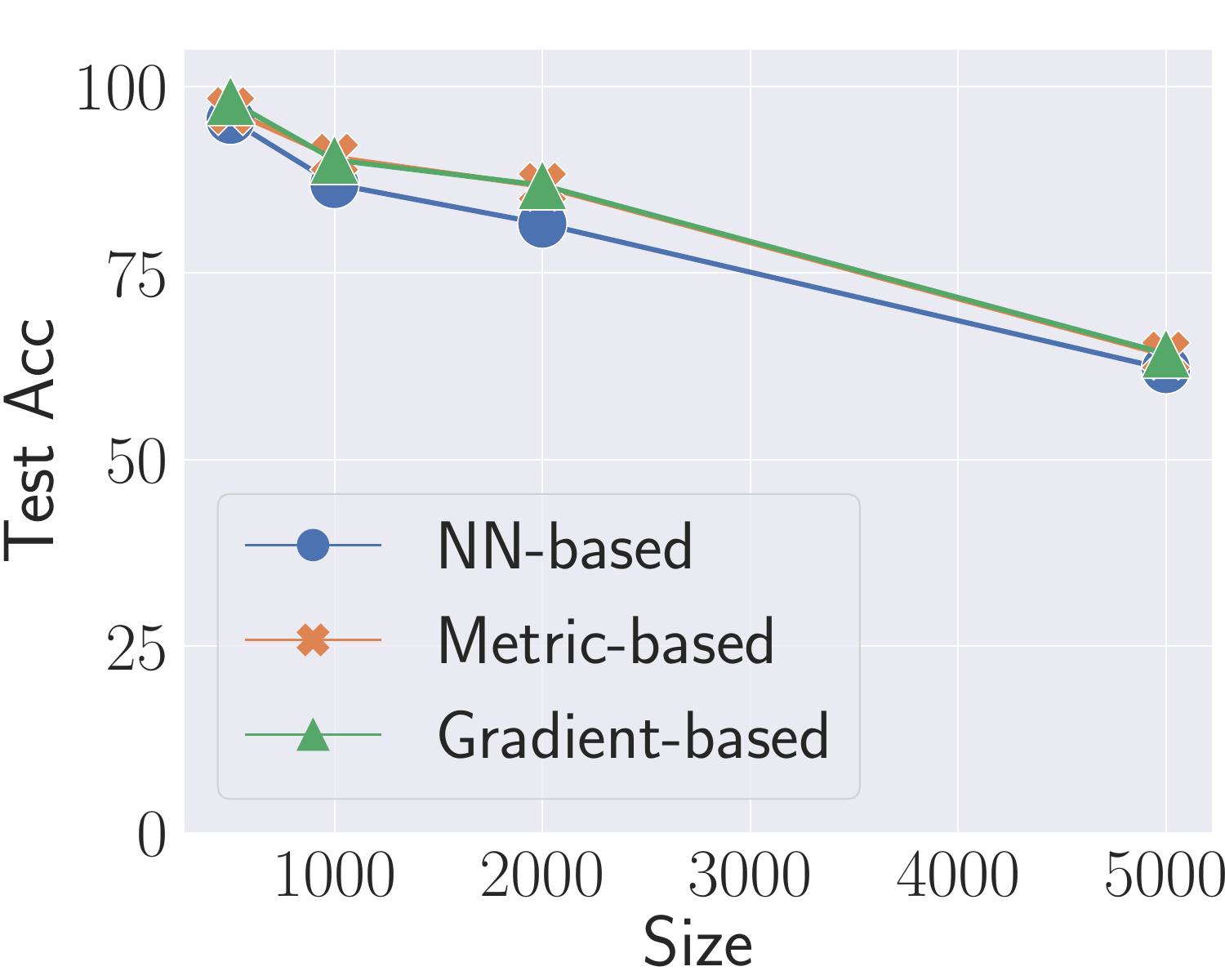}
\caption{CIFAR10}
\label{figure:size_effect_cifar10}
\end{subfigure}
\begin{subfigure}{0.4\columnwidth}
\includegraphics[width=\columnwidth]{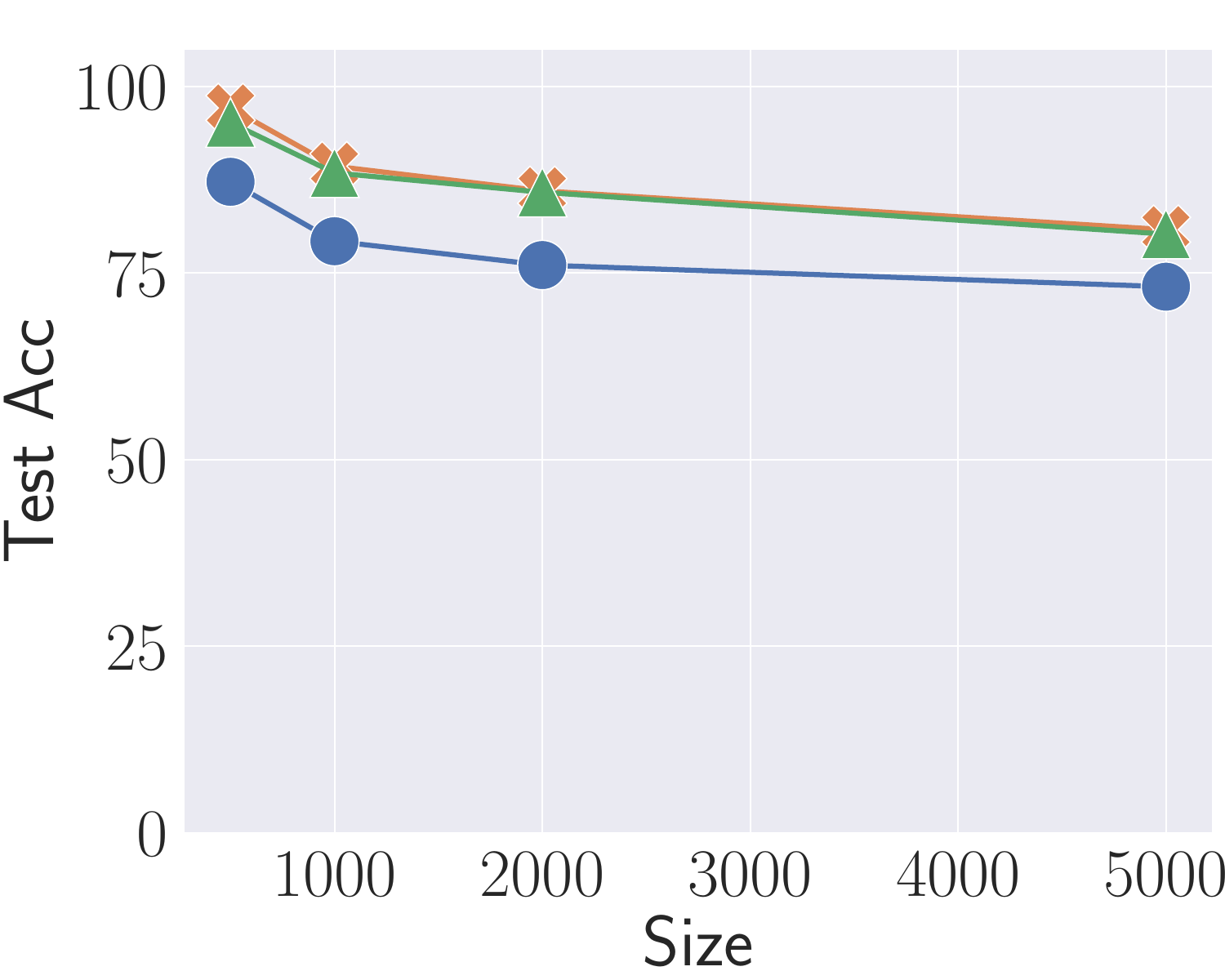}
\caption{CelebA}
\label{figure:size_effect_celeba}
\end{subfigure}
\begin{subfigure}{0.4\columnwidth}
\includegraphics[width=\columnwidth]{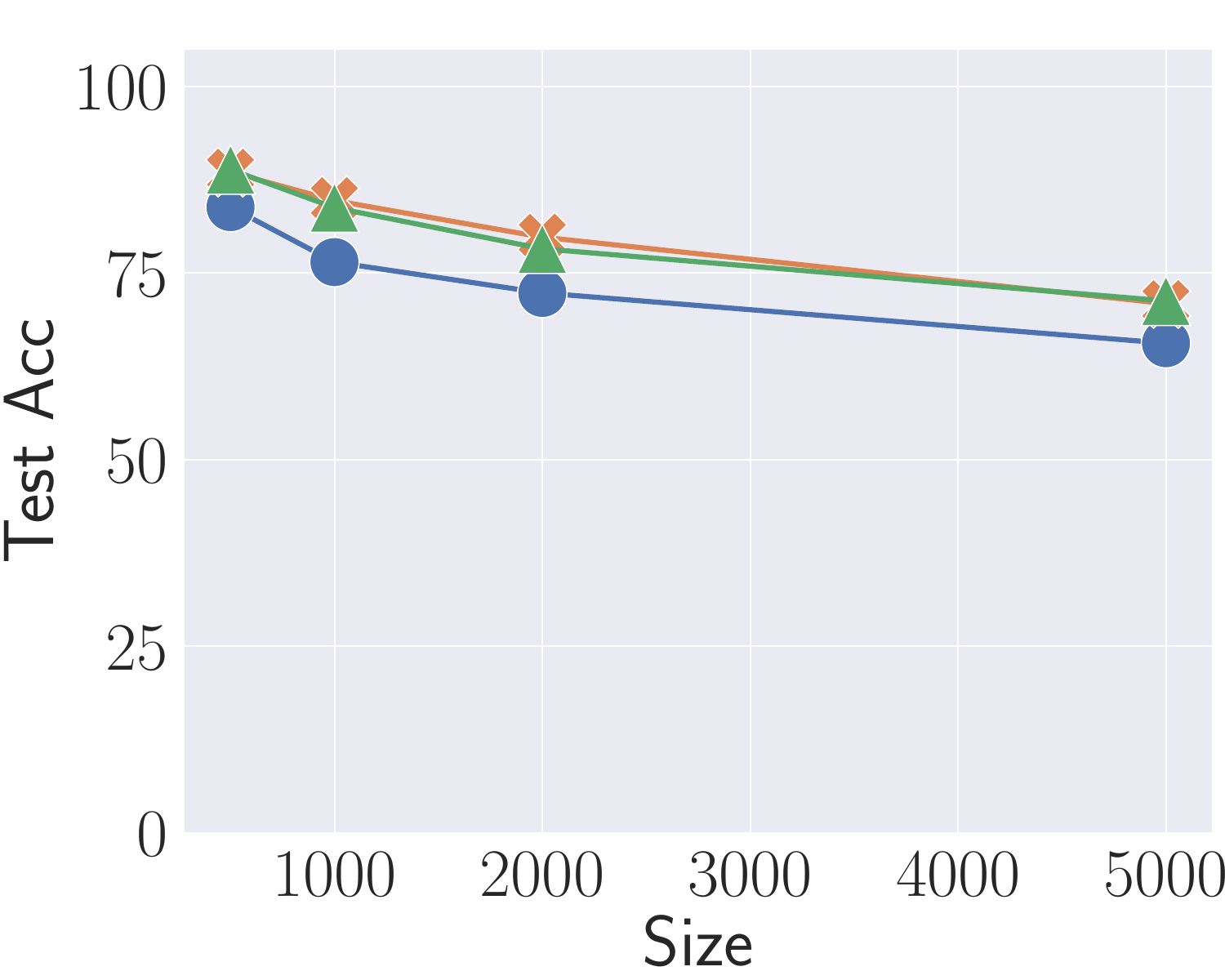}
\caption{UTKFace}
\label{figure:size_effect_utkface}
\end{subfigure}
\begin{subfigure}{0.4\columnwidth}
\includegraphics[width=\columnwidth]{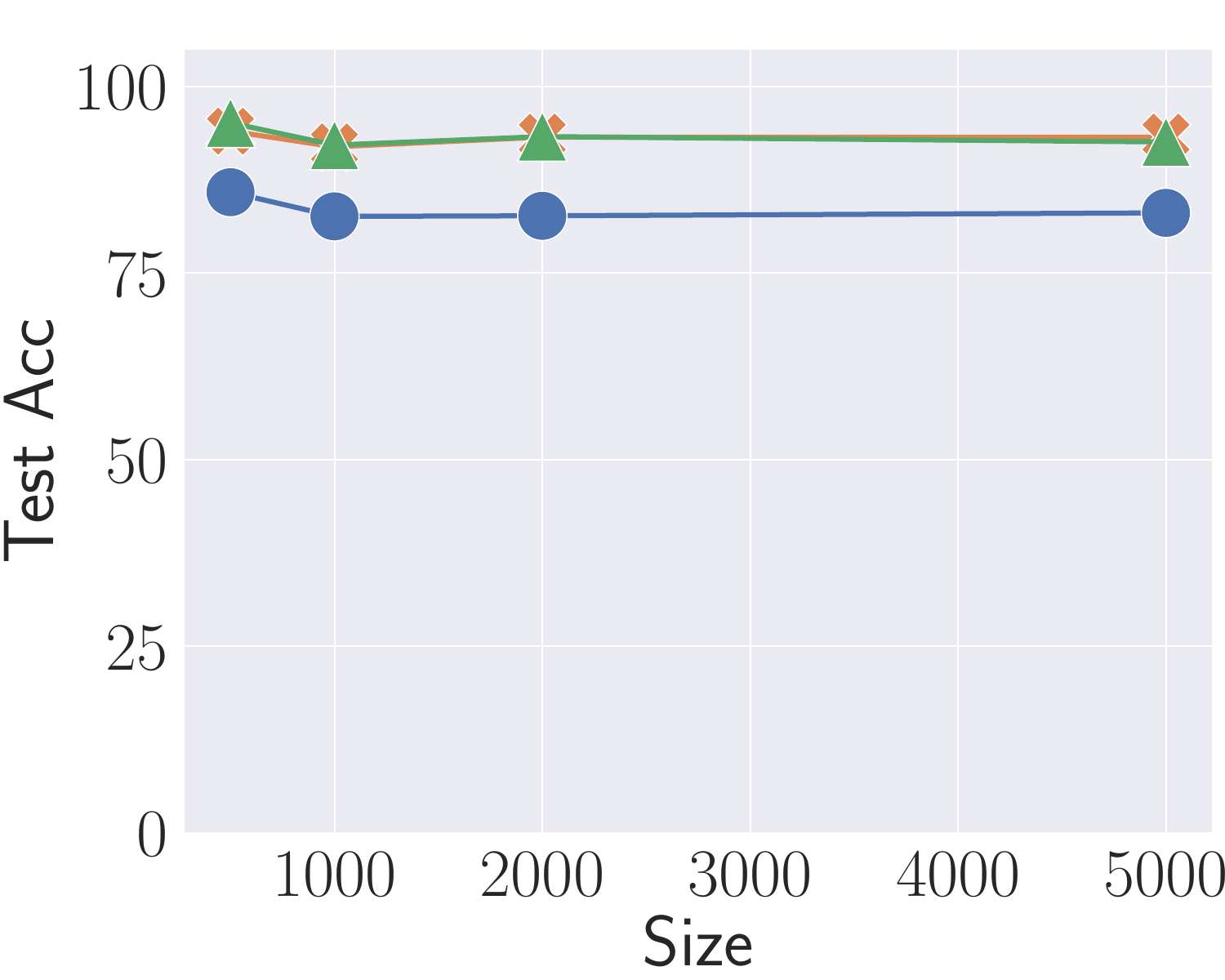}
\caption{AFAD}
\label{figure:size_effect_afad}
\end{subfigure}
\caption{Attack performance of three membership attacks with different sizes of \dttr, using ViT-B as the pre-trained model.}
\label{figure:mia_size}
\end{figure*}

\subsection{Measurement Results}
\label{section:mia_results}

\mypara{Membership Inference Privacy Risks}
We report the performance of three membership inference attacks in~\autoref{figure:mia_main_result}.
We conduct three separate runs of each attack experiment and report the average values as the final results.
We observe that \metric attacks achieve the best performance in most cases, e.g., 93.20\% membership inference attack accuracy on AFAD.
Unless otherwise specified, we use metric-ment attacks as the representative of the \metric attacks, as they consistently achieve the best performance across all datasets (see~\autoref{appendix:diff_metric_attack}).
The \grad attacks also exhibit strong performance, with results that are closely comparable to those of the \metric attacks.
Song and Mittal~\cite{SM21} also report that the \metric attacks can outperform \trad attacks and have similar performance as \grad attacks.
\trad attacks perform worse than \grad attacks.
This is expected since the adversary leverages less information from the target prompt.

\mypara{Analysis}
\autoref{figure:mia_main_result} shows that the performance of three attacks varies on different pre-trained models and different datasets.
We hypothesize that the different overfitting levels may affect the attack performance.
Following previous work~\cite{HZ21, SS17}, we calculate the difference between train accuracy and test accuracy to measure the overfitting level of a given target prompt.
We train five target prompts with different random seeds for each experimental setting.
The relationship between overfitting levels and attack performance is illustrated in~\autoref{figure:overfit}.
We observe that different pre-trained models and different datasets have different overfitting levels.
Meanwhile, our results demonstrate that overfitting does have a significant effect on membership inference attacks.
The overall trend is that the higher the overfitting level, the better the attack performance.
To quantify this correlation, we calculate the Pearson correlation scores between the overfitting level and attack performance.
The result is 0.89.
Our finding is in line with previous analyses~\cite{SH21,LWHSZBCFZ22}.

\mypara{Takeaways}
We show that prompts can leak sensitive membership information of their training dataset.
Similar to previous analyses, overfitting is strongly correlated with the membership inference performance.

\subsection{Factors Affecting Membership Inference From the Victim's Side}

In this section, we measure the factors that may affect the membership inference privacy risks from the perspective of the victim.
As shown in~\autoref{figure:overfit_effect}, the number of epochs used to train the target prompt and its training dataset size (\dttr) are closely related to the overfitting level of the target prompt.
A larger number of epochs results in larger overfitting levels.
On the contrary, a larger training dataset size results in reduced overfitting levels.
Therefore, we explore how these two factors affect the attack performance.

\mypara{Number of Epochs} 
We set the number of epochs for the target prompt to \{200, 400, 600, 800, 1000\} and use the same number of epochs for the shadow prompt in each experiment.
The results are shown in~\autoref{figure:mia_epoch}.
We observe that, in general, more epochs lead to better attack performance, hence greater membership inference privacy risks.
The attack performance becomes steady after 500 epochs, while the overfitting level also becomes stable simultaneously in~\autoref{figure:overfit_epoch_effect}.

\mypara{Prompt Training Dataset Size}
We investigate the effect of the training dataset size on the attack performance by varying the size from $500$ to $5000$.
To control the variables, we always fix the other three sets (\dtte, \dstr, and \dste) to the same size as \dttr.
As illustrated in~\autoref{figure:mia_size}, the attack performance decreases as the dataset size grows.
The general trend of the attack performance is also consistent with the findings in~\autoref{figure:overfit_size_effect}.
That is, more training data reduces the overfitting level in most cases, leading to a decrease in the attack performance.
There is a significant drop in the overfitting level on CIFAR10 when increasing the size of \dttr from 2000 to 5000.
Therefore, the test accuracy of \metric attacks drops from 86.60\% to 64.00\%.

\mypara{Takeaways}
We perform an analysis of the relation between overfitting levels and attack performance.
Our results show that more epochs and fewer training data can aggravate overfitting and pose a more severe threat to membership privacy.

\subsection{Factors Affecting Membership Inference From the Adversary's Side}

We evaluate the factors that may affect the membership inference privacy risks from the perspective of the adversary.
In previous experiments, we made two assumptions: 1) the adversary has a dataset \ds that comes from the same distribution as \dt, and 2) the PaaS provider offers users the target prompt with white-box access to the pre-trained model.
Here, we evaluate if these two assumptions are needed to mount a successful membership inference attack.

\begin{figure*}[!t]
\centering
\begin{subfigure}{0.4\columnwidth}
\includegraphics[width=\columnwidth]{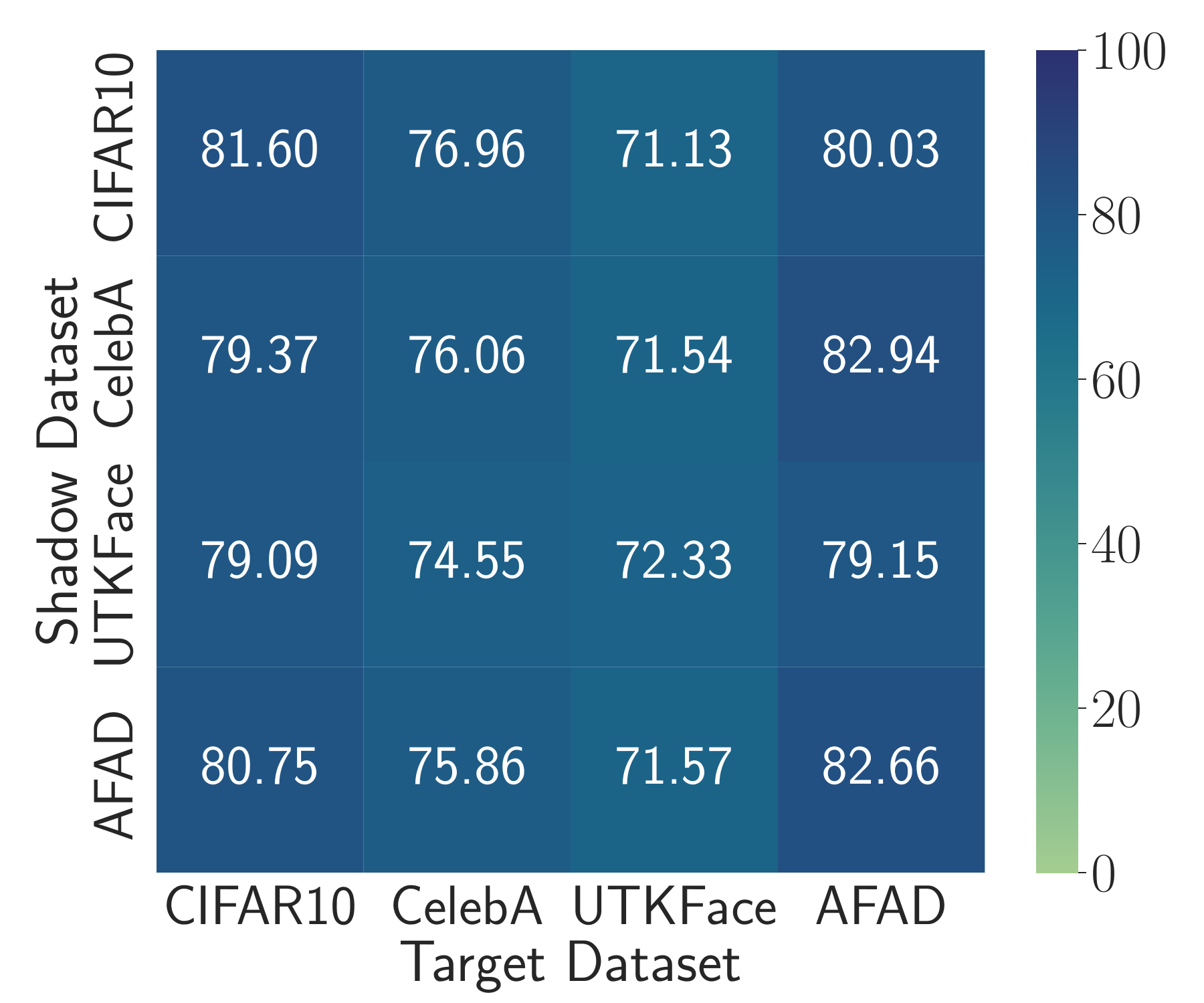}
\caption{NN-based}
\label{figure:diff_d_vit_trad}
\end{subfigure}
\begin{subfigure}{0.4\columnwidth}
\includegraphics[width=\columnwidth]{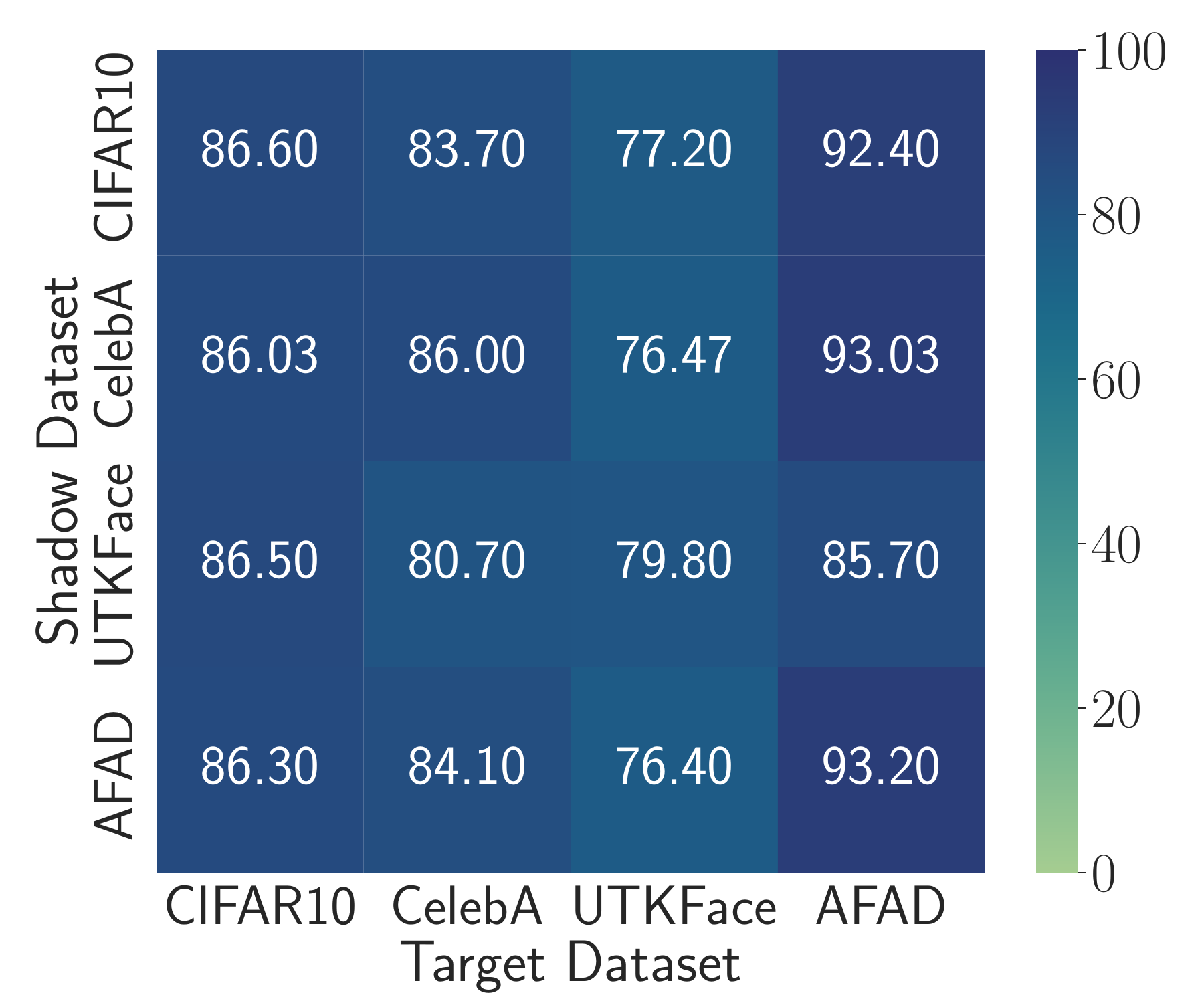}
\caption{Metric-based}
\label{figure:diff_d_vit_metric}
\end{subfigure}
\begin{subfigure}{0.4\columnwidth}
\includegraphics[width=\columnwidth]{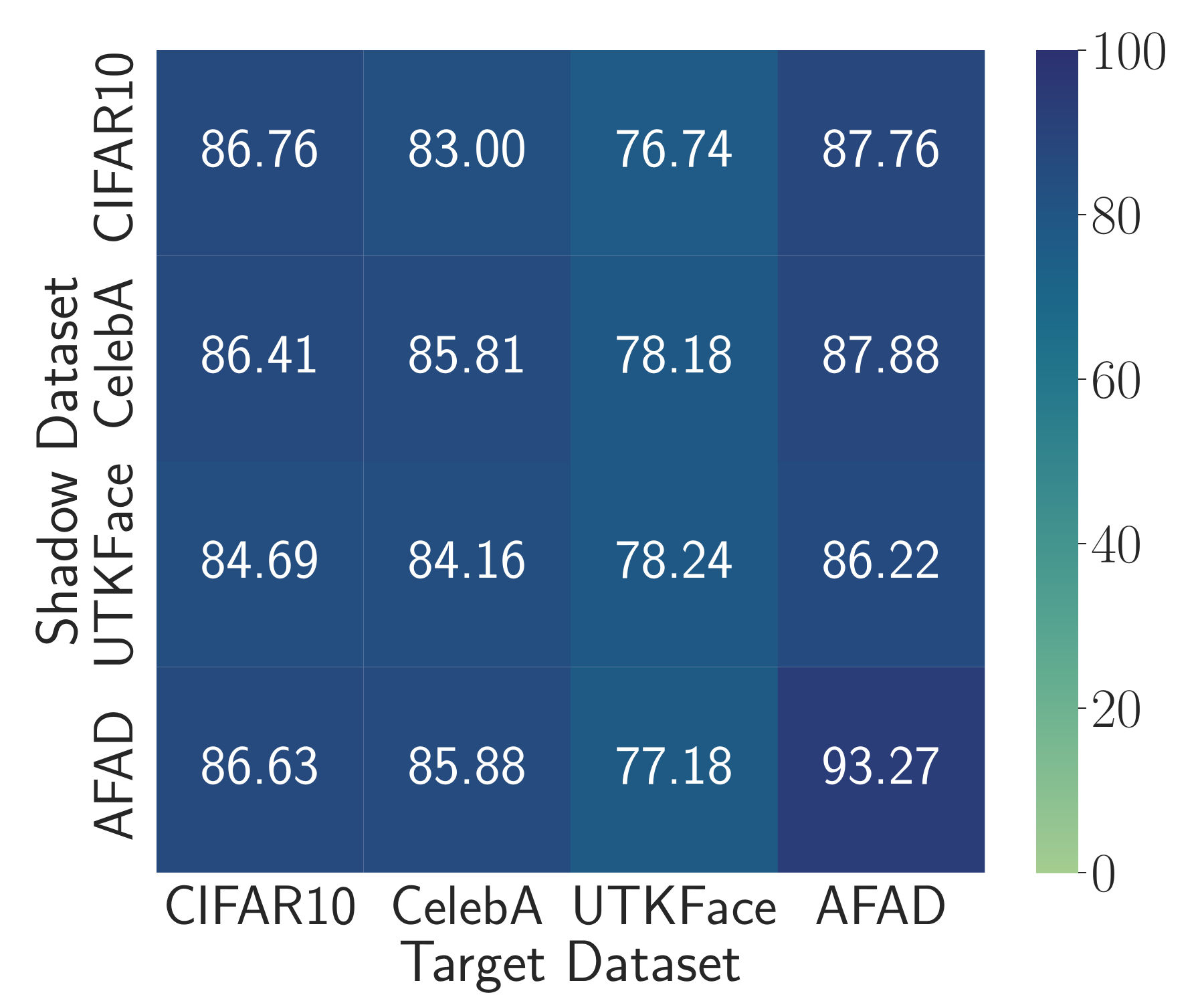}
\caption{Gradient-based}
\label{figure:diff_d_vit_white}
\end{subfigure}
\caption{Attack performance of three attacks after relaxing the dataset assumption, using ViT-B as the pre-trained model.}
\label{figure:diff_dataset}
\end{figure*}

\mypara{Dataset Assumption}
We relax the same distribution assumption by leveraging a shadow dataset that comes from a different distribution than \dt to train the shadow prompt; the results with three attack methodologies are shown in~\autoref{figure:diff_dataset}.
In the diagonal of the heatmaps, we show the results of the adversary having access to \ds that comes from the same distribution as \dt.
We observe that the performance of \trad, \metric, and \grad attacks is slightly reduced but remains effective.
For instance, as shown in~\autoref{figure:diff_d_vit_metric}, using any one of the four datasets as the shadow dataset to launch the \metric attack can achieve a test accuracy of around 86.00\%, when the target dataset is CIFAR10.
Interestingly, CIFAR10 contains images of 10 classes such as cars and trucks, but the other three datasets only include facial images.
This supports the findings of Salem et al.~\cite{SZHBFB19} and Li et al.~\cite{LLHYBZ22}, which also report the effectiveness of membership inference using shadow datasets from different domains.

Moreover, we present the average test accuracy and the average drop in accuracy of three attacks on different pre-trained models in~\autoref{appendix:mia_average}.
The results show that the \metric and \grad attacks achieve the best attack performance on average, while the \trad and \grad attacks, in general, are more robust than \metric attacks.
Hence, we conclude that the \grad attacks exhibit superior performance in terms of both utility and robustness after relaxing the dataset assumption.
However, it should be noted that the \grad attacks come at the cost of high computational resources and a significant amount of information needed.
Overall, our findings suggest that we can relax the assumption of the same-distribution shadow dataset, implying greater membership inference privacy risks of prompts.

\begin{figure*}[!t]
\centering
\begin{subfigure}{0.4\columnwidth}
\includegraphics[width=\columnwidth]{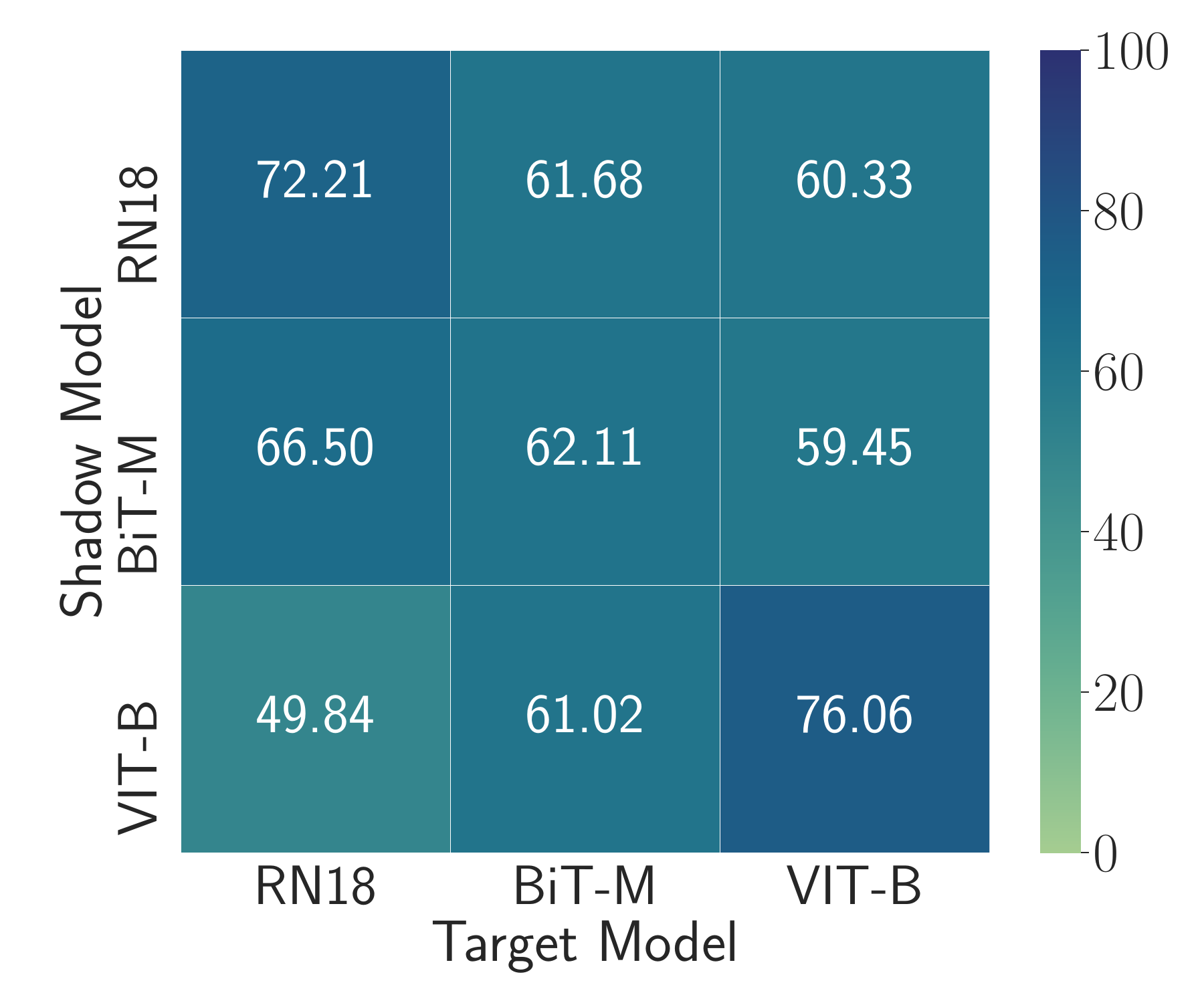}
\caption{NN-based}
\label{figure:diff_m_trad_celeba}
\end{subfigure}
\begin{subfigure}{0.4\columnwidth}
\includegraphics[width=\columnwidth]{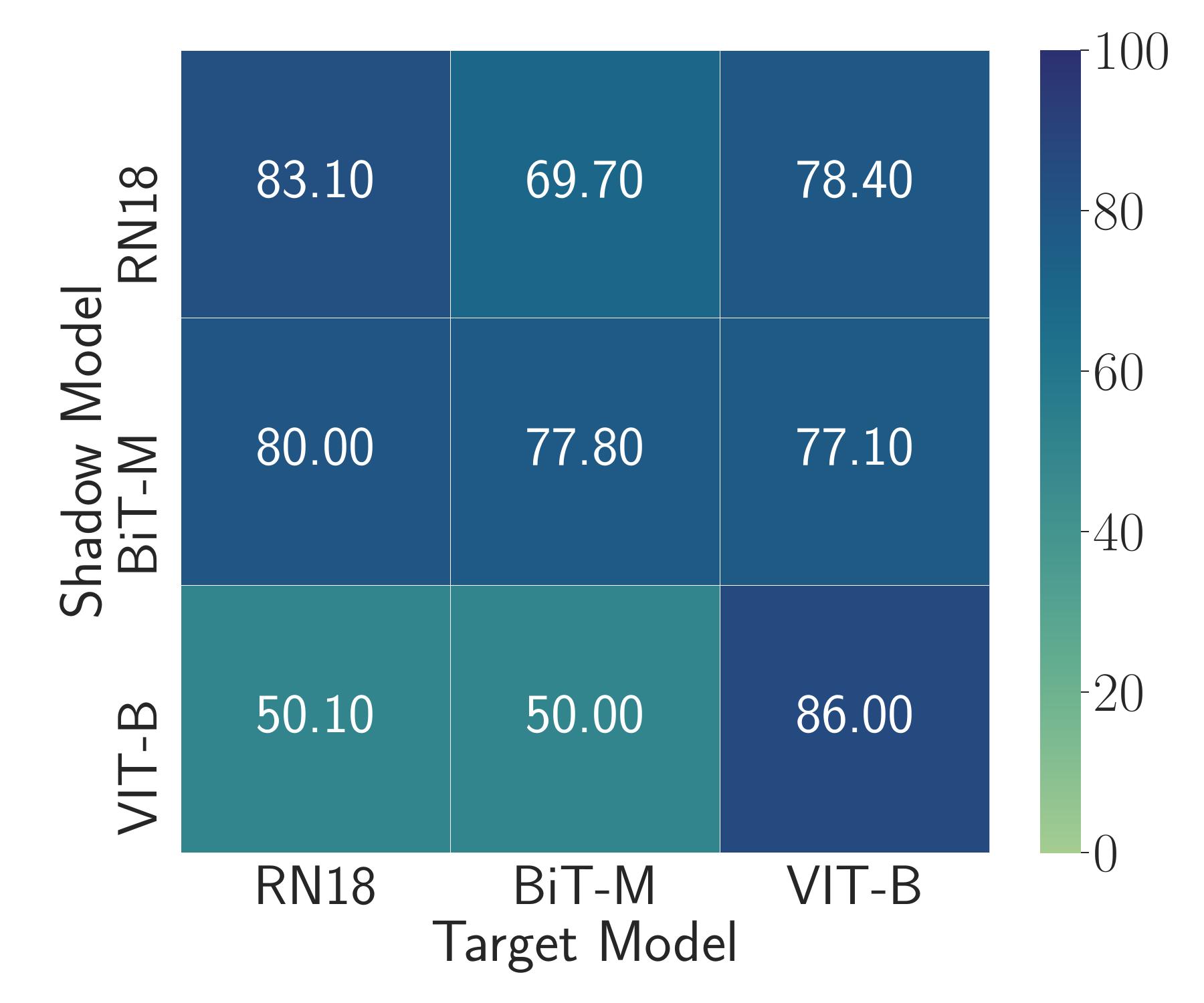}
\caption{Metric-based}
\label{figure:diff_m_metric_celeba}
\end{subfigure}
\begin{subfigure}{0.4\columnwidth}
\includegraphics[width=\columnwidth]{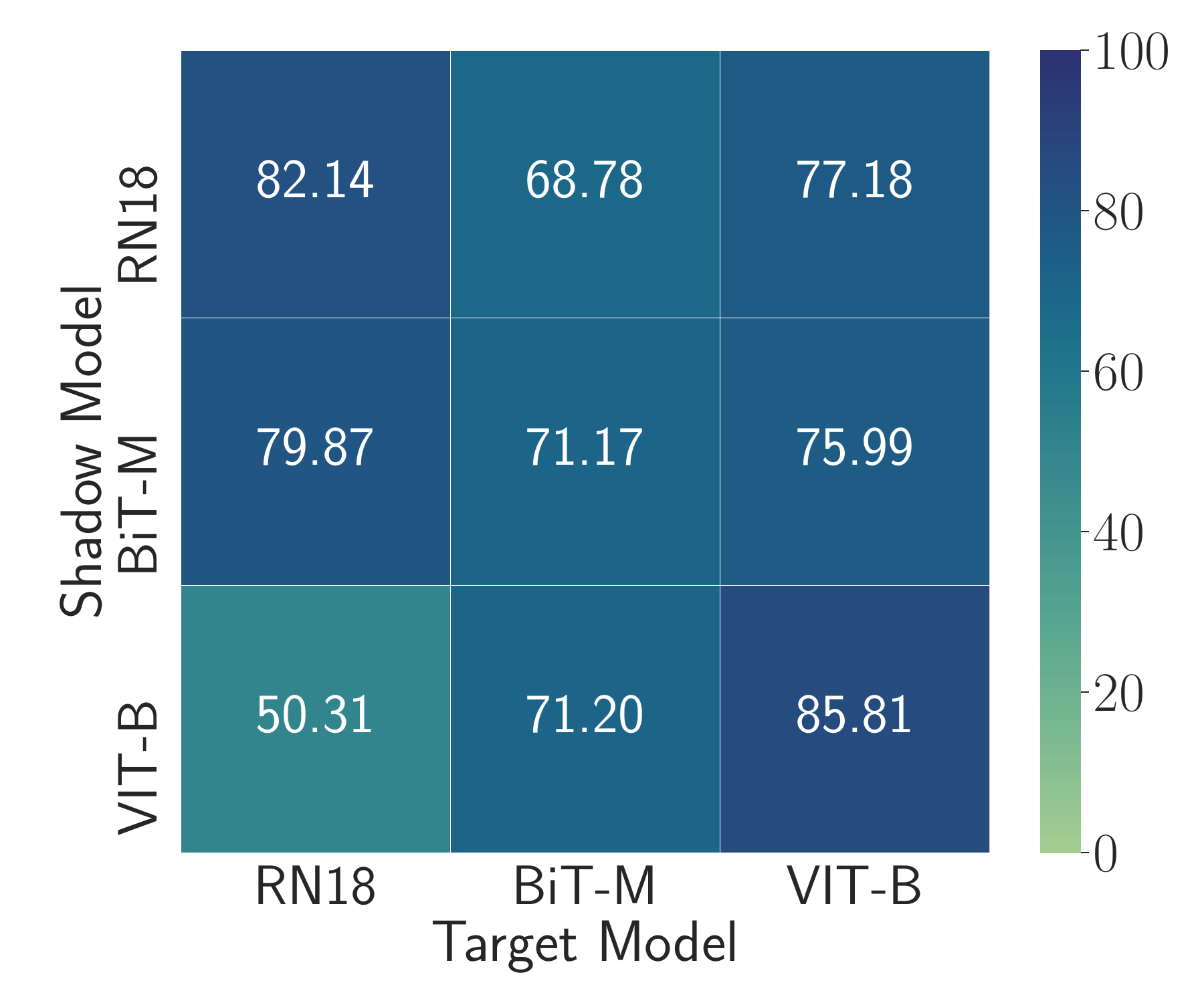}
\caption{Gradient-based}
\label{figure:diff_m_white_celeba}
\end{subfigure}
\caption{Attack performance of three attacks after relaxing the pre-trained model assumption on CelebA.}
\label{figure:diff_model_celeba}
\end{figure*}

\begin{figure}[!t]
\centering
\begin{subfigure}{0.4\columnwidth}
\includegraphics[width=\columnwidth]{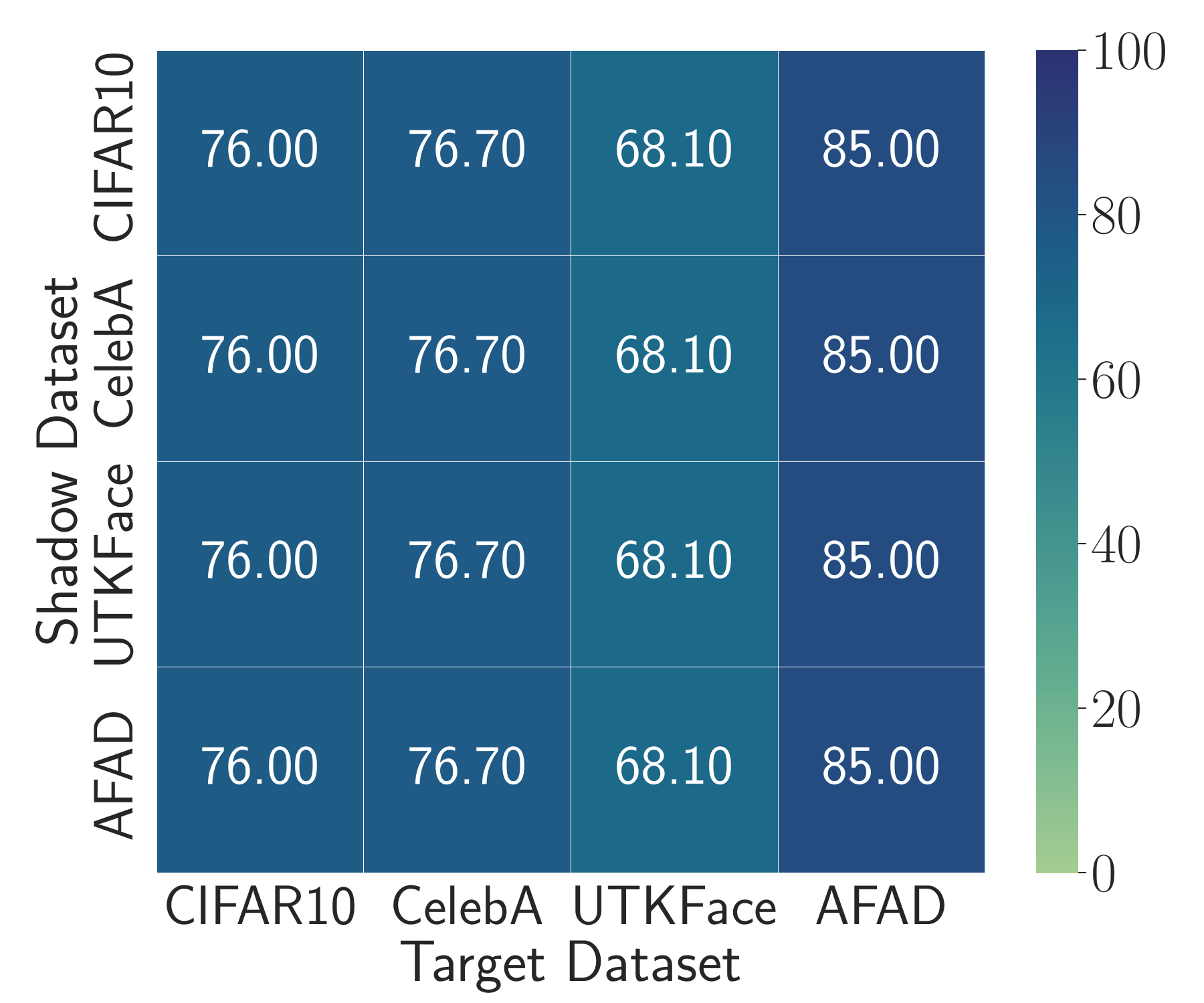}
\end{subfigure}
\begin{subfigure}{0.4\columnwidth}
\includegraphics[width=\columnwidth]{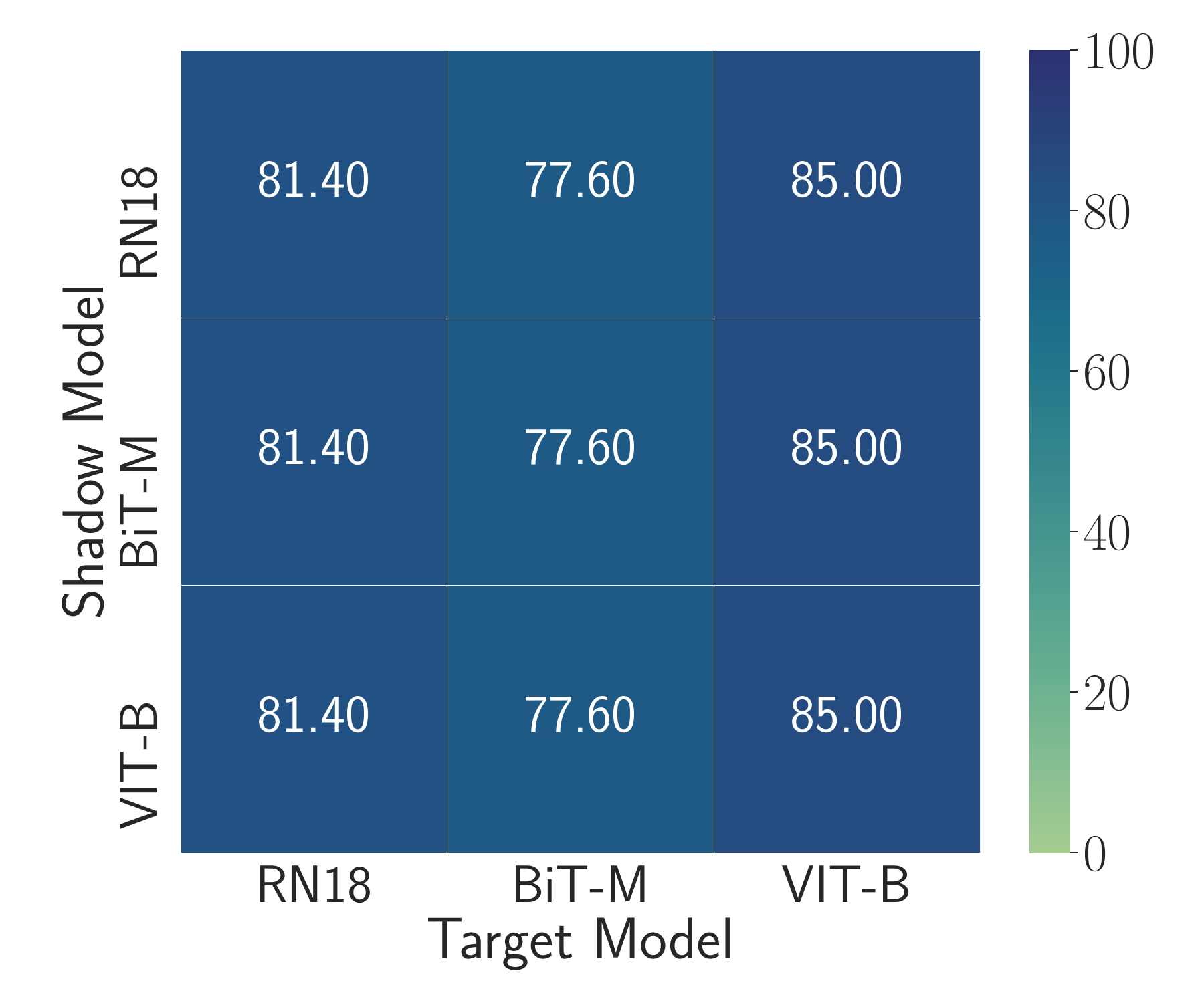}
\end{subfigure}
\caption{Attack performance of metric-corr attacks after relaxing (a) dataset assumption using ViT-B as the pre-trained model and (b) pre-trained model assumption on AFAD.}
\label{figure:metric_corr_relax_assumption}
\end{figure}

\mypara{Pre-trained Model Assumption}
In the previous evaluation, we assume the adversary has white-box access to the pre-trained model.
However, the PaaS provider may only allow users to submit prompted images and receive corresponding results, thus limiting access to the pre-trained model.
The adversary has to develop their pre-trained models, which may be different from the pre-trained models used to train the target prompts (abbreviated as the target model).
We, therefore, measure the impact of the discrepancy in architecture between the pre-trained model used to train the shadow prompts (abbreviated as shadow model) and the target model on the attack performance.
The results of three attacks are shown in~\autoref{figure:diff_model_celeba}.

In the diagonal of the heatmaps, we show the results of the adversary having white-box access to the same pre-trained model used to train the target prompt.
We observe that, in some cases, the attack performance decreases noticeably but remains effective.
For example, when the pre-trained model of the target prompt is ViT-B on CelebA, the performance of \metric attacks drops from 86.00\% to 78.40\% (77.10\%) when using RN18 (BiT-M) as the pre-trained model for the shadow prompt.
However, in certain cases, all three attacks fail completely, i.e., they become random guesses.
For instance, when the adversary uses ViT-B to attack the target prompt trained on RN18, these three methodologies become random guesses.
We also present the average test accuracy and the average drop in accuracy of three attacks on different datasets in~\autoref{appendix:mia_average}.
The \grad and \metric attacks achieve the best attack performance, and the \grad attacks are more robust than the \metric attacks.
However, the average drop in accuracy of all attacks after relaxing the pre-trained model assumption, in general, is larger than that of relaxing the dataset assumption.

\begin{figure*}[!t]
\centering
\begin{subfigure}{0.4\columnwidth}
\includegraphics[width=\columnwidth]{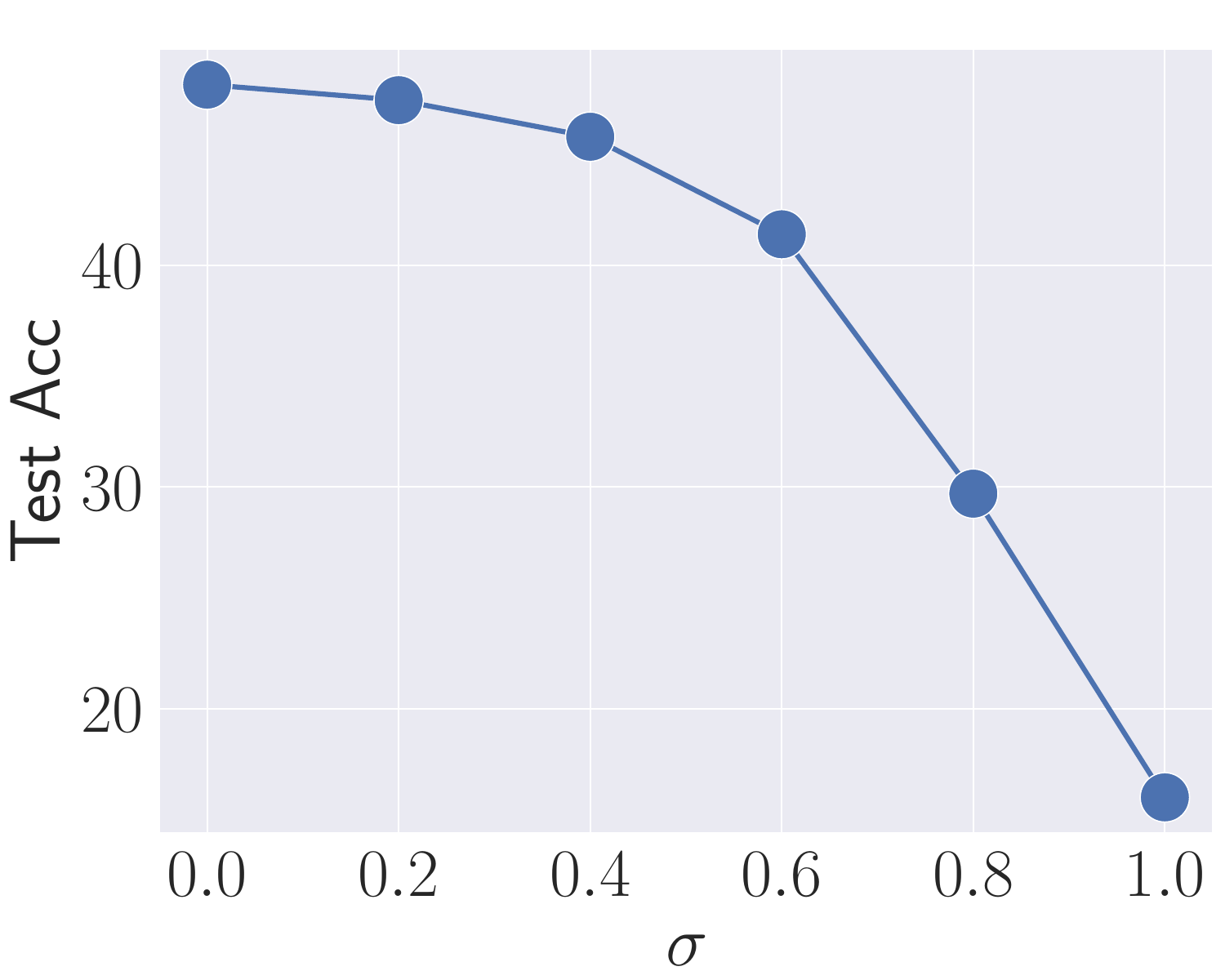}
\caption{Prompt Utility}
\label{figure:mia_utility_cifar10}
\end{subfigure}
\begin{subfigure}{0.4\columnwidth}
\includegraphics[width=\columnwidth]{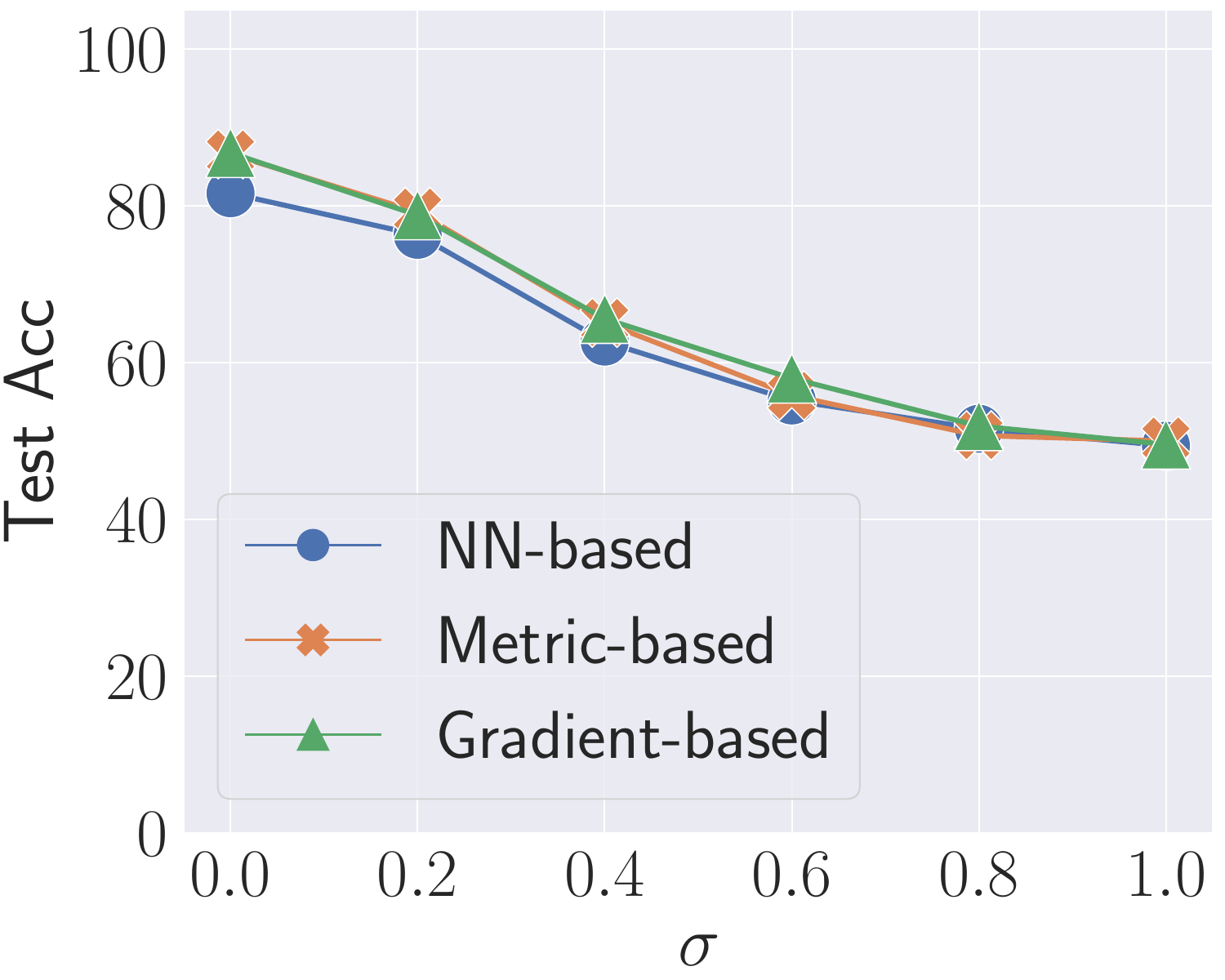}
\caption{Naive Attacks}
\label{figure:mia_noise_effect_cifar10}
\end{subfigure}
\begin{subfigure}{0.4\columnwidth}
\includegraphics[width=\columnwidth]{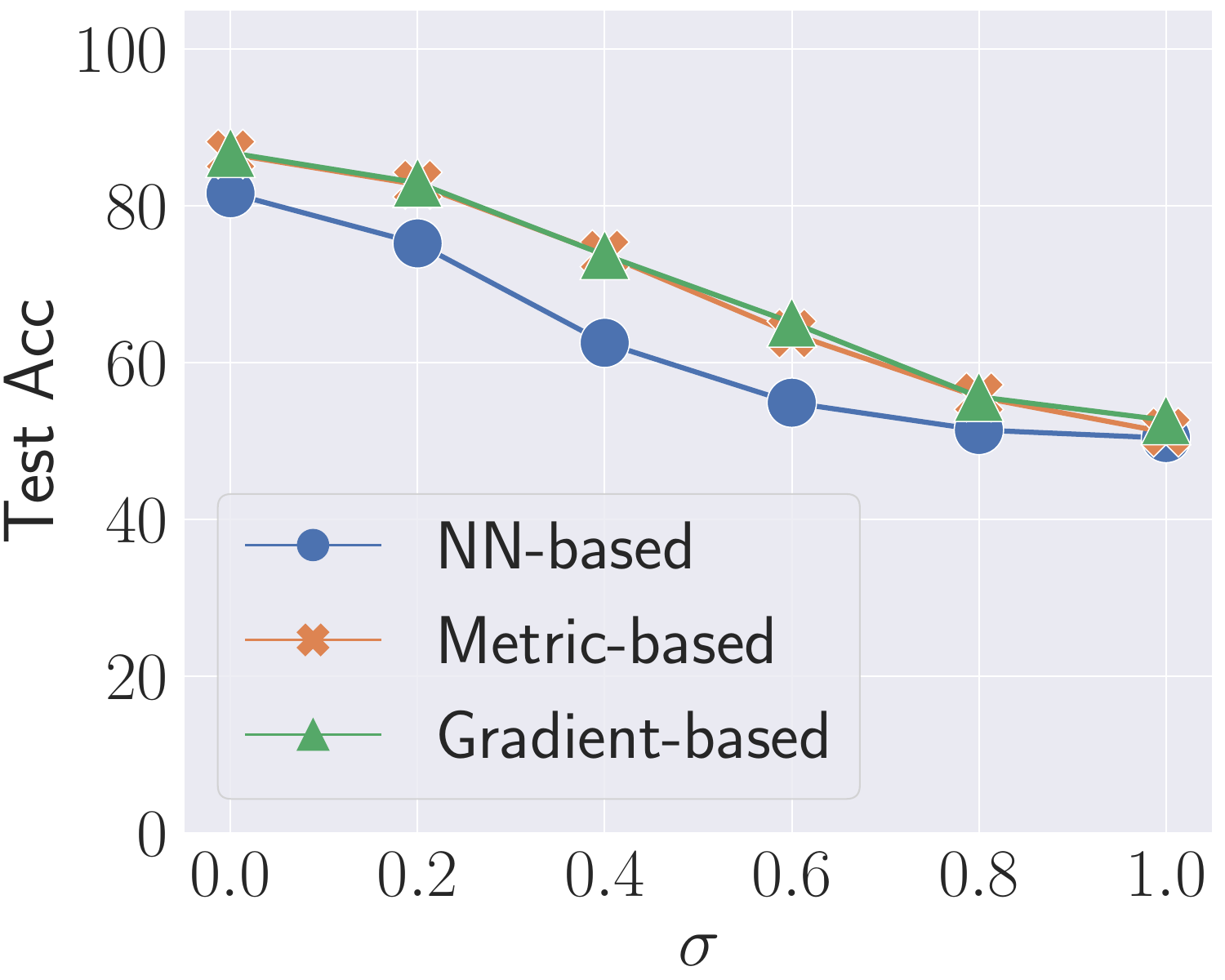}
\caption{Adaptive Attacks}
\label{figure:mia_noise_effect_cifar10_ad}
\end{subfigure}
\caption{Prompt utility and attack performance using the proposed defense on CIFAR10.}
\label{figure:mia_defense_cifar10}
\end{figure*}

\mypara{Discussion}
We have shown that all methodologies only have slight performance degradation after relaxing the dataset assumption.
Meanwhile, after relaxing the pre-trained model assumption, these attack methodologies are effective in some cases but fail to maintain high robustness, i.e., they fail in other cases.
Previous work~\cite{SZHBFB19, LLHYBZ22} on membership inference against traditional ML classifiers has shown that having shadow models with different architectures than the target models does not have a strong impact on the attack performance.
However, we do not observe the same in VPL.
One possible explanation is that a prompt is specific to the machine learning model it is trained on.
In other words, prompts from different models share less similarity, which makes the membership inference knowledge hard to transfer among them.
As illustrated in~\autoref{figure:metric_corr_relax_assumption}, we find that metric-corr attacks have no performance deterioration after relaxing these assumptions, as they do not rely on the shadow technique.
Thus, the adversary can leverage the metric-corr attacks when relaxing the data assumption and the pre-trained model assumption.

\mypara{Takeaways}
Our results show that the adversary can be data-free, as the attack performance only has a slight deterioration and remains effective.
The results also indicate that the adversary has some dependency on the knowledge of pre-trained models to steal private information, as not all attack methodologies can be successfully launched after relaxing the pre-trained model assumption.
However, we show that the adversary can still leverage the metric-corr attacks to obtain decent attack performance with high robustness, as they do not rely on the shadow technique.

\subsection{Defense}
\label{section:mia_defense}

\mypara{Gaussian Noise as Defense}
We have demonstrated that the prompts are also vulnerable to membership inference attacks.
Meanwhile, in the above experiments, we observe that the performance of membership inference is heavily related to the overfitting level of the target prompt.
Potentially, a defender can decrease the threat to membership privacy by reducing the overfitting level.
As shown in~\autoref{figure:mia_epoch} and~\autoref{figure:mia_size}, leveraging fewer epochs and more data to train the target prompt can decrease the attack performance to some extent.
However, using these methods comes at the cost of either the utility of the target prompt or the resource to collect and process data.
We also apply the widely adopted Differential Privacy-Stochastic Gradient Descent (DP-SGD)~\cite{ACGMMTZ16}, which involves adding noise to clipped gradients, as a defense mechanism.
However, the experimental results show that it is hard to maintain the prompt utility even with a larger privacy budget, e.g., $\epsilon=20$.
We hypothesize that DP-SGD may work on large datasets, but not on the data for prompt learning since it is relatively small.
Hence, we investigate if the defense mechanism used in~\autoref{section:pia_defense}, i.e., adding Gaussian noise to the prompts, can reduce the risks of membership leakage.
We set $\sigma \in \{0.2, 0.4, 0.6, 0.8, 1.0\}$.
We first report the target performance on CIFAR10 in~\autoref{figure:mia_utility_cifar10}.
The evaluation metric is the test accuracy of the target prompt.
We observe that the target performance, i.e., prompt utility, only decreases heavily when $\sigma \geq 0.6$.
For example, the prompt utility remains above 41.40\% when $\sigma \leq 0.6$ and then decreases heavily from 41.40\% to 29.70\% on CIFAR10 when increasing $\sigma$ from 0.6 to 0.8.
We then present the attack performance where the adversary is unaware of the defense mechanism in~\autoref{figure:mia_noise_effect_cifar10}.
We can observe that all attacks are close to random guesses when $\sigma \geq 0.6$, showing that there is a practical utility-defense trade-off when $\sigma=0.6$.

\mypara{Adaptive Attacks}
We further consider an adaptive adversary~\cite{JSBZG19} who is aware of the defense mechanism.
Hence, the adversary can craft their attack training datasets using the shadow prompt with Gaussian noise.
We set $\sigma \in \{0.2, 0.4, 0.6, 0.8, 1.0\}$ for both shadow and target prompts.
We report the performance of adaptive attacks on CIFAR10 in~\autoref{figure:mia_noise_effect_cifar10_ad}.
The results show that the attack performance is still close to random guess when $\sigma \geq 0.6$.

\mypara{Takeaways}
When defending against membership inference attacks, the proposed defense mechanism can achieve a decent utility-defense trade-off when setting $\sigma=0.6$.
A similar conclusion can be drawn from the other three datasets.

\section{Related Work}
\label{section:related_work}

\mypara{Property Inference Attacks}
Property inference~\cite{AMSVVF15,GWYGB18,ZCSZ22,ZCBSZ22,MGC22,CAOJTU23} aims to extract sensitive global properties of the training data distribution from an ML model that the model owner does not want to share.
It is an important privacy attack against ML models, as it can violate prompt owner's privacy, i.e., proprietary information about the dataset, and enable attackers to perform tailored attacks, e.g., enhancing membership inference attacks~\cite{ZCSZ22}.
The main approach for launching these attacks is building a meta-classifier on a large number of shadow models~\cite{AMSVVF15}.
Existing work focuses on deep neural networks, including fully connected neural networks~\cite{GWYGB18}, generative adversarial networks (GANs)~\cite{ZCSZ22}, and graph neural networks (GNNs)~\cite{ZCBSZ22}.

\mypara{Membership Inference Attacks}
Membership inference~\cite{SSSS17,SZHBFB19,SM21,NSH19,LF20,LZ21,LJQG21} is another important type of privacy attack against ML models, where the adversary aims to infer whether the given data sample was involved in a target model's training dataset.
Shokri et al.~\cite{SSSS17} propose the first membership inference attack which depends on training multiple shadow models for developing their attack models.
Salem et al.~\cite{SZHBFB19} then relax assumptions proposed by Shokri et al.~\cite{SSSS17}.
Yeom et al.~\cite{YGFJ18} attribute the vulnerability of membership inference to the overfitting of ML models.
Song and Mittal~\cite{SM21} propose metric-based attacks that rely on pre-calculated thresholds over shadow models to determine the membership status.
Nasr et al.~\cite{NSH19} perform a thorough investigation of membership privacy in both black-box and white-box settings for both centralized and federated learning scenarios.
More recently, Liu et al.~\cite{LZBZ22} leverage the loss trajectory to further enhance the attack performance.
Most recent work focuses on deep neural networks, including GNNs~\cite{WYPY21, HWWBSZ21}, multi-modal models~\cite{HWSWX22}, and multi-exit networks~\cite{LLHYBZ22}.

Previous work has demonstrated that the fine-tuning paradigm is vulnerable to these privacy attacks~\cite{CTCP21,MUWEB22}.
The privacy risk in the fine-tuning paradigm resides at the model level, as the privacy information is leaked through fine-tuned models.
This differs from the privacy risk associated with the prompt learning paradigm, where the risk lies at the input level, as the prompt exists in the pixel space.

\section{Limitation and Future Work}
\label{section:limitation}

\mypara{Efficacy of VPL}
VPL is an emerging ML paradigm.
Although its current performance cannot rival that of a fine-tuned model, an increasing number of studies are attempting to enhance its performance through various approaches, e.g., label mapping~\cite{CYCZL22} and better data homogeneity~\cite{HDCZWHY23}.
Since we are the first to explore the vulnerabilities of the visual prompt, we have focused on the widely recognized VPL paradigm and followed their default training settings~\cite{BJSI22}.
We anticipate that as VPL with enhanced performance are introduced in the future, it will be straightforward for us to extend our measurement, and thus we recognize this as a promising avenue for future research.

\mypara{Defense} 
In the evaluation, we show that adding Gaussian noise to the prompt can mitigate the membership inference attacks with a decent utility-defense trade-off but fails to defend against property inference attacks.
DP-SGD fails to preserve the original prompt utility.
Since the privacy risk in the prompt learning paradigm is at the input level, devising diverse defense mechanisms for it is more challenging compared to addressing the privacy risk at the model level.
We leave it as a future work to explore effective defenses against property inference attacks.

\mypara{NLP Prompt Learning}
Another interesting future work is to apply the two proposed privacy attacks along with their motivation to prompt learning in the NLP domain~\cite{LAC21, LL21}, as the NLP prompt is essentially a (soft) token that can be added to the text input, operating at the input level.

\section{Conclusion}
\label{section:conclusion}

In this paper, we conduct the first privacy assessment of prompts learned by VPL through the lens of property inference attacks and membership inference attacks.
Our empirical evaluation shows that prompts are vulnerable to both of these attacks.
Moreover, we have discovered that an adversary can successfully mount the property inference attacks by training only a few shadow prompts.
They can also relax the dataset assumption to achieve effective membership inference attacks.
We further make some initial investigations on possible defenses.
Experiments show that our method, i.e., adding Gaussian noise to prompts, can mitigate the membership inference attacks with a decent utility-defense trade-off but fails to defend against property inference attacks.
We hope our results can raise the awareness of the stakeholders when deploying prompt learning in real-world applications.
Moreover, we will share our code and models to facilitate research in this field.

\medskip
\mypara{Acknowledgements}
We thank all anonymous reviewers for their constructive comments.
This work is partially funded by the European Health and Digital Executive Agency (HADEA) within the project ``Understanding the individual host response against Hepatitis D Virus to develop a personalized approach for the management of hepatitis D'' (D-Solve) (grant agreement number 101057917).

\bibliographystyle{plain}
\bibliography{normal_generated_py3}

\appendix
\section*{Appendix}
\label{section:appendix}

\section{Details of Pre-trained Models}
\label{appendix:detail_models}

\autoref{table:pretrained_models} shows details about the architectures and pre-trained datasets of the pre-trained models used in the paper.

\begin{table}[!ht]
\caption{Overview of the pre-trained models.}
\label{table:pretrained_models}
\centering
\renewcommand{\arraystretch}{1.1}
\scalebox{0.6}{
\begin{tabular}{c|c|c|c}
\toprule
Model & Architecture & Pre-trained Dataset & \# Parameters \\
\midrule
RN18 &  ResNet-18 & 1.2M ImageNet-1k & 11,173,962 \\
BiT-M & ResNet-50 & 14M ImageNet-21k & 23,520,842 \\
ViT-B & ViT-B/16 & 14M ImageNet-21k & 86,567,656 \\
\bottomrule
\end{tabular}}
\end{table}

\section{Relax Assumptions of Property Inference Attacks}
\label{appendix:pia_append}

\begin{figure*}[!ht]
\centering
\begin{subfigure}{0.4\columnwidth}
\includegraphics[width=\columnwidth]{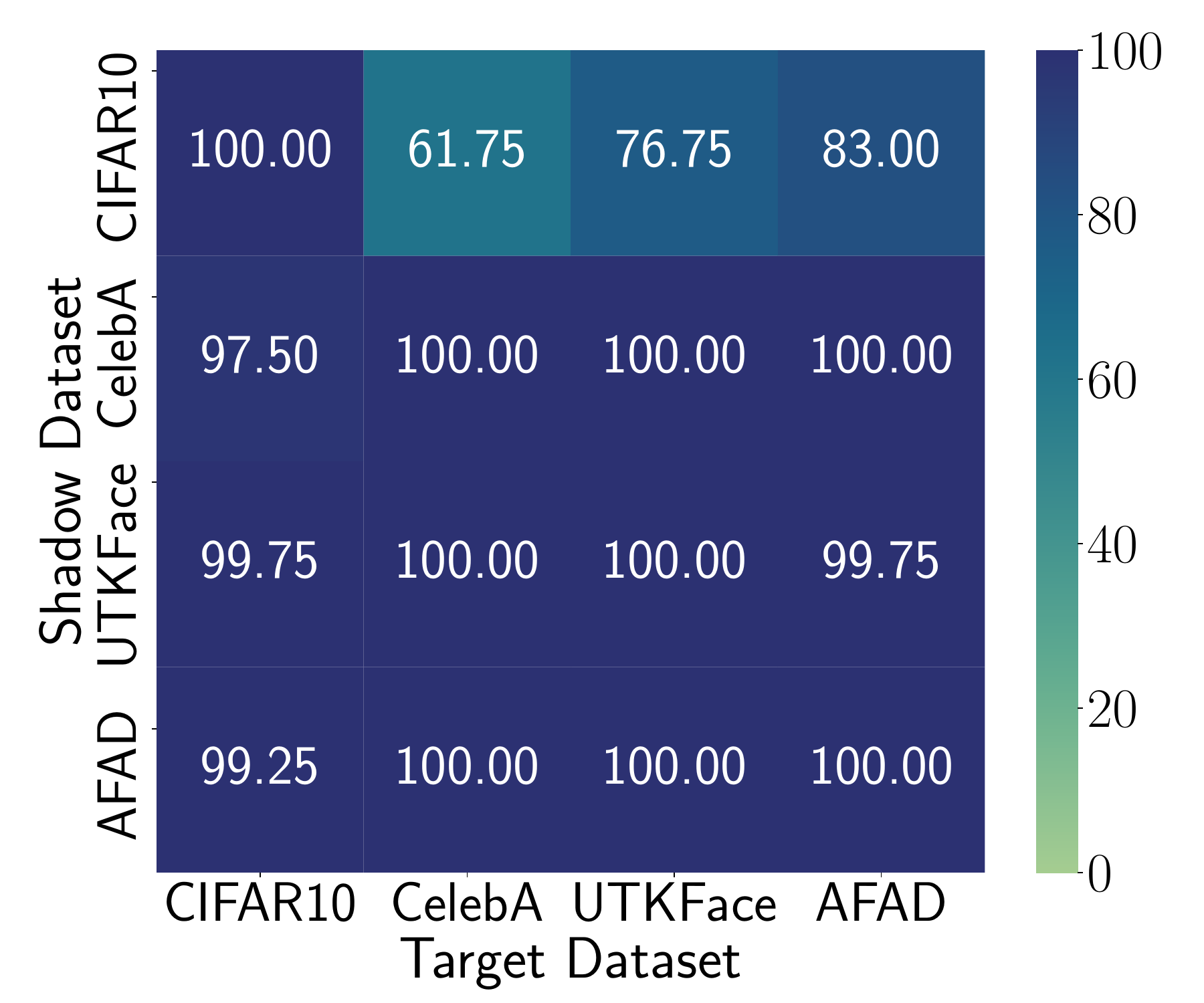}
\caption{Size}
\label{figure:pia_diff_d_size}
\end{subfigure}
\begin{subfigure}{0.4\columnwidth}
\includegraphics[width=\columnwidth]{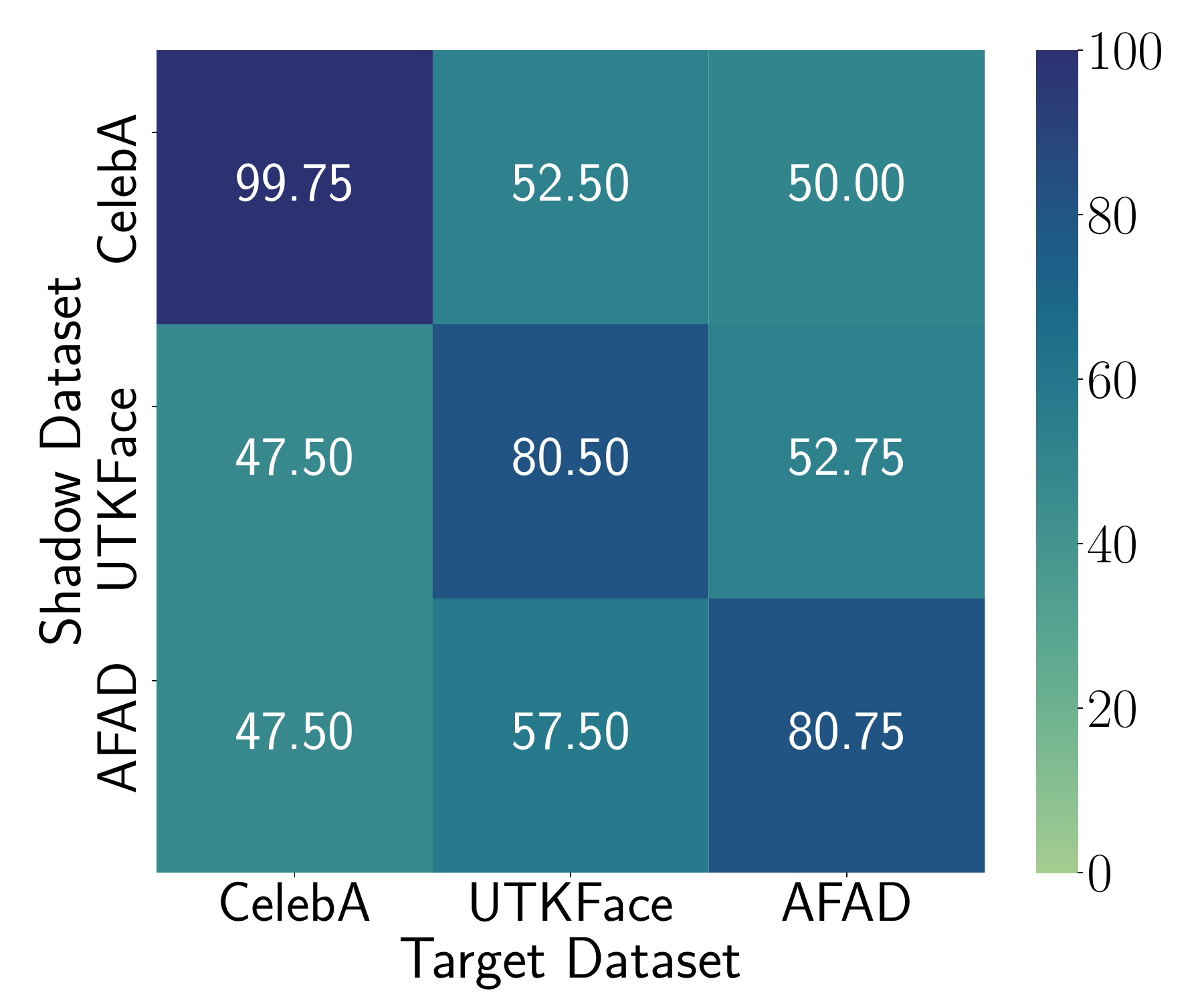}
\caption{Proportion of Males}
\label{figure:pia_diff_d_male}
\end{subfigure}
\begin{subfigure}{0.4\columnwidth}
\includegraphics[width=\columnwidth]{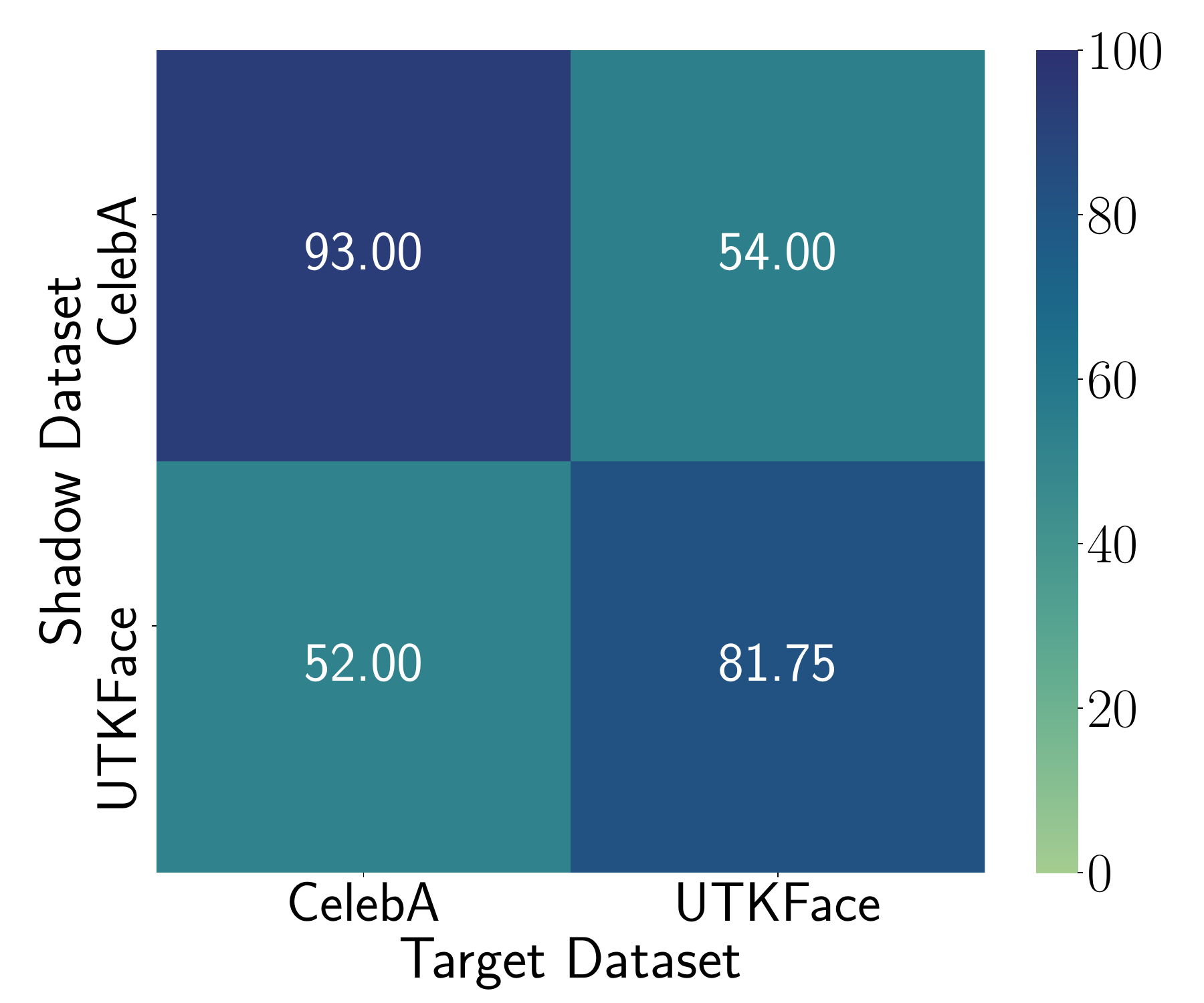}
\caption{Proportion of Youth}
\label{figure:pia_diff_d_youth}
\end{subfigure}
\caption{Attack performance of property inference attacks after relaxing the dataset assumption, using RN18 as the pre-trained model.}
\label{figure:pia_diff_d}
\end{figure*}

\begin{figure*}[!ht]
\centering
\begin{subfigure}{0.4\columnwidth}
\includegraphics[width=\columnwidth]{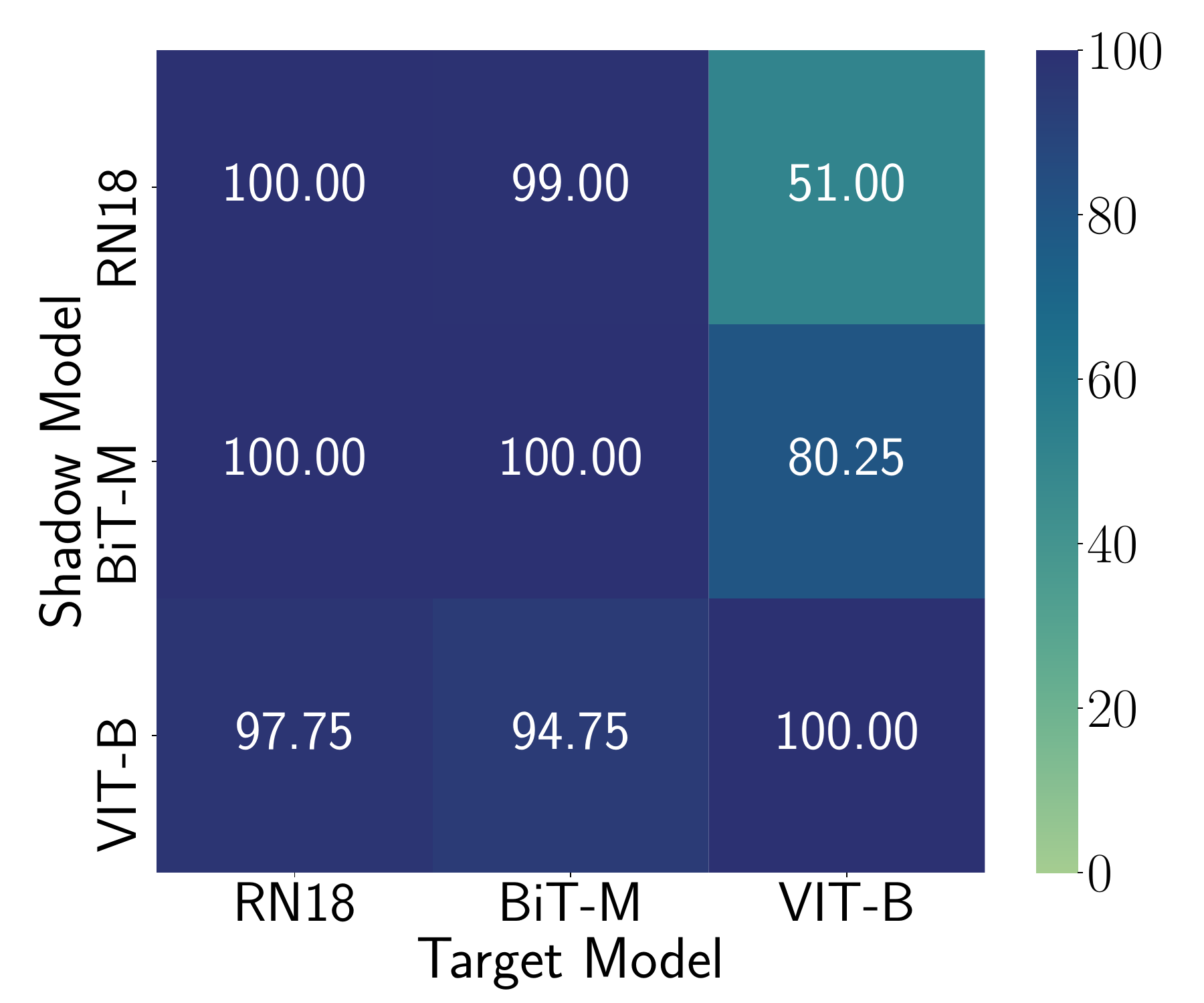}
\caption{Size}
\label{figure:pia_diff_m_size}
\end{subfigure}
\begin{subfigure}{0.4\columnwidth}
\includegraphics[width=\columnwidth]{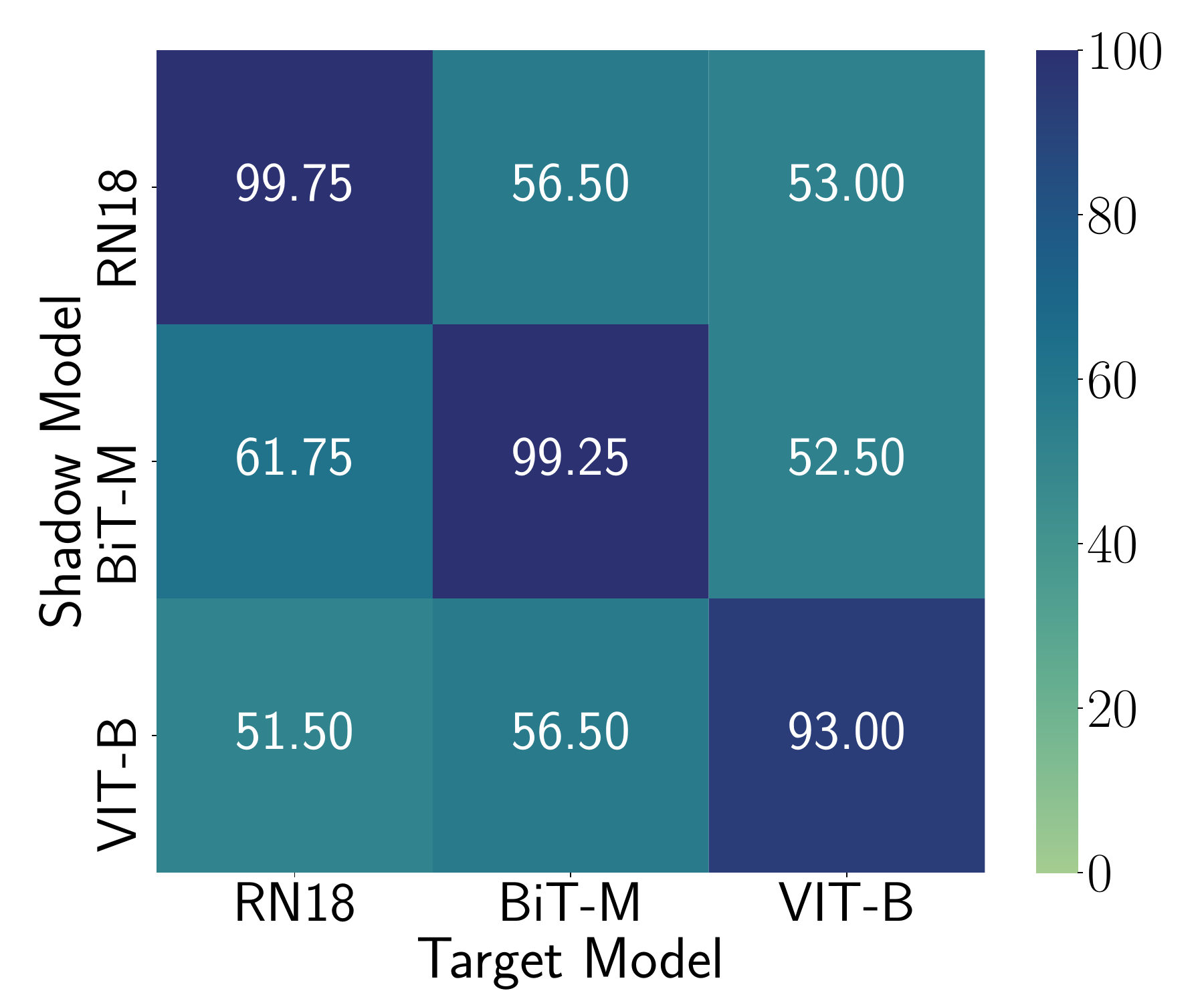}
\caption{Proportion of Males}
\label{figure:pia_diff_m_male}
\end{subfigure}
\begin{subfigure}{0.4\columnwidth}
\includegraphics[width=\columnwidth]{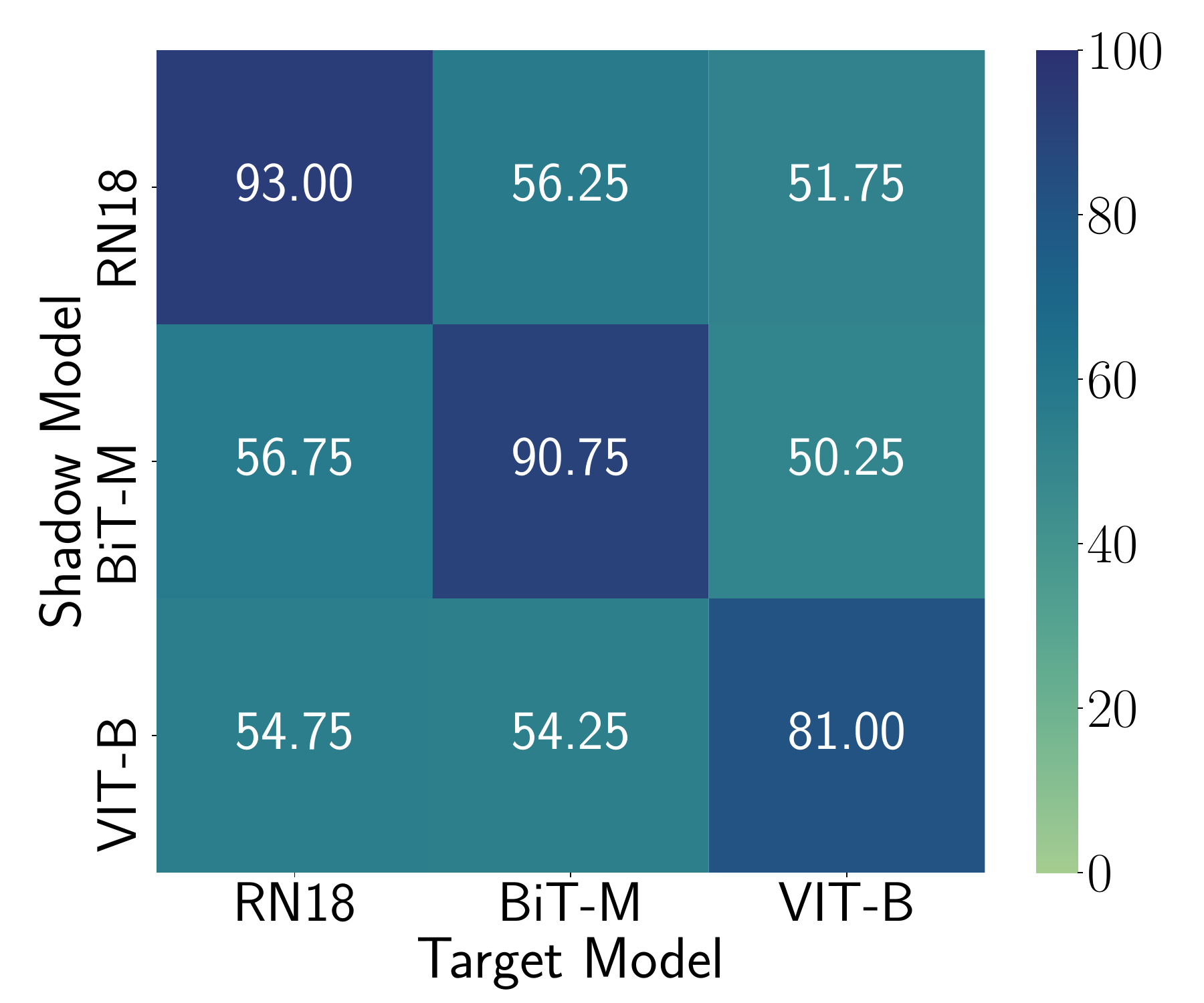}
\caption{Proportion of Youth}
\label{figure:pia_diff_m_youth}
\end{subfigure}
\caption{Attack performance of property inference attacks after relaxing the pre-trained model assumption on CelebA.}
\label{figure:pia_diff_m}
\end{figure*}

We first relax the assumption that the adversary has a shadow dataset of similar distribution as the target dataset, i.e., the dataset assumption.
As shown in~\autoref{figure:pia_diff_d}, we observe that the performance of the property inference attack is close to a random guess, considering the proportion of males and youth as the target properties.
When inferring the prompt training dataset size, the property inference attacks maintain high accuracy in most cases.
However, the attack performance decreases significantly when leveraging CIFAR10 as the shadow dataset.
We reason this is because the CIFAR10 datasets contain objections such as cars and birds, whereas other datasets only contain facial images.
We then relax the assumption that the pre-trained model used to train the shadow prompts and target prompts are the same, i.e., the pre-trained model assumption.
As illustrated in~\autoref{figure:pia_diff_m}, we notice that the attack performance has a significant degradation when inferring the proportion of males and youth.
When inferring the prompt training dataset size, the property inference attacks are effective in some cases.
However, they are not robust, as they become random guesses in certain cases.
For example, when the target prompt is trained on ViT-B and the shadow prompt is trained on RN18, the property inference attacks fail.
Thus, we conclude that it is necessary to leverage a shadow dataset of similar distribution as the target dataset and the same pre-trained model to train the shadow prompt.

\section{Performance of Metric-based Attacks With Different Metrics}
\label{appendix:diff_metric_attack}

\begin{figure*}[!ht]
\centering
\begin{subfigure}{0.4\columnwidth}
\includegraphics[width=\columnwidth]{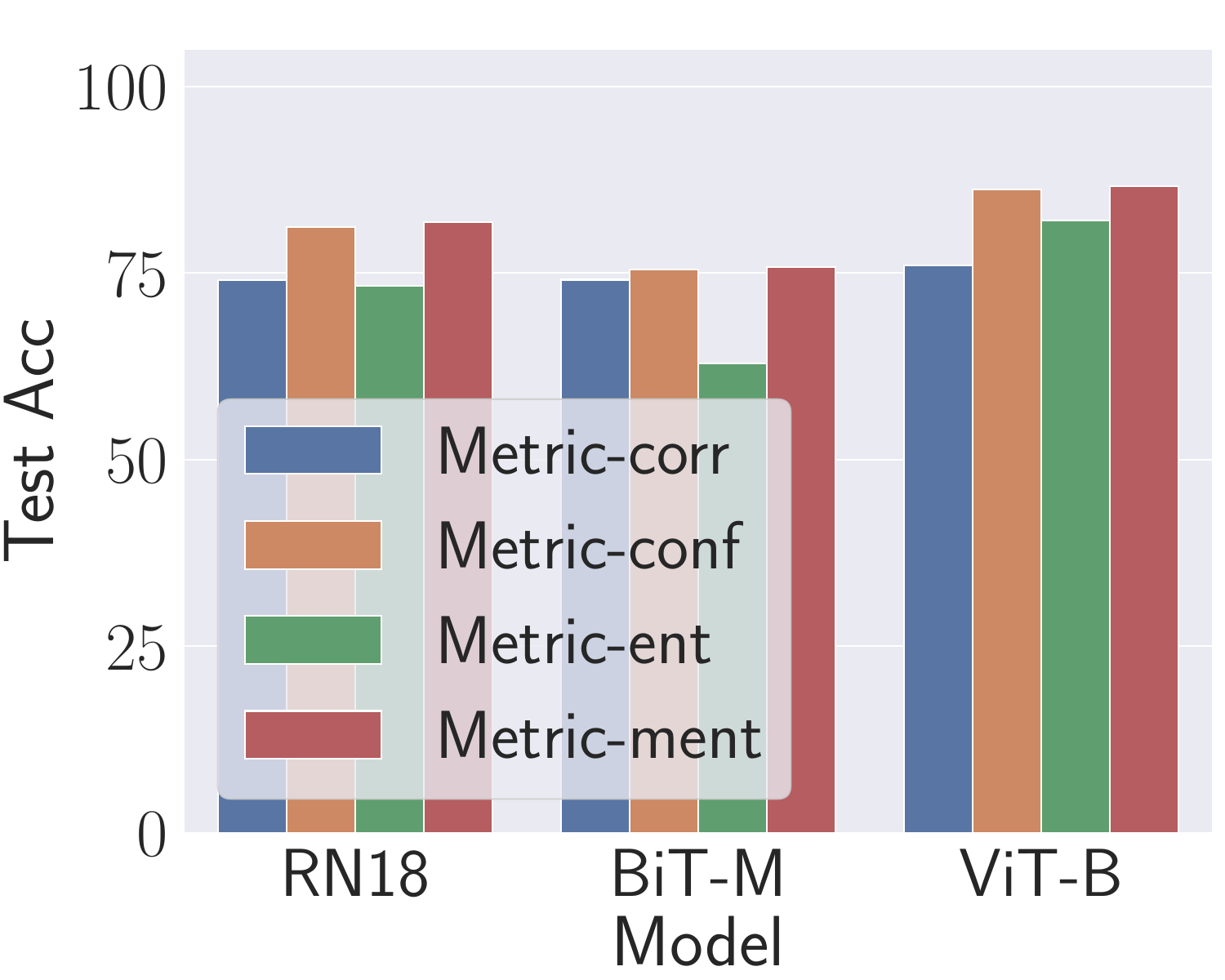}
\caption{CIFAR10}
\label{figure:metric_main_result_cifar10}
\end{subfigure}
\begin{subfigure}{0.4\columnwidth}
\includegraphics[width=\columnwidth]{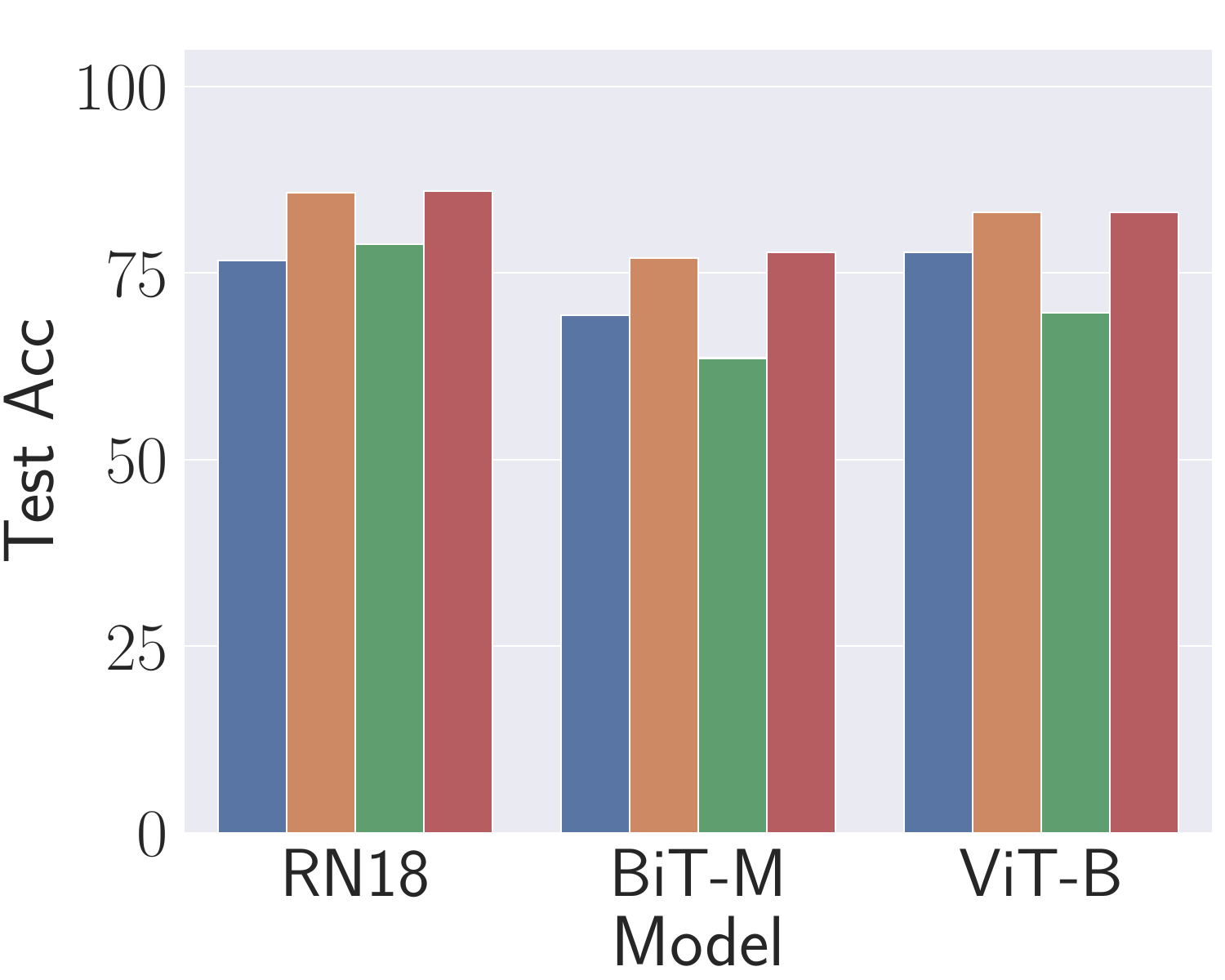}
\caption{CelebA}
\label{figure:metric_main_result_celeba}
\end{subfigure}
\begin{subfigure}{0.4\columnwidth}
\includegraphics[width=\columnwidth]{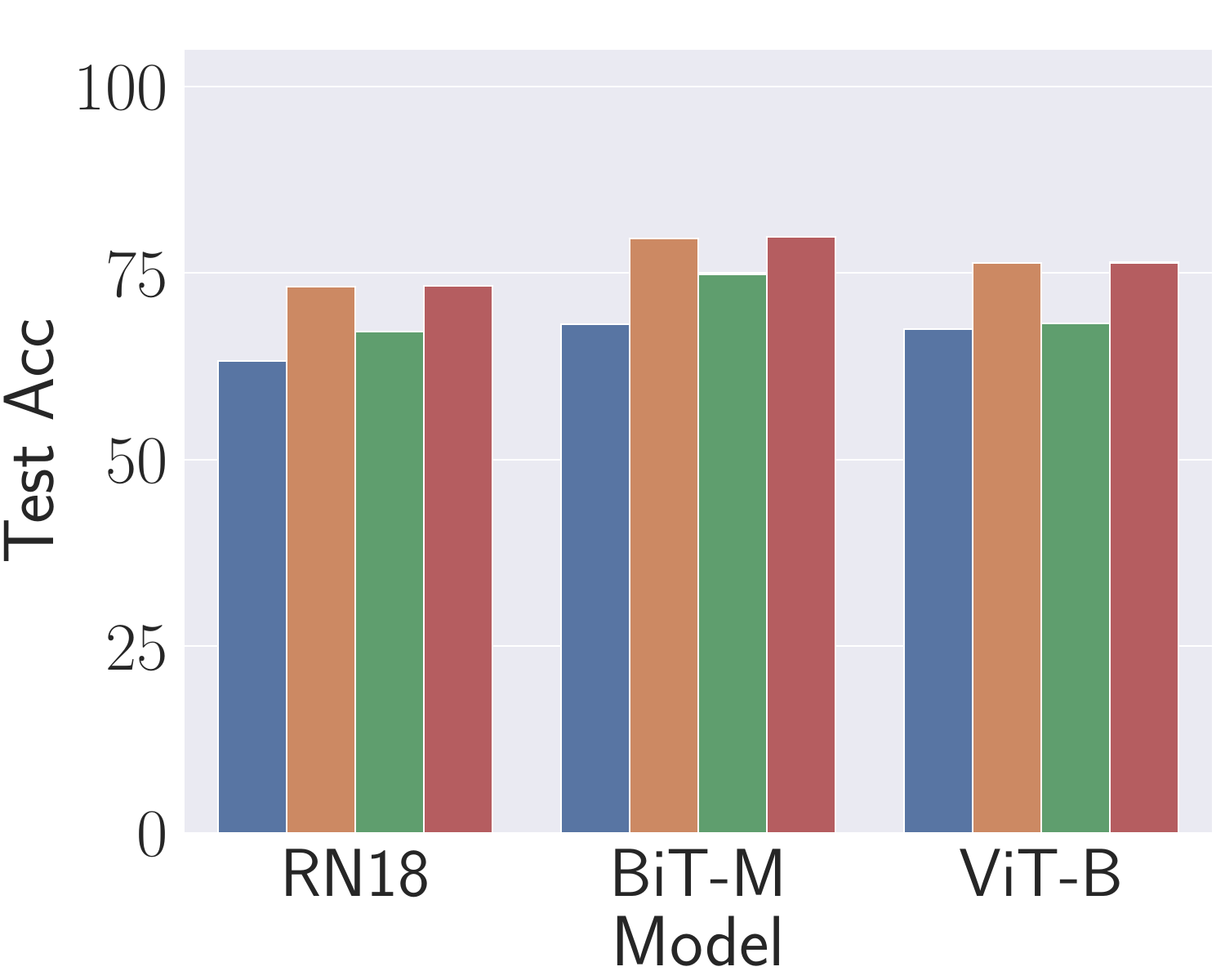}
\caption{UTKFace}
\label{figure:metric_main_result_utkface}
\end{subfigure}
\begin{subfigure}{0.4\columnwidth}
\includegraphics[width=\columnwidth]{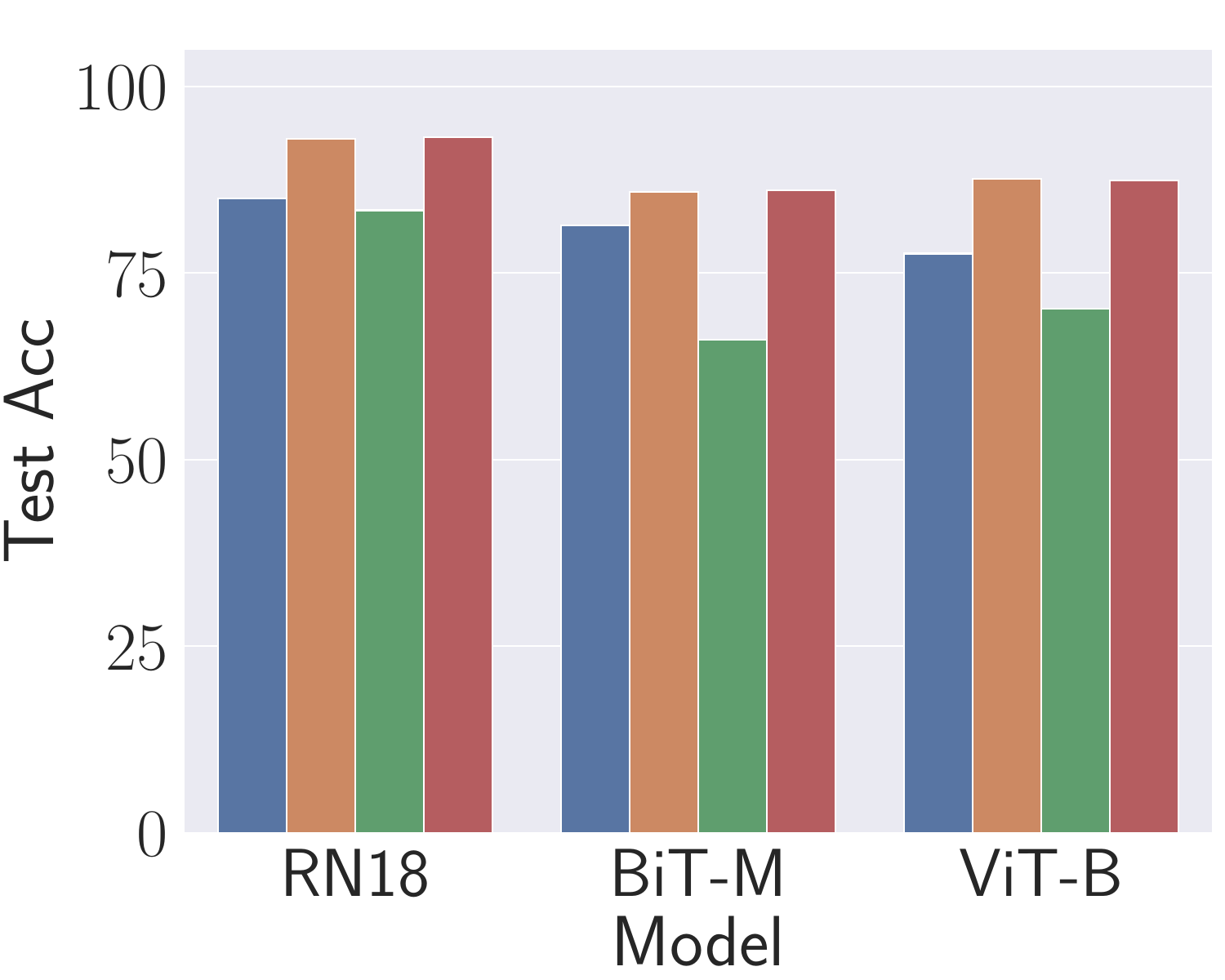}
\caption{AFAD}
\label{figure:metric_main_result_afad}
\end{subfigure}
\caption{Attack performance of four \metric attacks on four datasets.}
\label{figure:metric_mia_main_result}
\end{figure*}

As shown in~\autoref{figure:metric_mia_main_result}, metric-conf and metric-ment attacks achieve the best performance in all cases.
The reason why they work better than the other two metrics is that they take both prediction correctness and confidence into account, while the other two metrics only consider prediction correctness.

\section{Average Test Accuracy and Drop in Accuracy}
\label{appendix:mia_average}

\begin{figure}[!ht]
\centering
\begin{subfigure}{0.4\columnwidth}
\includegraphics[width=\columnwidth]{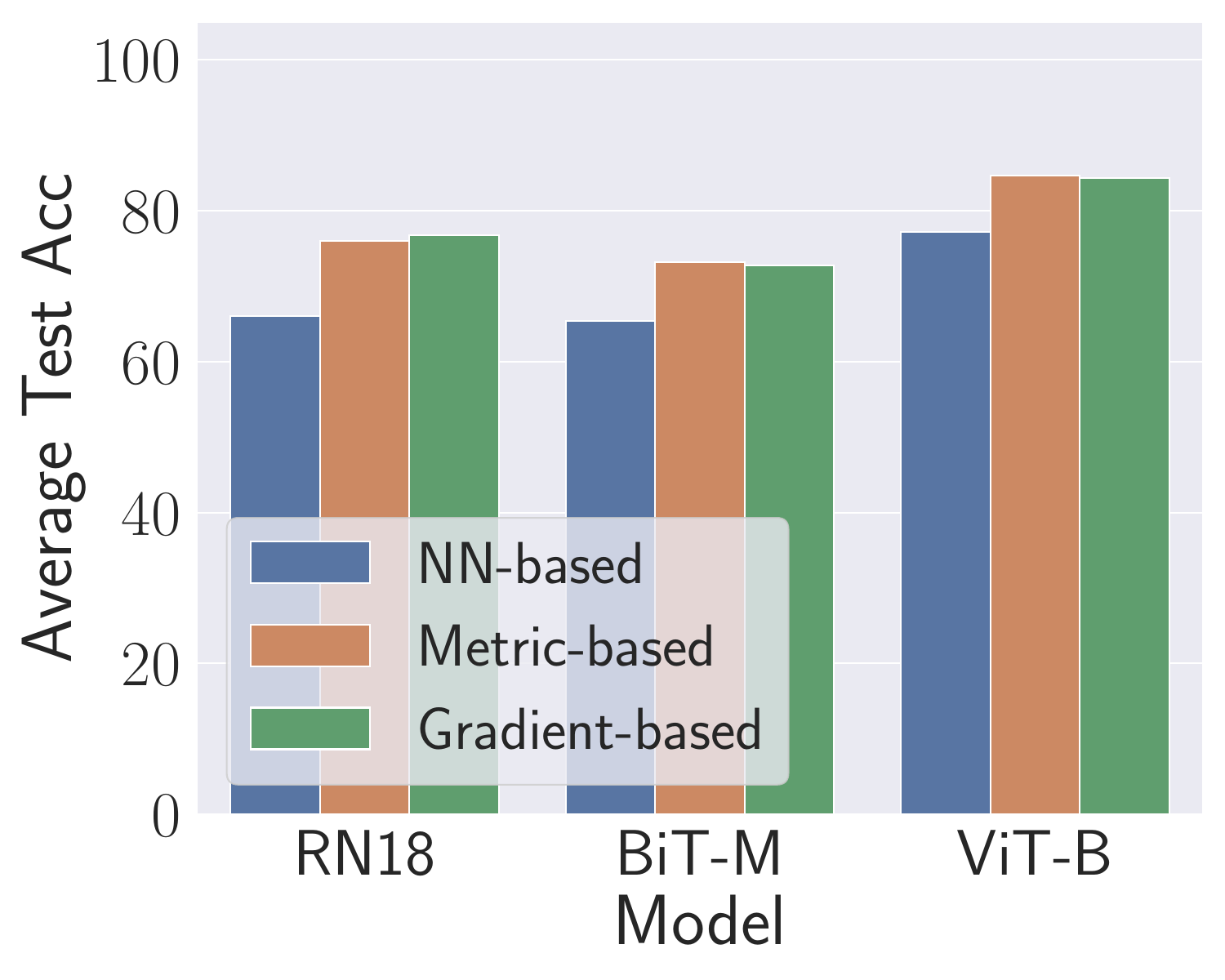}
\caption{Test Accuracy}
\label{figure:mia_diff_d_effect_average_test_acc}
\end{subfigure}
\begin{subfigure}{0.4\columnwidth}
\includegraphics[width=\columnwidth]{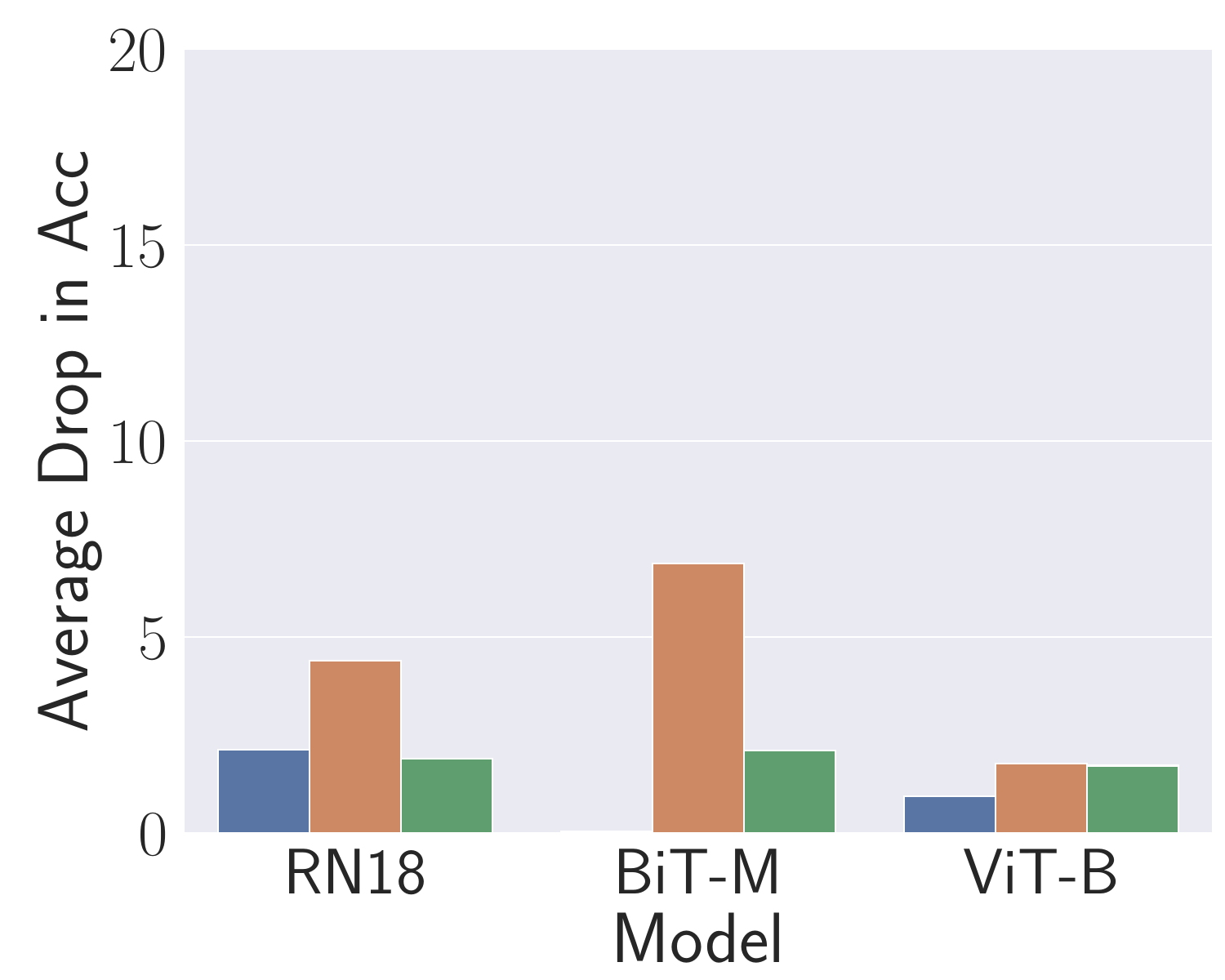}
\caption{Drop in Accuracy}
\label{figure:mia_diff_d_effect_average_drop_acc}
\end{subfigure}
\caption{Average test accuracy and drop in accuracy of three attacks after relaxing the dataset assumption.}
\label{figure:mia_diff_d_effect_average}
\end{figure}

\begin{figure}[!ht]
\centering
\begin{subfigure}{0.4\columnwidth}
\includegraphics[width=\columnwidth]{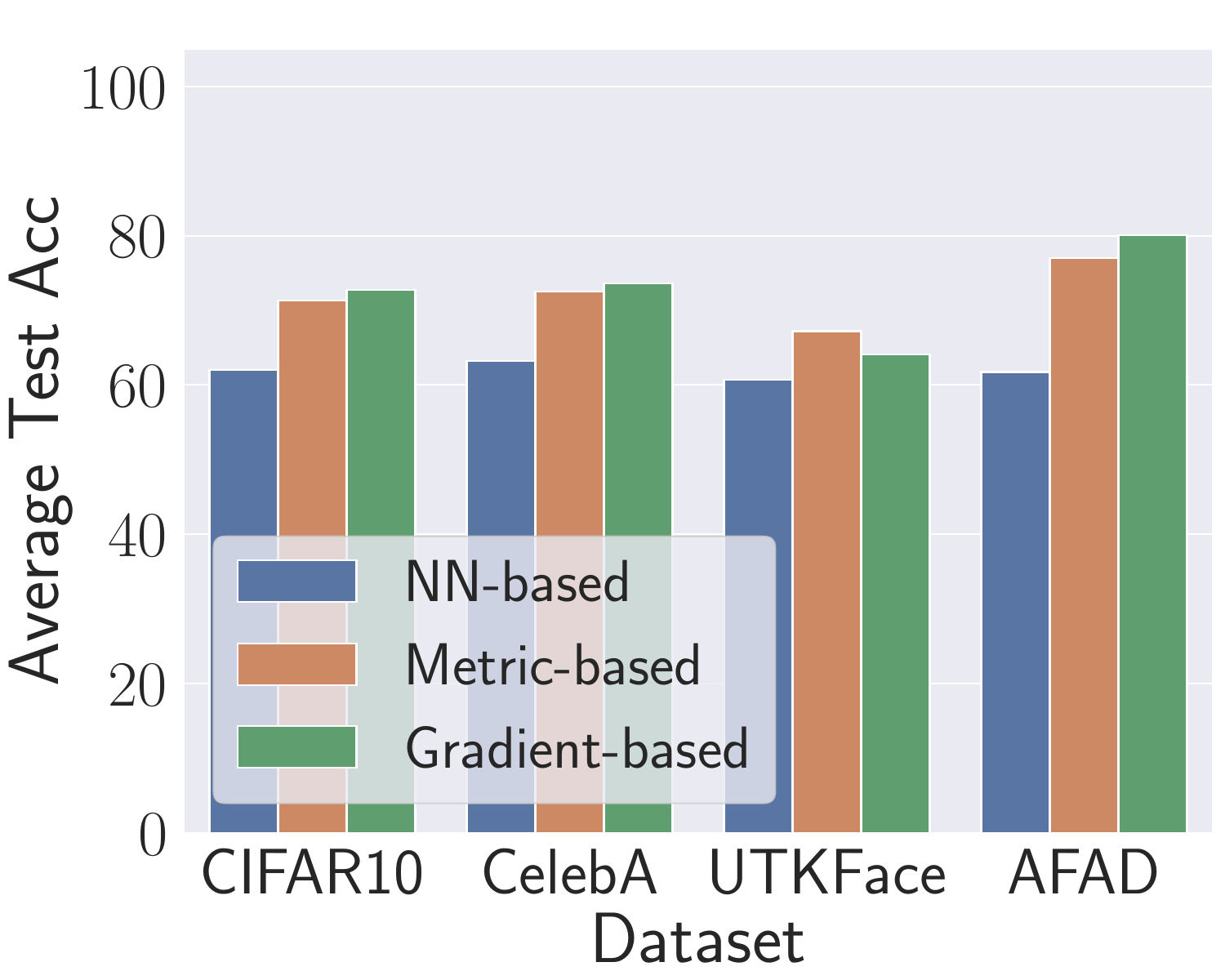}
\caption{Test Accuracy}
\label{figure:mia_diff_m_effect_average_test_acc}
\end{subfigure}
\begin{subfigure}{0.4\columnwidth}
\includegraphics[width=\columnwidth]{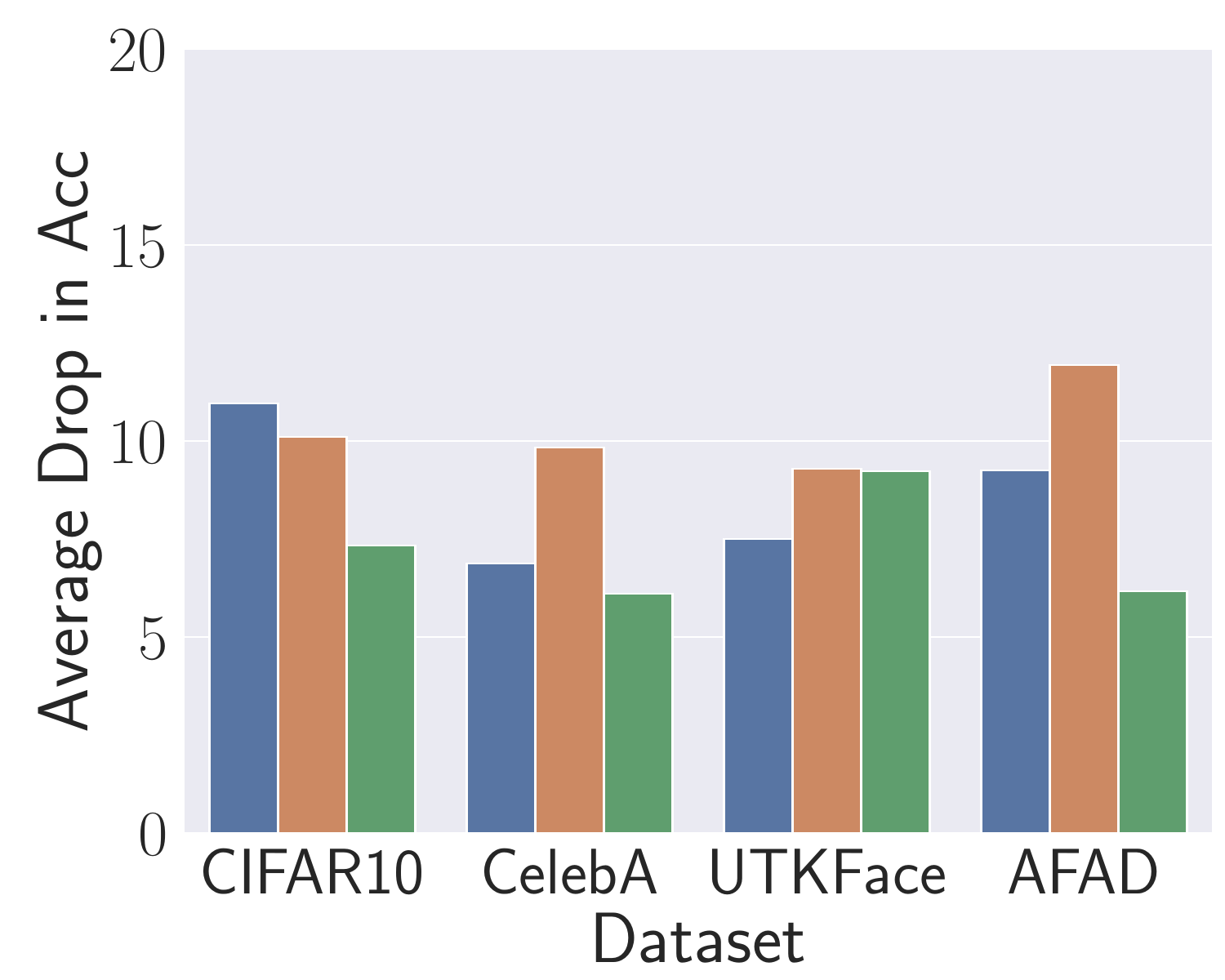}
\caption{Drop in Accuracy}
\label{figure:mia_diff_m_effect_average_drop_acc}
\end{subfigure}
\caption{Average test accuracy and drop in accuracy of three attacks after relaxing the pre-trained model assumption.}
\label{figure:mia_diff_m_effect_average}
\end{figure}

We present the average test accuracy and drop in accuracy of three attacks on different pre-trained models in~\autoref{figure:mia_diff_d_effect_average}.
We calculate these values based on the heatmap in~\autoref{figure:diff_dataset}.
Specifically, for each heatmap, we take the average of all values as the average test accuracy for each attack methodology on a pre-trained model.
We calculate the difference between each cell value and the diagonal value in the corresponding column and take the average as the average drop in accuracy.
Basically, a lower drop accuracy value means a more robust attack when relaxing the dataset assumption.
As illustrated in~\autoref{figure:mia_diff_d_effect_average_test_acc}, \metric and \grad attacks achieve the best attack performance on average.
Meanwhile, as shown in~\autoref{figure:mia_diff_d_effect_average_drop_acc}, we observe that the average drop in accuracy is smaller than 5.00\% in most cases.
The \trad and \grad attacks, in general, are more robust than \metric attacks.
We also present the average test accuracy and drop in accuracy of three attacks on different datasets in~\autoref{figure:mia_diff_m_effect_average}.
The \grad and \metric attacks achieve the best attack performance, and the \grad attacks are more robust than the \metric attacks.

\end{document}